\documentclass[11pt,a4paper,twocolumn]{article}
\pdfoutput=1
\usepackage[utf8]{inputenc}
\usepackage[margin={2.6cm,2.5cm}]{geometry}
\usepackage{color}
\usepackage[dvipsnames]{xcolor}
\usepackage{hyperref}
\usepackage{tikz}
\usepackage{eso-pic}
\usepackage[toc,page]{appendix}
\hypersetup{colorlinks=true,linktoc=all,linkcolor=Violet,citecolor=Blue,}
\hypersetup{linktocpage}
\usepackage{amsmath}
\usepackage{graphicx}
\usepackage{transparent}
\usepackage{lipsum}
\usepackage{subcaption}
\usepackage{listings}
\graphicspath{ {C:/Users/Luke/Downloads/mphys/} }
\usepackage{amsfonts}
\usepackage{amssymb}
\AddToShipoutPicture*{
	\put(0,0){
		\parbox[b][\paperheight]{\paperwidth}{%
			\vfill
			\centering
			{\transparent{0.15}\includegraphics[scale=1.5]{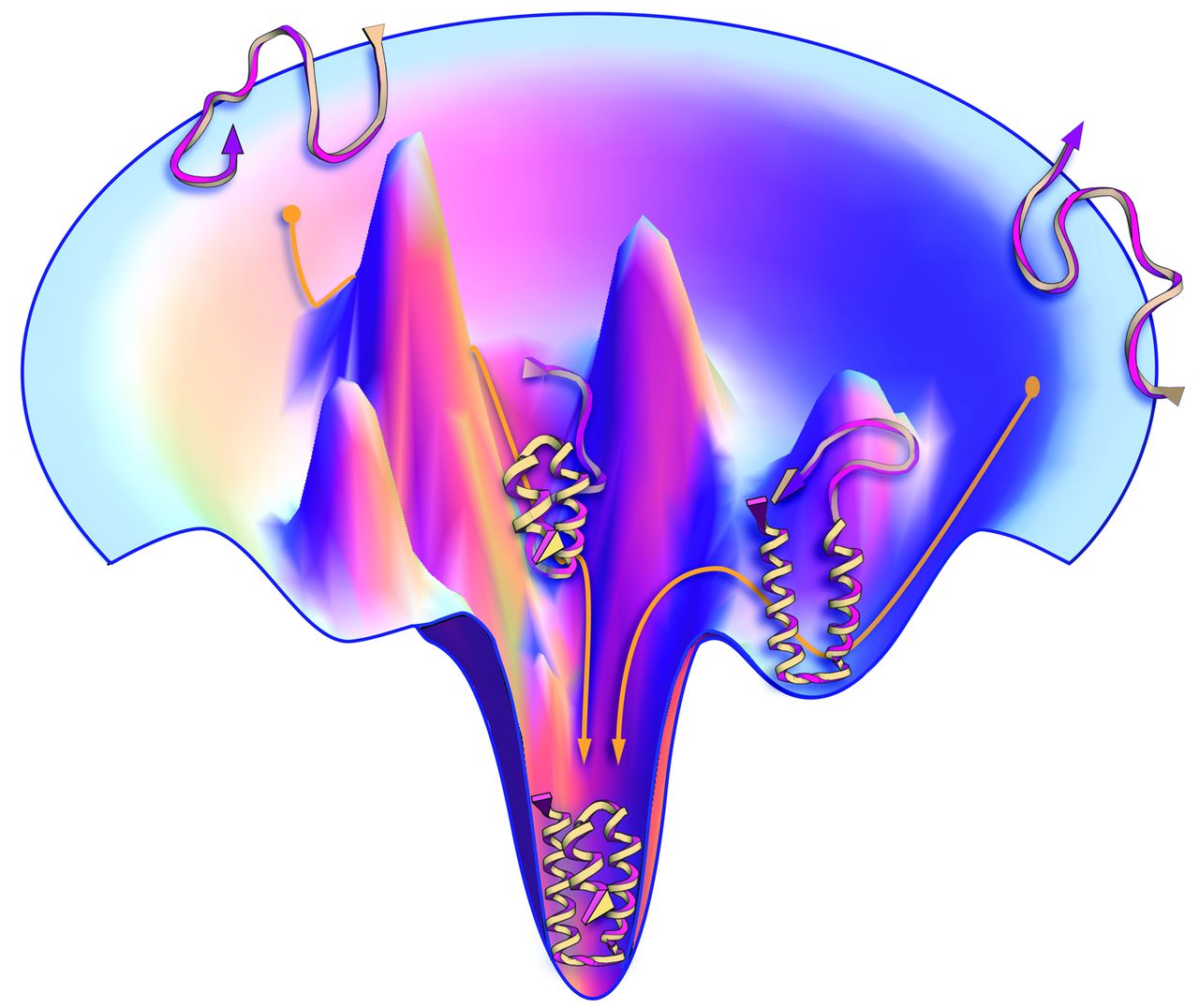}}
			\vfill
		}
	}
}

\definecolor{mygreen}{rgb}{0,0.6,0}
\definecolor{mygray}{rgb}{0.5,0.5,0.5}
\definecolor{mymauve}{rgb}{0.58,0,0.82}

\newcommand*{\titleGP}{\begingroup 
\centering 
\vspace*{\baselineskip} 

\rule{\textwidth}{1.6pt}\vspace*{-\baselineskip}\vspace*{2pt} 
\rule{\textwidth}{0.4pt}\\[\baselineskip] 

{\LARGE \textbf{A Parallel Trajectory Swapping \\ Wang - Landau Study Of The HP Protein Model}}\\[0.2\baselineskip] 

\rule{\textwidth}{0.4pt}\vspace*{-\baselineskip}\vspace{3.2pt} 
\rule{\textwidth}{1.6pt}\\[\baselineskip] 

\scshape 
A computational approach for investigating lattice polymers \\ 
and the thermodynamics of protein folding \\[\baselineskip] 
Department of Physics Swansea, Wales\par 

\vspace*{2\baselineskip} 

Written by \\[\baselineskip]
{\LARGE {\textbf{{Luke Kristopher Davis}}}\par} 
{\itshape  \textbf{Supervisor}: Professor Biagio Lucini \\ \par} 

\vspace*{3\baselineskip}

\vspace*{6\baselineskip}

{\scshape \textit{2016}} \\[0.3\baselineskip] 
For the \large MPHYS programme\par 
\endgroup}

\begin{document}
\onecolumn

\titleGP

\begin{center}
\includegraphics[scale=0.2]{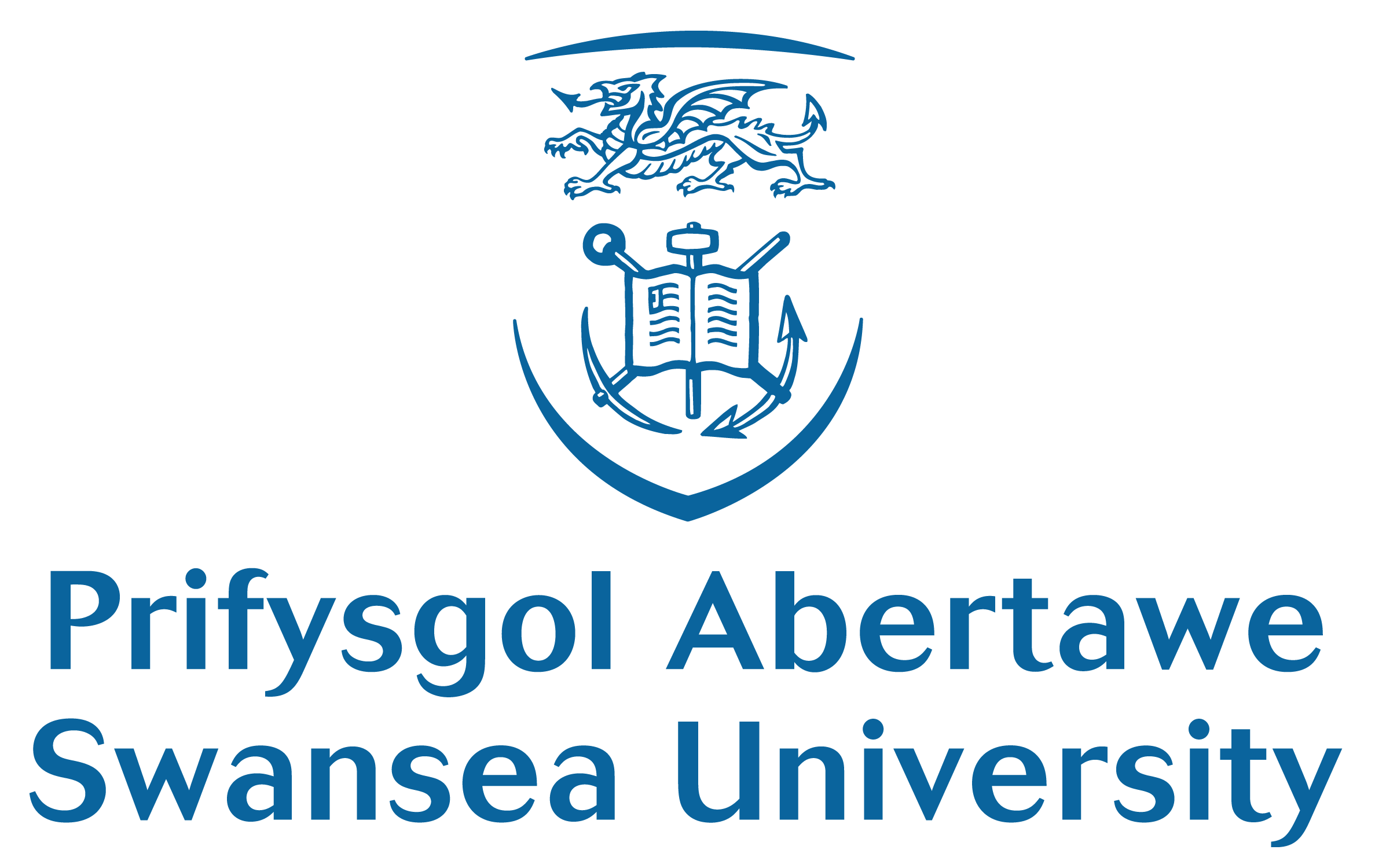}
\end{center}

\vspace*{\fill}
\begingroup

\centering
\par 
\par 
\par
\clearpage

\vspace*{\fill}
\begin{center}
\textbf{"Everything should be made as simple as possible, but not simpler."} \newline -\textit{Albert Einstein}
\end{center}
\AddToShipoutPicture*{
	\put(0,0){
		\parbox[b][\paperheight]{\paperwidth}{%
			\vfill
			\centering
			{\transparent{0.3}\includegraphics[scale=1]{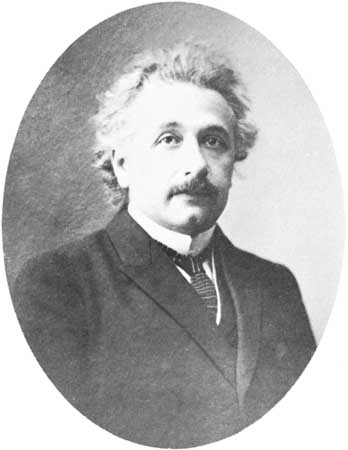}}%
			\vfill
		}
	}
}
\endgroup
\vspace*{\fill}
\clearpage
\vspace*{6\baselineskip}

\begin{center}
\begin{abstract}

The HP model of protein folding, where the chain exists in a free medium, is investigated using a parallel Monte Carlo scheme based upon Wang-Landau sampling. Expanding on the recent work of Wust and Landau \cite{oplandau} \cite{ISAW} by introducing a lesser known replica -exchange scheme between individual Wang- Landau samplers, the problem of dynamical trapping (spiking in the density of states) was avoided and an enhancement in the efficiency of traversing configuration space was obtained. Highlighting dynamical trapping as an issue for lattice polymer simulations for increasing lengths is explicitly done here for the first time. The 1/t scheme is also integrated within this sophisticated Monte Carlo methodology.

 A trial move set was developed which includes pull, bond re-bridging, pivot, kink-flip and a newly invented and implemented \textit{fragment random walk} move which allowed rapid exploration of high and low temperature configurations. A native state search was conducted leading to the attainment of the native states of the benchmark sequences of 2D50 (-21), 2D60 (-36)and 2D64 (-42), whilst attaining minimum energies close to the native state for 2D85(-52 NATIVE= -53), 2D100a (-47 NATIVE= -48) and 2D100b (-49 NATIVE=-50).

Thermodynamic observables such as $C_{V}/N$, $U/N$, $S/N$ and $F/N$ were computed for 2D benchmark sequences and folding and unfolding behaviour was investigated. Lattice polymers with monomeric hydrophobic structure were also studied in the same manner with the recording of minimum energy values and thermodynamic behaviour. The native results for the benchmark sequences and lattice polymers were compared with varying computational methods.

\end{abstract}
\rule{\textwidth}{1pt}

\textbf{Keywords:} HP model, Monte Carlo, Wang-Landau, 1/t, trajectory swapping, protein folding, lattice polymer, thermodynamics, biophysics, dynamical trapping, ISAW, fragment random walk.
\end{center}
\vfill

\onecolumn

\clearpage
\tableofcontents
\newpage
\onecolumn
\listoffigures
\newpage
\listoftables
\newpage

\section{Introduction}
\label{sec:intro}

Evolution has, through billions of years of selection mechanisms, formed a vast array of living organisms on this planet \cite{selfish}. These organisms have intricate internal machinery which allows them to persist through time and compete to pass their genetic information on to the next generation. Proteins are the main workers in all living organisms which fuel this internal machinery.

Proteins are the building blocks of cells and they also perform nearly all the cell's functions. For instance, enzymes provide the molecular surfaces in a cell that promote its multitude of chemical reactions \cite{molbio}. Some proteins send messages from one cell to another which is vital for large scale cellular activity. Yet others act as tiny molecular machines with moving parts \cite{molbio} \textit{kinesin}, for example, allows organelles to travel through the cytoplasm via propulsion; also \textit{topoisomerase} can unravel knotted DNA molecules.

The physics of proteins, ranging from folding mechanisms to calculations of specific binding energies of ligands, has developed rapidly over the last 50 years \cite{probdill} and is of great interest to computational physicists and those from a statistical mechanical background. The development of the HP model of proteins devised by K.Dill \cite{HPdill}, outlined in \ref{sec:hpmodel}, which provides a simple 'Ising-like' model  has enabled scientists to use computational techniques to explore the global transitions of proteins into their native state. \par It is imperative that scientists build a solid and comprehensive understanding of proteins in order to paint a complete picture of the mechanisms of life.

\subsection{Protein Structure and Function}

Proteins are chains which contain sequences of amino acids which, when one considers the protein as a polymer, act as the repeating subunits known as monomers. These monomers connect to one another via a peptide bond. There are 20 known amino acid bases which form an alphabet (see Appendix \ref{sec:A}) and the sequence of a protein chain consists of elements within this alphabet. An amino acid is a chemical group which is defined by its chain residue, differing residues have different chemical properties. 

\begin{figure}[h]
\begin{center}
\fbox{\includegraphics[scale=0.75]{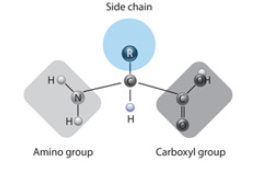}}
\caption{What defines the amino acid is its side chain (blue). (\textit{Thanks go to Nature Education in Protein Structure})}
\label{fig:amino}
\end{center}
\end{figure}

These small organic molecules (amino acids) consist of an alpha carbon atom connected to an amino group, a carboxyl group, a hydrogen atom and a variable side chain as shown in figure \ref{fig:amino}. 

Residues can be hydrophobic or polar (hydrophilic) (or degrees of both) \footnote{Hydrophobic effect arises from the fact that water molecules seek to form hydrogen bonds with each other and push non-polar material away to form these bonds. Polar material can form hydrogen bonds \cite{huangpro}}, vary in size and charge \cite{clote}.

The structure of a protein can be described in the following hierarchical way; \textbf{primary structure}: the sequence of amino acid bases, \textbf{secondary structure}: local formation of $\alpha$ helices and $\beta$ sheets, \textbf{tertiary structure}: typically the 3 dimensional structure of a protein domain in the native structure which is more irregular than the secondary structures \cite{huangpro} and the \textbf{quaternary structure}: the 3 dimensional, native structure of the fully functional protein \cite{clote}.

It is known that the sequence of amino acids determine the three dimensional structure of the protein which affects how it interacts with other molecules \cite{clote} \cite{huangpro}\cite{molbio}. A protein molecules physical interaction with other molecules determines its biological function \cite{molbio}. For example, antibodies in the human immune system recognize antigens by having a complementary surface to that of the antigen \cite{clote}. Also the enzyme \textit{hexokinase} binds glucose and ATP so as to catalyze a reaction between them.

All proteins bind to other molecules, where in some cases the binding is strong and in others weak. The binding always shows great specificity.

\subsubsection*{Hydrophobic Effect}
\label{sec:hydro}

\begin{figure}[h]
\begin{center}
\fbox{\includegraphics[scale=0.3]{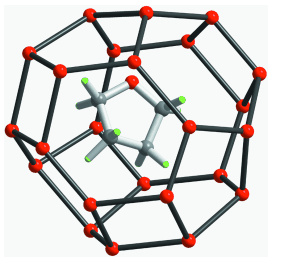}}
\caption{The hydrophobic material is surrounded by a three dimensional cage of bonded water molecules called clathrate structures.}
\label{fig:cage}
\end{center}
\end{figure}

It has been mentioned that a subset of amino acid bases are hydrophobic, which means they cannot form bonds with the surrounding water molecules, hence the water molecules prefer to bond with themselves and the hydrophilic bases. The water pushes these hydrophobic bases away in order to form these preferred bonds. This pushing is a direct consequence of the water molecules forming an ice-like structure around the hydrophobic material which drives the protein into a compact structure ( \textit{See} figure \ref{fig:cage}).

\subsubsection*{Local Forces}
\label{sec:forces}
There are forces which occur between the atoms and molecules of proteins and hence should, in principal, physically play a role in the folding mechanism and stability of the structure. \textit{Covalent bonds} involve the sharing of two electrons between the interacting partners. For example, the $H_{2}O$ water molecule has its hydrogen atoms bound to the oxygen atom via covalent bonding. The bonding leaves $H^{+}$ with an excess positive charge, and $O^{-}$ with an excess negative charge. \par \textit{Hydrogen bonds} involve sharing an H atom between the interacting partners. The bond has polarity with H covalently bonded to one partner and more weakly attached to the other through its excess charge \cite{huangpro}. \textit{Ionic bonds} arise from the exchange of one electron. There are also \textit{Van Der Waals} interactions which arise from temporary mutual electric polarization. Since molecules can have charge they must interact through the coulomb potential which is screened by the surrounding aqueous solution.

\subsection{One or Many Driving Forces?}

It was accepted that the mechanism of protein folding was a sum of the contributions of different local interactions as briefly described in section \ref{sec:forces}. The prevailing paradigm of the folding sequence asserted that the primary structure encoded the secondary structure which then determined the tertiary structure \cite{anfinsenpri}.

Sophisticated statistical mechanical simulations have unearthed a new view on the dominant driving component of protein folding. Making varying use of the HP model for proteins these simulations show that the hydrophobic effect outlined in section \ref{sec:hydro} is the dominant driving force and its effects, while non-specific in nature, are felt locally and non-locally in the sequence \cite{oplandau} \cite{domforces}.

\begin{figure}[h]
\begin{center}
\fbox{\includegraphics[scale=0.6]{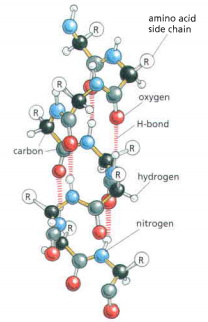}}
\caption{A diagram showing the basic structure and atomic composition of an alpha helix with H-bonds highlighted. \cite{molbio}}
\label{fig:alpha}
\end{center}
\end{figure}

Electrostatic interactions among the charged side chains are not likely to dominate the folding process. This is because most proteins have few charged residues which are concentrated in high-dielectric regions on the protein surface \cite{probdill}. Hydrogen bonding is a key element in the formation of the secondary structures in a protein state, for example hydrogen bonds between oxygen and hydrogen help to form the $\alpha$ helix (\textit{see} figure \ref{fig:alpha}). Also when the protein becomes increasingly more compact, Van der Waals interactions described in section \ref{sec:forces} play a significant role \cite{vanderwaals}.

However there is greater interest and importance attached to finding the dominant factor which distinguishes how two separate proteins fold into distinct native structures. There is considerable evidence, experimental and computational, that shows that the hydrophobic effect is the dominant driving force for the folding of proteins. 

For example model compound studies show 1-2 kcal/mol for transferring a hydrophobic side chain from water into oil-like media and there are a significant amount of them \cite{wolf} \cite{probdill}.

Sequences that keep there HP sequence but have there amino acids jumbled fold to their respective native conformations without the need to tamper with local interactions \cite{probdill} [and references therein].

\begin{figure}[h]
\begin{center}
\fbox{\includegraphics[scale=0.35]{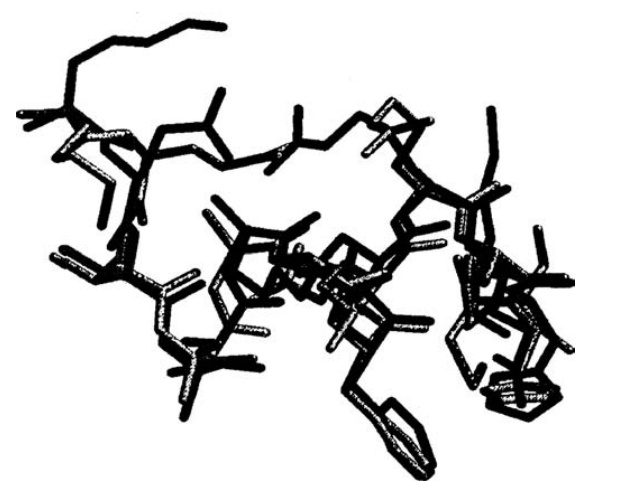}}
\caption{Tertiary structure determined by the lowest energy state computed in a multicanonical Monte Carlo study (black) superposed with structure found from X-ray crystallography (grey). \cite{hans}}
\label{fig:montefit}
\end{center}
\end{figure}

Also computational simulations using the HP model have reproduced tertiary structures of proteins very well, for example the tertiary structure of the C-peptide of ribonuclease A (\textit{see} figure \ref{fig:montefit}) \cite{landaubinder}. 

For free energy and thermodynamic calculations on simple HP models on square 3D lattices have also proven very successful \cite{oplandau} \cite{landaubinder}. Hence simulations and empirical investigations focusing on the hydrophobic nature of the bases to probe the global behaviour of folding are well founded.

\subsubsection*{Stabilization of secondary structures}

Studies of proteins, namely of the lattice  and tube variety, have revealed that secondary structures of the protein are stabilized due to the compactness of the conformation which is a direct consequence of the hydrophobic effect in action \cite{probdill}. 

\subsubsection*{Folding Into The Native State}

Proteins fold into conformations which minimize the entropy, they are guided into this structure by the non-local hydrophobic force and the secondary, tertiary and quaternary structures were thought to be stabilized solely by local hydrogen bonding \cite{molbio}. However, it has been argued \cite{probdill} that the secondary structures become more stabilized as the protein forms a tighter conformation and hence the tertiary structure directly controls this.

\begin{figure}[h]
\begin{center}
\fbox{\includegraphics[scale=0.4]{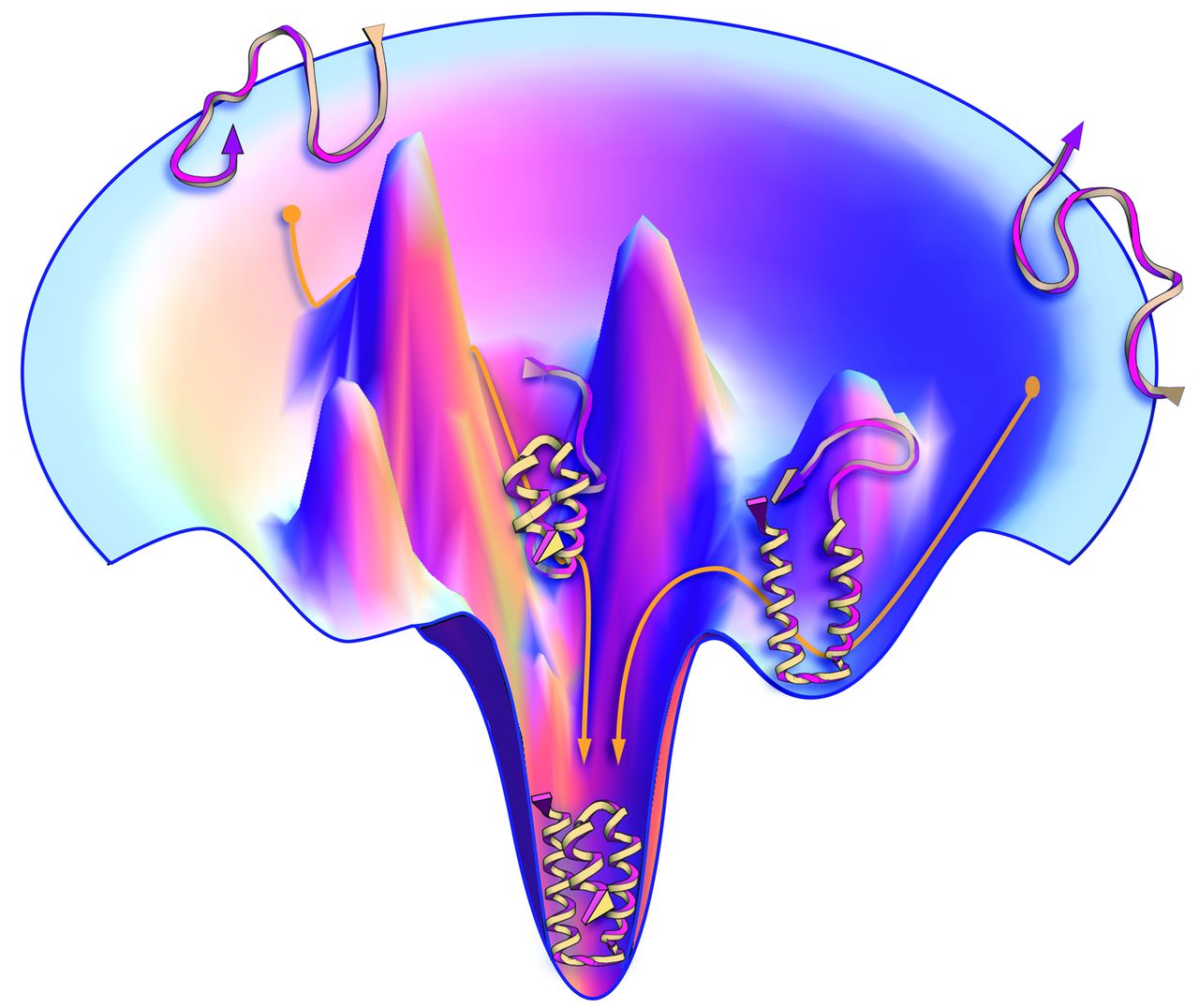}}
\caption{The protein traverses the energy landscape attaining a more stabilized structure as it ventures towards the global native minimum. Note the 'string-like' configurations at higher energies and the compact native structure at the minimum. }
\label{fig:landscape}
\end{center}
\end{figure}

The energy landscape of proteins is normally rough and complex, hence there is no absolute minimum but rather a group of minima or constraint minimum which define the preferable conformations of proteins \cite{huangpro}. As the protein changes conformation its energy states glide along the energy landscape and are guided towards the most stable state (\textit{see} figure \ref{fig:landscape}). The protein is encouraged into this native state due to the aqueous solution surrounding it which sparks the hydrophobic force to act. 
Each protein normally folds up into a single stable conformation. Due to interactions with other molecules in the cell the native conformation changes. These changes in structure however are usually crucial to how the protein functions.  
 
\subsection{The Protein Folding Problem}
\label{sec:pfp}

The birth of a protein folding problem arose, in 1961, from the experimental results of Anfinsen on ribonuclease \cite{anfinsen}. The conclusions drawn from these results show that the sequence of amino acids is enough information to dictate the native conformation of the protein in a specific solution. The native structure is then indifferent to how the polypeptide chain is synthesized in the first place, say if it was synthesized on a ribosome or in a test tube \cite{probdill}. 

This spurred biologists and biochemists to conduct experimental work on the amino acid sequence in light of the fact that a protein in a test tube environment could convincingly replicate its behaviour in an organism (there are rare exceptions for example see \cite{probdill}). 

Following these results one can then ask: what thermodynamic and kinetic interactions take the string like protein with its amino acid sequence to a compact native structure? Various approaches, experimental and computational, have been devised to tackle this question. For example NMR (nuclear magnetic resonance) imaging is used to probe the details of folding and misfolding \footnote{ Misfolding of proteins occurs when the protein does not find its most thermodynamically stable state but is fixed in its partially folded state by thermodynamic means or interactions with other molecules. The misfolding of proteins leads to modified functionality which can be toxic \cite{misfolding}. } , allowing characterization of the molecular structure and dynamics of folding. \par Molecular dynamics simulations, which invoke various force fields to mimic inter-atomic forces (\text{see \ref{sec:forces}}) and Monte Carlo methods are recently proving successful in characterising properties of folding \cite{oplandau} \cite{accmd}. 

However there is no unified framework, analytical or computational, which can adequately describe and explain the complete mechanisms of protein folding. 

It is of my opinion that there are essentially two problems that the scientific community needs to address in order to form a complete understanding of protein folding. 
\newline
\newline
\textbf{Problem A :} What are the \textit{detailed} physical mechanisms, atomistic and statistical mechanical, of protein folding? 
\newline
\newline
\textbf{Problem B :} Can we efficiently and consistently predict, via computational simulation, the native structure of any amino acid sequence?
\newline
\newline
Problem A appeals to our desire to understand the basic mechanisms of nature whilst problem B is more focused on medical, biological and pharmaceutical applications. A solution to problem B in the form of a universal computer program (UCP) could allow the quick prediction of new proteins from amino acid sequences which are not found in nature. This could lead to artificial proteins bettering those carved out by natural selection and further advance the battle with dominant diseases. 

It is obvious that these two problems are not independent and that success in problem A will further success in problem B and vice versa. As resources and techniques in high performance computing have, without doubt, increased in power we will no doubt see rapid progress on these problems. 

\subsection{Introduction to the HP Model}
\label{sec:hpmodel}

Inspired by the accumulation of evidence affirming that the hydrophobic effect is the globally dominant driving force in the folding process for globular proteins K.A Dill proposed a simplified model to characterize this striking behaviour (\textit{a reveiw is given by \cite{HPdill}}). He proposed that the alphabet of 20 amino acids should each be labelled as 'hydrophobic' or 'polar' (\textit{see} appendix \ref{sec:A} for a conversion table for all the amino acids). Then the monomeric sequence of the protein is a sequence of (H)'s and (P)'s. 

There are four simple rules for a HP protein:

\begin{enumerate}
\itemsep0em
\item Monomers have uniform size.
\item The peptide bond length between monomers is uniform.
\item Positions of the monomers are restricted to positions on a regular lattice.
\item No two monomers can occupy the same position and overlapping of bonds is forbidden.
\end{enumerate}

There are only nearest-neighbour interactions and there is an attractive potential $\epsilon_{HH}$ between two H monomers which are topologically connected i.e. not direct neighbours on the sequence. 

Hence the Hamiltonian of this simple system is given by:
\begin{equation}
\label{eq:Hamiltonian}
H = -\epsilon_{HH} n_{HH}
\end{equation}

\begin{figure}[h]
\begin{center}
\fbox{\includegraphics[scale=0.4]{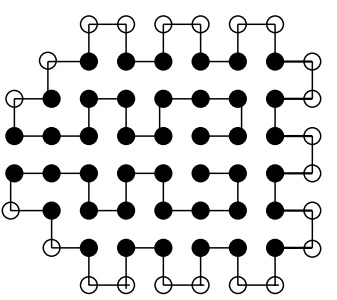}}
\caption{A 2D native state of the protein sequence S1-8 (2D64) in the 2D HP model found by the ACO method. Black beads are hydrophobic amino acids and white beads are polar. Thick black lines represent the covalent bonds between sequence amino acids. \cite{ACO}}
\label{fig:hpexample}
\end{center}
\end{figure}

Where $n_{HH}$ is the number of nearest neighbour topologically connected H-H monomers. The general energy function has values obeying table \ref{table:energy}.

\begin{table}[h]
\centering
\begin{tabular}{c|c|c}
  & H & P \\
 \hline
 H & -$\epsilon_{HH}$ & 0 \\
 \hline
 P & 0 & 0 \\
\end{tabular}
\caption{Contributions to the energy.\label{table:energy}}
\end{table}

A native state is a conformation having a minimum Hamiltonian energy value. Despite the simplicity of the model it is difficult, for long chain lengths, to compute the lowest possible energy for the folded chain \cite{wust}.

In 2D and more especially in 3D one of the drawbacks of this model is that it is highly degenerate. This is emphasized for long sequences of protein chains. The degeneracy is normally very low in the low temperature ranges \cite{oplandau}.
\subsubsection{Introduction to The Self-Avoiding Walk}

The conditions of the protein chain in the HP model outlined in section \ref{sec:hpmodel} exactly match the conditions of a length conserving self-avoiding walk (LCSAW)\cite{huangpro}\cite{oplandau}\cite{wust}\cite{clote}.

In general, A self-avoiding walk is a path on a lattice that does not visit the same site more than once \cite{madras}. It sits on an undirected graph which is a collection of points, with a collection of pairs of points known as \textit{edges}. The basic undirected graph which is used here and in the literature is the \textit{d-dimensional hypercubic lattice} \textbf{$Z^d$}. The points of this lattice are of the \textit{d}-dimensional Euclidean space \textbf{$R^d$} in which all the components are all integers, and the edges are given by the set of all unit line segments as \textit{nearest-neighbour bonds}. The LCSAW is defined formally as follows:
\newline
\begin{description}
\item \textbf{Definition (LCSAW)} Let d $\geqslant$ 1. An \textit{n} - step \textit{self-avoiding walk} from x   $\in$ $Z^d$ to y $\in$ $Z^d$ is a map \textit{w}: [0,\textit{n}] $\rightarrow$ $\in$ $Z^d$ with:
\begin{enumerate}
\item \textit{w}(0) = x and \textit{w}(\textit{n}) = y
\item $|\textit{w}(i+1) - \textit{w}(i)|$ = 1 (unit length bonds)
\item $\forall$ i, j $\in$ [0,\textit{n}], i $\neq$ j $\Rightarrow$ \textit{w}(i) $\neq$ \textit{w}(j)
\item $|\textit{w}|$ is a constant.
\end{enumerate}
\end{description}

The main idea here is that the movements of the protein chain respect the self-avoiding conditions and each move which obeys these generates a new LCSAW. 

While physicists and biologists using the HP model merely make use of the properties of SAWs and algorithms for move sets on them, the mathematics behind SAWs is rich and contains many open basic questions. For a rigorous introduction and overview see Madras and Slade \cite{madras}.

\subsubsection*{NP Completeness}

\textbf{Problem B}, which is the problem of predicting the native conformation of a protein chain defined by a sequence of amino acids, can be stated formally as a combinatorial optimization problem in the HP model \cite{geometric}: 

\begin{description}
\item \textbf{Optimal Folding Problem}: Given a sequence of (H)'s and (P)'s, find a LCSAW on the 2D or 3D lattice which maximises $n_{HH}$ (the number of H-H contacts).
\end{description}

It has been proven that this problem is NP - complete in 2D and 3D \cite{NP}. Their proof revolves around asking whether the graph representing the HP model contains a Hamiltonian cycle. This means that finding the conformation which minimizes the energy cannot be done in polynomial time.

The proof, although initially far removed from proteins, does emphasize the need for the protein chain to form a compact cubic shape in 3D \cite{NP}.

There is an interesting question whether all HP models on all lattices are NP-complete, since it has not been formally shown that the problem is NP-complete for triangular or non-square lattices and that different computational methods may affect the exact nature of the problem \cite{NP?}. 

\subsubsection*{The need for computational studies}

Recalling the two specific sub-problems associated with the general protein folding problem stated in \ref{sec:pfp} it appears that both of them cannot be solved analytically. Since finding native conformations has been mathematically proven to be NP-hard and that the complicated nature of atomistic dynamics seems to evade purely pen and paper advances, it seems inevitable that the community will need to make use of sophisticated computational simulations.

This not only means carefully constructed models, simulation techniques and software but advances in computational hardware are needed to meet the computational demands.

Also as protein folding and protein structure prediction are interdisciplinary areas of study, it is necessary for those using computational techniques to directly work in unison with those conducting experiments.

\subsection{What has been done already? Computational}
\subsubsection*{Molecular Dynamics}

Molecular dynamics (MD) concerns itself with simulating the physical system by keeping track of all the coordinates of the constituent particles. The system then evolves in time obeying equations of motion (usually Newtonian) which are integrated numerically. Simulating a protein molecule, which is a macroscopic system relative to simple  atomic systems, is extremely computationally demanding if one uses atoms as the basic constituent particles. Hence coarse grained approaches are used. For example the Go model is a popular coarse graining approach where the protein is represented as a chain of one-bead amino acids whose structure is biased toward the native configuration \cite{GO}.

An example of an established MD approach was proposed by Sugita and Okamoto \cite{sugita} which is a replica-exchange method for protein folding. The appeal of their approach was that it could overcome the multiple minima problem by exchanging non-interacting replicas of the system at several temperatures \cite{sugita}. Their insight was to take random walks in \textit{energy space} not \textit{probability space} by avoiding the use of Boltzmann weighting.

Their methodology for the replica exchange method consists of M non-interacting replicas of the original system (of N atoms) in the canonical ensemble at different temperatures. The replicas are arranged such that there is always exactly one copy of the system at each temperature and then there is a one-to-one correspondence between replica systems and temperatures \cite{sugita}.

However, even if their weighting is known \textit{a priori}, they still need to determine the optimal temperature distribution \cite{sugita}. Also the method, like any MD approach, is computationally demanding since it requires the simulation of many atomistic systems at a wide range of temperatures.

While computational power is on the increase there exists other regimes which are becoming more successful in protein structure prediction and folding, which are computationally cheaper. For example Monte Carlo methods are playing an increased role in these areas for which the explicit time dependence is not the ultimate goal \cite{landaubinder}.

\subsubsection*{Protein Threading}
\label{sec:threading}

Suppose we have a sequence \textit{s} of known structure, can we determine the structure of a sequence \textit{s$'$} that is homologous\footnote{\textit{Homologous}: having the same relation, relative position or structure} to \textit{s}? The fact that \textit{s} and \textit{s$'$} are homologous could be derived from experimental biological data or alignment distances \cite{clote}.

This is essentially the protein threading problem which relates to problem B in section \ref{sec:pfp}. The basic idea is to use the known structure of \textit{s} to guide the secondary and tertiary structure prediction for \textit{s$'$}.

It is another optimization problem and was shown to be NP-complete by R.H.Lathrop in 1994 \cite{lathrop}. 

Through the use of experimental data or a protein data bank such as RCSB PDB \footnote{www.rcsb.org/pdb/home/home.do} it is possible to construct a program which automatically searches this databank for homologous sequences to \textit{s$'$} and then predict its structure to some degree of error. There are many programs and methods for doing this for example see \cite{raptor} \cite{threading}.

While this approach has its successes it is not completely blind, as it depends on currently known structures, and hence is in some sense scientifically incomplete. It is my opinion that it is ultimately more satisfying to find and understand the mechanisms of the system and then use this knowledge to make predictions. 

\subsubsection*{Other Algorithmic Approaches}
\label{sec:otherapp}
The protein folding problem has attracted many computational approaches, some being very sophisticated. For example sequential importance sampling, PERM \footnote{Pruned-enriched Rosenbluth method.}\cite{nPERMIS} and other chain growth methods are in use in exploring the energy space of proteins in the HP model \cite{frmc}.

Also as the protein folding problem can be formed as an combinatorial optimization problem, ingenious and unexpected methods have come from the fields of mathematics and computer science \cite{oplandau}. Some of these include; genetic algorithms, ant colony models \cite{ACO} and constraint-based algorithms.

While these approaches, as with the protein threading paradigm, will advance our ability to predict the quaternary structure of proteins and help solve problem B (\textit{see} \ref{sec:pfp}), it is however not attacking the essence of the physical problem at hand. This physical problem being of a statistical mechanical nature. 

\subsubsection{The Work of Wust and Landau}

A very successful regime for the study of the HP model was presented to the arXiv community in 2012 by Thomas Wust and David P.Landau \cite{oplandau}. Their work is mentioned here as it describes the only generic and fully blind Monte Carlo sampling scheme that can reproduce all known ground state energies and bettering one (for 3D103)\cite{oplandau}. Their scheme also allows the computation of thermodynamic and structural quantities at any temperature such as the specific heat capacity and the radius of gyration \cite{oplandau}. Their approach has also proven powerful for exploring the low-temperature behaviour of the self avoiding proteins even for $N \gg 1000$ \cite{wustlandau}.

\begin{figure}[h]
\begin{center}
\fbox{\includegraphics[scale=0.6]{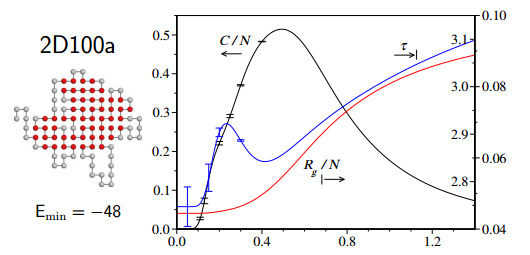}}
\caption{The computation of observables at different temperatures for sequence 2D100a. Specific heat capacity $C/N$ is shown in black (left ordinates), root mean squared radius of gyration $R_g / N$ is shown in red (outer right ordinates) and tortuosity $\tau$ is shown in blue (inner right ordinates).}
\label{fig:wlobserv}
\end{center}
\end{figure}

They use Wang Landau sampling described in section \ref{sec:method} and detailed in \cite{wanglandau} to compute the density of states. Since the density of states does not depend on temperature in the canonical ensemble (\textit{see} section \ref{sec:wlmc}) they were able to compute observables over the entire temperature range as shown in figure \ref{fig:wlobserv} \cite{oplandau}. 

Their implementation is continuously compared to its close competitors, nPERMis\footnote{'New' Pruned Enriched Rosenbluth Method 'Importance Sampling'} and FRESS\footnote{Fragment Re-growth via Energy-guided Sequential Sampling (FRESS)}, and seems to have beaten them in the quality of the ground state energies and in computational efficiency. However direct comparison with other algorithmic methods briefly outlined here \ref{sec:otherapp} was not done. 

Their general paradigm, and the specific conclusion that implementation of trial move sets is vital to a well-performing simulation of HP lattice proteins \cite{oplandau}, seems a valid foundation to focus ones attention on in venturing into this rapidly growing area of biophysical simulation. 

\subsection{Aims of this project}

The general aim of this project is to investigate the behaviour of heteropolymer chains with protein-like sequences in a computational manner. More specifically to compute thermodynamic observables which will give insight into the folding/un-folding behaviour of protein chains. 
To achieve the native states of typical 2D benchmark sequences and make a clear comparison with other methods.
Another aim is to achieve some sort of parallelism within the simulation, using a new or unknown scheme which has not been explicitly implemented to this particular problem.
I would also like to obtain broad knowledge of this niche field in quantitative biology and understand the various approaches used to obtain understanding of protein folding.
\clearpage
\section{Necessary Theory}
\label{section:theory}

\subsection{Statistical Mechanics}
\label{sec:statmech}

\subsubsection{Canonical Ensemble}
\label{sec:canonical}

The protein chain, which consists of connected particles (monomers), can be assumed to exist in an aqueous solution which acts as a heat reservoir \cite{huangpro}. So that the protein chain is a small subsystem within the heat reservoir. Since the particles of the protein chain remain part of the subsystem we can use canonical ensemble theory to describe it statistical mechanically.

Therefore the protein chain is described by an ensemble of fixed temperature instead of fixed energy, since it exchanges energy with the solution around it.

Let the labels $1$ and $2$ denote the protein subsystem and the heat reservoir respectively. Working in the micro-canonical ensemble for the whole system, the total number of particles and energy are simply the sums of those in the system $1$ and $2$:

\begin{equation}
N = N_{1} + N_{2}
\end{equation}
\begin{center}
$E= E_{1} + E_{2}$
\end{center}

where

\begin{equation}
\label{eq:large}
N_{2} \gg N_{1}
\end{equation}
\begin{center}
$E_{2} \gg E_{1}$
\end{center}

It is reasonable to assume that both systems are macroscopically large and that $N_{1}$ and $N_{2}$ remain fixed. The energies $E_1$ and $E_2$ fluctuate because the boundaries between the two subsystems allow energy exchange.

The goal is to find the phase-space density $\rho_{1}(s_{1})$ for system 1 in its own phase space. It is directly proportional to the probability of finding system 1 in the state $s_{1}$ with no regard of the state of system 2. We see that it is proportional to the phase-space volume of system 2 in its own phase space with an energy $E_2$. Taking the proportionality constant $= 1$ we have:

\begin{equation}
\rho_{1}(s_{1}) = \Gamma_{2}(E_{2}) \equiv \Gamma_{2}(E - E{1})
\end{equation}

Since equation \ref{eq:large} is assumed it is appropriate to Taylor expand the entropy of system 2 for small $E_1$:

\begin{equation}
k_B \cdot ln [\Gamma_{2}(E-E_{1})] \approx S_{2}(E) - \frac{E_{1}}{T}
\end{equation}

where $k_B$ is Boltzmann's constant and $T$ is the temperature of system 2. In the thermodynamic limit for system 2 i.e. $N_{2} \rightarrow \infty$ the density function for system 1 becomes:

\begin{equation}
\label{eq:densityfunc}
\rho_{1}(s_{1}) = exp[S_{2}(E)/k_B] \cdot exp[-E_{1}/(k_B T)]
\end{equation}

The first factor in equation \ref{eq:densityfunc} is a constant and hence can be omitted after a normalization procedure. The energy of system 1, $E_1$, can be replaced by the Hamiltonian for the protein chain in the HP model using equation \ref{eq:Hamiltonian}:

\begin{equation}
E_{1} = H_{1}(s_{1})
\end{equation}

Therefore, omitting indices since system 2 is no longer relevant, the Boltzmann weight for a system at temperature $T$ is

\begin{equation}
\rho(s)= exp[-H(s)/(k_{B} T)]
\end{equation}

which defines the \textit{canonical ensemble}.

The partition function, with which all thermodynamic quantities can be derived from, is introduced as:

\begin{equation}
\label{eq:partition}
Z \equiv \sum_{s} exp[-H(s)/(k_{B} T)]
\end{equation}

where the summation is over all states. The partition function can also be written as:

\begin{equation}
\label{eq:partitionWL}
Z \equiv \sum_{E} g(E) \cdot exp[-E/(k_{B} T)]
\end{equation}
 
where the sum runs over all energy values and $g(E)$ is the density of states. Here $E$ is the energy value of the protein chain computed via equation  \ref{eq:Hamiltonian}.
\subsubsection{\textcolor{black}{Energy Fluctuations and Observables}}
\label{sec:energyfluc}

Let $U$ be the mean internal energy of system 1 (the protein) which is given by the ensemble average of the Hamiltonian. Using equation \ref{eq:partitionWL} as the partition function $U$ is

\begin{equation}
\label{eq:internalenergy}
U = \frac{\sum_{E} E \cdot g(E) \cdot exp[-E \beta]}{Z} \equiv \frac{\sum_{E} E \cdot g(E) \cdot exp[-E \beta]}{\sum_{E} g(E) \cdot exp[-E \beta]}
\end{equation}

where $\beta = \frac{1}{k_{B} T}$. Differentiating $U$ w.r.t $\beta$ we obtain:

\begin{equation}
\frac{\partial U}{\partial \beta} = - \frac{\sum_E E^{2} \cdot g(E) \cdot exp[-E \beta]}{\sum_E g(E) \cdot exp[-E \beta]} + \frac{(\sum_{E} E \cdot g(E) \cdot exp[-E \beta])^2}{(\sum_{E} g(E) \cdot exp[-E \beta])^2}
\end{equation}

changing variables:

\begin{equation}
\frac{\partial U}{\partial \beta} = \frac{\partial U}{\partial T} \cdot \frac{\partial T}{\partial \beta} = -k_{B} \cdot T^{2} \cdot \frac{\partial U}{\partial T}
\end{equation}

where we recognise that the last partial derivative w.r.t $T$ can be replaced with, $C_{V}$, the specific heat capacity. 

So $C_{V}$ is expressed as:

\begin{equation}
\label{eq:specificheat}
C_{V} = (\frac{\sum_E E^{2} \cdot g(E) \cdot exp[-E \beta]}{\sum_E g(E) \cdot exp[-E \beta]} - \frac{(\sum_{E} E \cdot g(E) \cdot exp[-E \beta])^2}{(\sum_{E} g(E) \cdot exp[-E \beta])^2}) \cdot \frac{1}{k_{B} T^{2}}
\end{equation}

The free energy $F(T)$ is defined as:

\begin{equation}
F(T)= -k_{B} \cdot T \cdot ln[Z]
\end{equation}

where $Z$ is the canonical partition function. Therefore to compute the free energy within the simulation one notes the more explicit form:

\begin{equation}
\label{eq:freeenergy}
F(T)= - k_{B} \cdot T \cdot ln[\sum_{E} g(E) \cdot  exp[-E\beta]]
\end{equation}

Then the entropy, $S(T)$ is then easily computed as:

\begin{equation}
S(T)= \frac{U(T)-F(T)}{T}
\end{equation}
\subsection{Probability Theory}
\label{sec:probtheory}
\subsubsection{Markov Chains} 
\label{sec:MarkovChains}

If we let the process that evolves the system be a stochastic one, so at discrete times  $t_{1},t_{2},t_{3},......,$ the system is in a state $W_{t}$ at time t which belongs to the set of all possible states denoted $\lbrace S \rbrace$.
The conditional probability that $X_{t_{n}} = S_{i_{N}}$ is given by: 

\begin{equation}
P(X_{t_{n}} = S_{i_{n}} | X_{t_{n-1}} = S_{i_{n-1}}, X_{t_{n-2}} = S_{i_{n-2}},..., X_{t_{1}} = S_{i_{1}})
\end{equation}

assuming that the state of the system was, in the previous time, in state $S_{i_{n-1}}$. If the immediate state only depends on the preceding state i.e.

\begin{equation}
\label{eq:markov}
P(X_{t_{n}} = S_{i_{n}} | X_{t_{n-1}} = S_{i_{n-1}})
\end{equation}

the stochastic process is then a Markov process and the set of states $X_{t}$ is known as a Markov chain. Equation \ref{eq:markov} is also referred to as the transition probability to go from one state to the next. 

\subsubsection{Non- Markovian Schemes}
\label{sec:nonmarko}
As explained in \cite{nonmark} Markov processes are the exception. Most stochastic systems and simulation models are intrinsically non- Markovian. A Markovian system is one where the distributional functions are solely given in \ref{sec:MarkovChains}, however in general one needs a different mathematical scheme to define the distribution of states.

Since Wang Landau sampling \ref{sec:wlmc} has inherent history, through collected histograms and knowledge of old paths through energy space, it is a non- Markovian scheme (as noted in \cite{oplandau}).

\subsubsection{Ergodic Process}
In statistical theory a stochastic process is \textbf{ergodic} if its statistical properties can be deduced from a random sample of that process. The idea is that the random sampling of the process meaningfully represents the average statistical properties of the entire process \cite{ergodictheory}. 

\subsubsection{Ergodic Hypothesis}

In computational physics it is more practical to view ergodicity as the ability to sample all of configuration space \cite{landaubinder}. In the case of finite protein folding there does not exist the phenomenon of spontaneous symmetry breaking, so the entire phase space is reachable at all times. This means there will be no intrinsic ergodicity breaking. 

In relation to simulations, it is of utmost importance that the operations which evolve the system can in principle take it through all of phase space in a finite amount of time. Since polymer dynamics require specialised and non-trivial move algorithms (see section \ref{sec:trialmove}) it is a danger that the simulation becomes non-ergodic and yields incorrect statistical results.
\newpage
\section{Methodology}
\label{sec:method}
\subsection{Monte Carlo Methods}
\label{sec:wlmc}
\subsubsection*{Reminder of Metropolis}

Monte Carlo simulations are used extensively in science when the system at hand is sufficiently complex enough to be intractable analytically. The key to Monte Carlo simulation is to use sequences of random numbers to evolve the system or to sample integrals.

The workhorse of Monte Carlo simulations has been the Metropolis-Hastings importance sampling scheme, a good general review is given here \cite{methast} and for applications for statistical physics here \cite{landaubinder}.

The Metropolis scheme can be briefly stated as follows:

\rule{\textwidth}{1pt}
\begin{center}
\textcolor{blue}{\textbf{METROPOLIS SCHEME}}
\end{center}

\begin{enumerate}
\item Choose an initial state of the chain.
\item Propose a trial move selected at random from the set.
\item Compute the energy change $\delta E$ which results from the conformational change.
\item Generate a random number $ran$ where $0< ran < 1$.
\item If $ran <$  exp[$-\delta E \cdot \beta$]  accept the move.
\item Go to step 2 and repeat $n$ times. 
\end{enumerate}
\rule{\textwidth}{1pt}
\subsubsection{\textcolor{black}{Wang Landau Sampling}}

Contrary to the Metropolis Hastings scheme in which the acceptance criterion is based on the difference in energy via Boltzmann weighting, Wang- Landau sampling has its acceptance criterion based on the inverse of the density of states \cite{wanglandau}. 

Say, for example, a protein chain in configuration $a$ has some energy $E_{a}$ computed using equation \ref{eq:Hamiltonian}. If we make a move on the chain i.e. perturb its configuration such that it now has energy $E_b$  where the configuration has gone from $a \rightarrow b$. Moves are accepted according to the probability:

\begin{equation}
\label{eq:wlprob}
p(E_{a} \rightarrow E_{b}) = min(\frac{g(E_{a})}{g(E_{b})}, 1)
\end{equation}

We want to ultimately compute the canonical partition function as shown in equation \ref{eq:partitionWL}, which entails approximating the DOS $g(E)$. For equation \ref{eq:wlprob} to work, we start with a simple guess of the DOS at each discrete energy level. This is because $g(E)$ is not known a priori but it is possible to iteratively refine the initial guess such that it converges to the correct DOS for the system. 

Let the initial guess be simple i.e. $g_{0} (E) = 1$ for all $E_{1}, ..., E_{n}$. Then following each move, whether accepted or rejected, we update the DOS for the resultant energy level $E$ via:

\begin{equation}
\label{eq:refineDOS}
g(E) \rightarrow g(E)\cdot f_{i}
\end{equation}

The modification factor, $f$, is also modified according to a flatness criterion for the collected histogram of energies. The factor starts out as  $3 > f_{0} > 1$\footnote{In the literature (see \cite{oplandau} and \cite{wust}) normally $f_{0} = e^{1}$ however there is no systematic way to determine the most efficient starting modification factor. } and if the histogram is flat, up to some pre-determined standard, $f$ is reduced: $f_{n+1} = (f_{n})^{\frac{1}{2}}$. The histogram entries are then reset to zero and the process begins again but with a reduced modification factor.

The aim is to have $\lim_{f \rightarrow 1} g_{approx}(E) = g_{exact}(E)$. Since this limit converges it is appropriate to foster an accuracy cut-off for the modification factor. This can be chosen to be $f_{final} \approx e^{10^{-8}}$ \cite{oplandau}.

In this simulation the DOS spans many orders of magnitude and hence may lead to numeric overflow in the 'long double' data type in C/C++ (as happened during initial stages). This leads to '-nan' for the thermodynamic observables. It is preferable to work with the natural logarithm of the DOS where initially $log[g_{0}(E)] = 0$ and the update procedure is then:

\begin{equation}
\label{eq:logupdate}
log[g(E)] \rightarrow log[g(E) \cdot f] \equiv log[g(E)] + log[f]
\end{equation}

and it is still reasonable to keep reducing $f$ directly.

The detailed step-by-step Wang-Landau scheme for this simulation is as follows:
\rule{\textwidth}{1pt}
\begin{center}
\textcolor{red}{\textbf{WANG-LANDAU SCHEME}}
\end{center}

\begin{enumerate}
\item Set a pre-defined range of discrete energies (not too large to be cumbersome) that the protein may take.
\item Initialise: $X(E_{i}) = 0$, $H(E_{i})=0$ and $F = 1$ (where $X(E)$ and $F$ represents $log[g(E)]$ and $log[f]$ respectively).
\item Initialise the chain positions.
\item Perform a random move but remember to store the previous energy and positions.
\item Compute $\eta = exp[X(E_{1}) - X(E_{2})]$ and generate a random $\# = ran$ between 0 and 1.
\item \textbf{IF} ($\eta > ran$) accept the move  \textbf{ELSE} return to the old configuration.
\item Update the Histogram and the DOS: $H(E)_{n+1} = H(E)_{n} + 1$, $X(E)_{n+1}=X(E)_{n} + F$.
\item \textbf{IF} $H(E_{i})> q \cdot \langle H(E) \rangle$ for all visited energies \textbf{DO} $F_{n+1} = {\frac{F_{n}}{2}}$. Reset the histogram.
\item \textbf{ELSE} Go to step 4.
\item Repeat until $f \approx {10^{-8}}$ or after a certain amount of time t.
\item Compute thermodynamic observables using $log[g(E)]$ etc.

\end{enumerate}
\rule{\textwidth}{1pt}
For step 8 a flat histogram occurs when the histogram value in each energy bin is above $q \cdot \langle H(E) \rangle$. The parameter $q$ can be set to any value $< 1$, although, as will be discussed later, the precision of the histogram directly affects WL convergence and will have to depend on the chain length. 

For new protein sequences, where the energy minimum is not known, and for existing sequences the Wang Landau scheme requires an energy range to sample the DOS from. Since this energy range is not known a priori it seems useful to conduct a pre-WL-run to ascertain the energy ranges. This is a time consuming procedure because many low energies are only visited during the final stages of the simulation. 

It seems more viable, also retaining the blindness of the approach, to only have the DOS and histogram updated for visited energy sites. So the Wang Landau algorithm and code is modified so that it checks whether the energy has been visited before. In this work another array called 'visited[ENERGY]' is initialised to zero at the beginning of the simulation and once the energy is visited its corresponding array value will be set equal to 1. This value will remain $=1$ for the rest of the Monte Carlo iterations. Once a new energy has been found the histogram (not the modification factor) is reset to zero. 

To accompany this, the flatness checking of the histogram occurs every $10^{6}$ iterations so that the modification factor isn't updated too prematurely for few visited energies.

 This will provide a quicker and easier way to attain the energy range without performing previous simulations.

It is also worth emphasizing that the DOS and histogram of a resulting configuration, which occurred due to a rejection of a proposed one, must be updated accordingly to ensure correct sampling of phase space. Not doing so would result in an incorrect estimate of the density of states and hence any observables derived from it would be devoid of physical meaning.

\subsubsection{\textcolor{black}{1/t algorithm}}

It has been shown and argued that the WL procedure presented above does not converge asymptotically to the correct density of states of the system \cite{fastwl} \cite{fastwlan}. This is due to the saturation in the modification factor which occurs for high MC iterations, the cause of the saturation is due to the function which reduces the modification factor. 

The cure for this which is presented in \cite{fastwl} is to have the reduction of the modification factor take on a functionality which depends on the MC time , $t$, only if all the relevant states of the system have been visited and that the modification factor is smaller than the current MC time. 

Using the Ising model Belardinelli and Pereyra defined the MC time to be $t=\frac{j}{N}$, where $j$ is the number of iterations attempted and $N$ is the number of energy states available to the system. Following in a similar fashion the MC time in this simulation is defined as $t= \frac{M}{\delta}$ where $M$ is the number of iterations attempted and $\delta$ is the energy range of the WL sampling scheme.

The step by step algorithm which alters the WL algorithm in the previous subsection is as follows:

\rule{\textwidth}{1pt}
\begin{center}
\textbf{\textcolor{cyan}{1/t Scheme}}
\end{center}

\begin{enumerate}
\item Set a pre-defined range of discrete energies (not too large to be cumbersome) that the protein may take.
\item Initialise: $X(E_{i}) = 0$, $H(E_{i})=0$ and $F = 1$ (where $X(E)$ and $F$ represents $log[g(E)]$ and $log[f]$ respectively).
\item Initialise the chain positions.
\item Perform a random move but remember to store the previous energy and positions.
\item Compute $\eta = exp[X(E_{1}) - X(E_{2})]$ and generate a random $\# = ran$ between 0 and 1.
\item \textbf{IF} ($\eta > ran$) accept the move  \textbf{ELSE} return to the old configuration.
\item Update the Histogram and the DOS: $H(E)_{n+1} = H(E)_{n} + 1$, $X(E)_{n+1}=X(E)_{n} + F$.
\item After some fixed sweeps (100000 iterations in this case), if, for all $E$, $H(E) \neq 0$ then $F_{n+1} = {\frac{F_{n}}{2}}$. Reset the histogram.
\item IF $F_{n+1} \leqslant t^{-1}$ DO $F_{n+1}= t^{-1}$ and in what follows $F$ is updated at each MC time for the rest of the simulation run (Step 8 is no longer used).
\item Stop the simulation after a fixed elapsed time or until the modification factor is small enough to warrant convergence.
\item Compute thermodynamic observables.
\end{enumerate}
\rule{\textwidth}{1pt}
\begin{center}
\textbf{Problems In Implementation}
\end{center}

This scheme, while seemingly optimal in abstraction, is difficult to implement. Runs were performed for sequences with $N < 20$ and many converged via this scheme however for benchmark sequences convergence via (t) functionality is almost impossible under the current definition of MC time. This is due to the slower rate of modification factor reduction which occurs as $N \rightarrow \infty$.

It is difficult to consider what changes can be made to the MC time without making it too sequence and simulation dependent (non-blind). This is noted in \cite{oplandau} where the tweaking procedures to ensure this scheme works may be too costly in time and effort. However WL sampling still estimates the DOS well if the histogram flatness criterion and final modification factor threshold are stringent enough.

This scheme is still embedded within the code in case the convergence rate increases, however to make this algorithm live up to its potential requires dedicated testing and tweaking of code.

\subsubsection{Detailed Balance}

WL sampling is a non- Markovian process \footnote{(see \ref{sec:nonmarko} for description)}  where it has been shown to provide a valid estimation of the density of states \cite{zhousu} \cite{zhoubhatt} without depending on detailed balance. However it is still vital that the trial moves respect detailed balance to avoid troubling systematic errors \cite{oplandau}. I ensure detailed balance by choosing trial moves at random but with constant probability. Since detailed balance is guaranteed if a trial move is reversible and the reverse/ original move have the same probability.

In every trial move if there is a choice to go in multiple directions they are chosen with equal probability. The only preference of 'choice' are the trial move ratios which are fixed through out a run of the simulation.

As the modification factor, $f$, converges to 1 detailed balance is recovered since:

\begin{equation}
\frac{1}{g(E_{1})} \cdot \pi (E_{1} \rightarrow E_{2}) = \frac{1}{g(E_{2})} \cdot \pi (E_{2} \rightarrow E_{1})
\end{equation}
\clearpage
\subsection{Lattice System}

\label{sec:latticesystem}

A protein chain, in this implementation, exists on a square 2D lattice of length $L$ where monomers can be located via column, $i$, and row, $j$, coordinates stored as $\epsilon_{ij}$ (see equation \ref{eq:coord}).

\begin{equation} \label{eq:coord}
\epsilon_{ij} = (i, j)
\end{equation}

It is computationally cheaper to work using a 1D array which maps onto the 2D lattice. Let $\epsilon$ $\in$ $\mathbb{Z}$ be an element of such an array and impose that $\epsilon$ $\in$ $\lbrace 0,...,L^{2}-1 \rbrace$. For example a 2D lattice in which $L = 4$ is pictured in table \ref{table:examplelattice}.

\begin{table}[h]
\centering
\begin{tabular}{c || c | c | c | c}
 & 0 & 1 & 2 & 3 \\
 \hline\hline
 0 & 0 & 1 & 2 & 3 \\
 \hline
 1 & 4 & 5 & 6 & 7 \\
 \hline
 2 & 8 & 9 & 10 & 11 \\
 \hline
 3 & 12 & 13 & 14 & 15 \\
 
\end{tabular}
\caption{An example 2D lattice with locations stored in a 1D array. \label{table:examplelattice}}
\end{table}

One can retrieve the row and column values from any $\epsilon$ using:

\begin{equation}\label{eq:row}
i = \epsilon \mod L
\end{equation}
\begin{equation}\label{eq:col}
j = \frac{\epsilon}{L} 
\end{equation}

where in equation \ref{eq:col} the value is rounded down to the nearest integer. 

It will be conducive to outline the relationships between neighbouring lattice sites and to define the values of $\epsilon$ which form the boundary.

The element, $\epsilon'$, directly above a given $\epsilon$ is given by $\epsilon' = \epsilon - L$ and the element directly below is given by $\epsilon' = \epsilon + L$. The element directly to the right of a given $\epsilon$ is given by $\epsilon' = \epsilon + 1$ and to the left is given by $\epsilon' = \epsilon - 1$. 

The values of $\epsilon$ which lie on the upper boundary satisfy: $\epsilon < L$. The values of $\epsilon$ which lie on the lower boundary satisfy: $L(L-1) \leq \epsilon \leq L^{2} - 1$. 

The values of $\epsilon$ which lie on the right boundary satisfy: $ \epsilon = aL -1$ for $a \in \lbrace 1,2,....,L \rbrace$. The values of $\epsilon$ which lie on the left hand boundary satisfy: $\epsilon = bL$ for $b \in \lbrace 0, 1,..., L-1 \rbrace$.

\begin{table}[h]
\centering
\begin{tabular}{c || c | c | c | c}
 & 0 & 1 & 2 & 3 \\
 \hline\hline
 0 & $\bullet$ &  & &  \\
 \hline
 1 & $\bullet$ & $\bullet$ &  &  \\
 \hline
 2 &  & $\bullet$ & $\bullet$ &  \\
 \hline
 3 &  &  &  &  \\
 
\end{tabular}
\caption{An example of (H) residues on a 2D lattice of side length $L =4$. \label{table:examplemono}}
\end{table}

An example of how the amino acid residues would be placed onto the 2D lattice with location values stored in a 1D array is shown in table \ref{table:examplemono}. This description is, while trivial, essential to the programming of the simulation as all operations on a chain are essentially operations on this lattice system.

\subsection{Dynamical Trapping}

Recently (2015) it has been shown that Wang Landau sampling of continuous systems suffers from a phenomenon known as dynamical trapping \cite{dyntrap}. The trapping is when the WL sampler only updates the same density of states and histogram for many iterations. The trapping is caused by the random walker coming close to extrema on the energy landscape and should be distinguished from the critical slowing down in conventional MD or MC simulations \cite{dyntrap}.

\begin{figure}[h]
\begin{center}
\fbox{\includegraphics[scale=0.4]{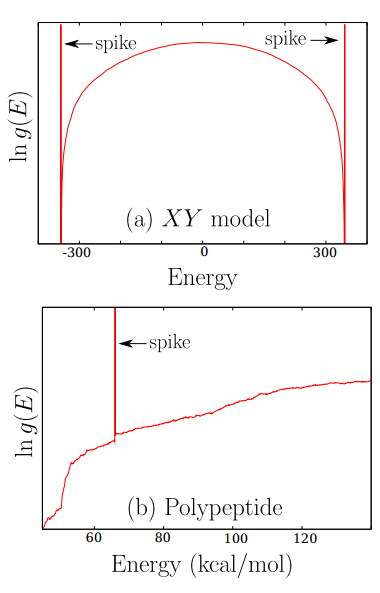}}
\caption{Spiked DOS for (a) the frustrated XY model and (b) the 8-mer poly-alanine. For details see \cite{dyntrap}.}
\label{fig:dosspike}
\end{center}
\end{figure}

The works mentioned in \cite{dyntrap} were all simulations of physical systems using continuous degrees of freedom. The problem of dynamical trapping can still be an issue for discrete models, as is used here, with rough energy landscapes. The compact configurations of proteins near the native region will increase the rejection rates of most moves within the trial set (see section \ref{sec:trialmove}), and whilst the FRW move does have the ability to escape these tight configurations (see section \ref{sec:frw}) it may not be the optimal solution on its own. Whilst the simulation is rejecting most local moves and some non-local moves on monomers that are completely surrounded by others, it will create spikes in the DOS and histogram (example shown in figure \ref{fig:dosspike}) which will greatly damage the accuracy of computed observables and convergence time. Also when trapped, the random walker misses entire or even several stages of Wang-Landau modification factor reduction, which leads to inadequate sampling of conformational space and a rough estimate of the DOS even if the modification factor is reduced to very small values \cite{dyntrap}.

To prevent dynamical trapping from occurring \textit{Koh, Sim and Lee} proposed a simple parallel trajectory-exchange scheme \cite{dyntrap}. This scheme consists of running multiple WL samplers for the system at hand and randomly swapping configurations with each other at regular intervals of MC time. This method is different to that proposed by \textit{Vogel, Li, Wust and Landau} \cite{REWL}(Replica-Exchange Wang Landau) which proposes the exchange of configurations existing within overlapping energy windows. 

\begin{figure}[h]
\begin{center}
\fbox{\begin{tikzpicture}[xscale=1,yscale=1]
\node () at (0,1) {$1$};
\node () at (2,1) {$2$};
\node () at (4,1) {$3$};
\node () at (6,1) {$4$};

\draw [ultra thick] (-1,0.5) -- (7,0.5);

\node () at (0,0) {$A$};
\node () at (2,0) {$B$};
\node () at (4,0) {$C$};
\node () at (6,0) {$D$};

\draw [->,blue] (0,0) -- (6,-1);
\draw [->,red] (4,0) -- (0,-1);
\draw [->,orange] (2,0) -- (4,-1);
\draw [->,purple] (6,0) -- (2,-1);

\node () at (0,-1) {$C$};
\node () at (2,-1) {$D$};
\node () at (4,-1) {$B$};
\node () at (6,-1) {$A$};

\draw [->,blue] (4,-1) -- (6,-2);
\draw [->,red] (2,-1) -- (0,-2);
\draw [->,orange] (6,-1) -- (4,-2);
\draw [->,purple] (0,-1) -- (2,-2);

\node () at (0,-2) {$D$};
\node () at (2,-2) {$C$};
\node () at (4,-2) {$A$};
\node () at (6,-2) {$B$};

\draw [->>,black] (8,1) -- (8,-2);

\end{tikzpicture}}
\caption{The numbers represent the thread ID's and letters represent arbitrary configurations. Trajectories are swapped through random shuffling. The arrow to the right represents the direction of MC time. \label{fig:traex}}
\end{center}
\end{figure}
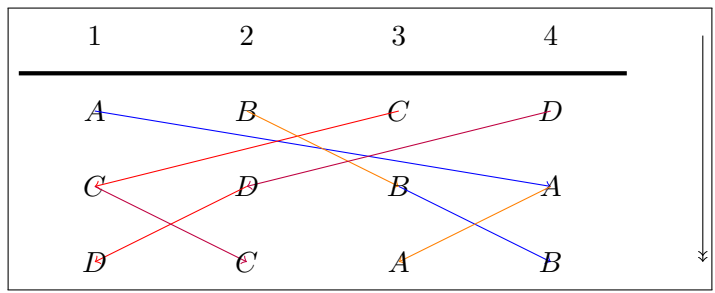

Each WL sampler in the trajectory-exchange scheme has its own private estimation of the DOS and thermodynamic observables, the scheme merely imposes the regular swapping of configurations. The main mechanics of the idea can be understood effectively through figure \ref{fig:traex}.

In \cite{dyntrap} they surmised that $T < 1000$ (where $T$ is the swapping period). In this work $N_P$, the number of processes/threads, is large enough to warrant the use of Gaussian statistics.

The most natural language to implement this scheme, in my opinion, is via MPI \footnote{For details on the MPI language and inherent library routines: 'Gropp, William., Lusk, Ewing. and Skjellum, Anthony. \textit{Using MPI: Portable Parallel Programming with the Message-Passing Interface (2nd Edition)'} is recommended by the author. }(Message Passing Interface) since it is very simple to adapt the serial code in order to impose trajectory swapping. In this work the master process produces a random source ID for every process ID such that the source process sends its trajectory to the destination process. In this way every process has its trajectory swapped in a random manner. 

Once the WL sampling routine is complete each process then computes thermodynamic quantities as outlined in section \ref{sec:energyfluc}, the results are then averaged and statistical error analysis is then conducted. 

A problem arises: If one process attains the native or a near native state (which is very compact), swapping the trajectories will not make it more likely to escape this configuration due to every process having the same trial move set ratios. So this could cause the downfall of the trajectory swapping method. Very low temperature configurations will eventually loosen since not every move will be rejected, but as this occurs using many processes they will not show extreme spikes in the DOS. The configuration will hop between processes whilst gradually unwinding. As long as swapping is very regular this problem does not pose any threat.

Also the fact that lower configurations may be shared by many processes before being completely changed helps each process explore the low temperature regions of phase space. 

Whilst the REWL (overlapping energy windows) scheme has been implemented successfully for a more sophisticated variant of the HP model \cite{HOP}, it is easier to augment existing serial code into a parallel framework using simple trajectory swapping whilst still making major improvements to the robustness of the simulation. Hence incorporating the trajectory-exchange parallel scheme allows for a simple and efficient way to explore the thermodynamics of the HP model.

\subsection{Logistics of  the parallel implementation}

After the threading environment has been created the root process reads in the (H)(P) sequence from a file and assigns the values to an array HP[ ], then it initialises the visited[ENERGY] array to 0. The size of the visited array, and any array which has an argument of ENERGY, can be set to the length of the protein chain for safety. The HP[ ] and initialised visited[ ] arrays are then sent to all processes.

Each process initialises the chain in the same manner but is assigned a personal seed number, $S$. A 'master' seed, $S_{M}$, is chosen from an external random number generator and the seed for each process is generated via:

\begin{equation}
S = (S_{M} + myid)\cdot a \cdot (myid+b)
\end{equation}

where $myid$ is the process id and $a$ and $b$ are arbitrary positive integers. 

The snippet of code which implements the trajectory swapping is shown in appendix \ref{sec:C}. A random source process is chosen to send its trajectory to a destination process. The destination process goes from 0 to $numprocs-1$ so that each process has a new configuration.

The minimum energy from all simulations were found via a MPI Reduce() function. Each process then computes thermodynamic quantities in a pre-defined temperature range. Statistical analysis was then conducted externally.

\clearpage
\subsection{\textcolor{black}{Trial Move Implementation}}
\label{sec:trialmove}

\begin{figure}[h]
\begin{center}
\fbox{\begin{tikzpicture}[xscale=0.7,yscale=0.7]
\draw [ultra thick](-1,0) -- (0,0);
\draw [black,fill=black](0,0) circle [radius=0.20];
\draw [ultra thick](0,0) -- (1,0);
\draw [black,fill=white](1,0) circle [radius=0.20];
\draw [ultra thick](1,0) -- (1,-1);
\draw [black,fill=black](1,-1) circle [radius=0.20];
\draw [ultra thick](1,-1) -- (1,-2);

\draw [ultra thick](2,0) -- (3,0);
\draw [black,fill=black](3,0) circle [radius=0.20];
\draw [ultra thick](3,0) -- (3,-1);
\draw [black,fill=white](3,-1) circle [radius=0.20];
\draw [ultra thick](3,-1) -- (4,-1);
\draw [black,fill=black](4,-1) circle [radius=0.20];
\draw [ultra thick](4,-1) -- (4,-2);

\end{tikzpicture}}
\caption{ An example of a kink flip move where the white monomer is at a corner between its sequential neighbours and 'flips' to the opposite corner if it is free. \label{fig:kink}}
\end{center}
\end{figure}
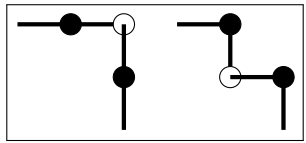

In polymer and past HP model simulations there are tried and tested trial move sets which respect the LCSAW condition. These include the end-bond flip(figure \ref{fig:end}), kink flip (figure \ref{fig:kink}) and the crankshaft move (not used in this simulation). A trial set consisting of these moves only, does not respect the ergodicity condition: that all of configuration space is reachable. However the inclusion of pull moves and pivot moves restores ergodicity \cite{oplandau} \cite{wust}.

\begin{figure}[h]
\begin{center}
\fbox{\begin{tikzpicture}[xscale=0.7,yscale=0.7]
\draw [ultra thick](0,0) -- (1,0);
\draw [black,fill=blue](1,0) circle [radius=0.20];
\draw [ultra thick](1,0) -- (1,-1);
\draw [black,fill=black](1,-1) circle [radius=0.20];

\draw [ultra thick](2,0) -- (3,0);
\draw [black,fill=blue](3,0) circle [radius=0.20];
\draw [ultra thick](3,0) -- (4,0);
\draw [black,fill=black](4,0) circle [radius=0.20];
\end{tikzpicture}

}
\caption{ An example of an end-bond flip move where the penultimate monomer or second monomer acts as an axis (blue) so that the 1st or last monomer can rotate about it. \label{fig:end}}
\end{center}
\end{figure}
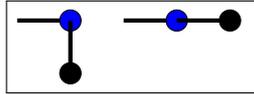

In this simulation the trial move set consists of pull, kink flip, pivot, bond rebridging and fragment random walk moves. As one shall see this trial move set necessitates the inclusion of the end-bond flip move. This move set is different to that used in \cite{oplandau} and \cite{wust}, in the fact that here the fragment random walk move is introduced and all moves in this simulation are coded originally and may be implemented slightly differently. 

\begin{figure}[h]
\begin{center}
\fbox{\begin{tikzpicture}[xscale=0.7,yscale=0.7]
\draw [black,fill=black](-1,0) circle [radius=0.20];
\draw [ultra thick](-1,0) -- (0,0);
\draw [black,fill=blue](0,0) circle [radius=0.20];
\draw [ultra thick](0,0) -- (1,0);
\draw [black,fill=purple](1,0) circle [radius=0.20];
\draw [ultra thick](1,0) -- (2,0);
\draw [black,fill=green](2,0) circle [radius=0.20];
\draw [ultra thick](2,0) -- (3,0);
\draw [ultra thick](2,0) -- (3,0);
\draw [black,fill=black](3,0) circle [radius=0.20];
\draw [ultra thick](3,0) -- (3,-1);
\draw [black,fill=black](3,-1) circle [radius=0.20];
\draw [ultra thick](3,-1) -- (4,-1);
\draw [black,fill=black](4,-1) circle [radius=0.20];
\draw [ultra thick](4,-1) -- (4,-2);
\draw [black,fill=black](4,-2) circle [radius=0.20];
\node () at (0,-1) [purple]{$P$};
\node () at (1,-1) [green]{$G$};

\draw [black,fill=black](5,0) circle [radius=0.20];
\draw [ultra thick](5,0) -- (6,0);
\draw [black,fill=blue](6,0) circle [radius=0.20];
\draw [ultra thick](6,0) -- (6,-1);
\draw [black,fill=purple](6,-1) circle [radius=0.20];
\draw [ultra thick](6,-1) -- (7,-1);
\draw [black,fill=green](7,-1) circle [radius=0.20];
\draw [ultra thick](7,-1) -- (7,0);
\draw [black,fill=black](7,0) circle [radius=0.20];
\draw [ultra thick](7,0) -- (8,0);
\draw [black,fill=black](8,0) circle [radius=0.20];
\draw [ultra thick](8,0) -- (9,0);
\draw [black,fill=black](9,0) circle [radius=0.20];
\draw [ultra thick](9,0) -- (9,-1);
\draw [black,fill=black](9,-1) circle [radius=0.20];

\end{tikzpicture}}
\caption{ An example of a pull move where the anchor monomer (blue) remains fixed and the primary monomer (purple) will move to $P$ and the secondary monomer (green) moves to $G$. The rest of the chain 'slithers' behind to keep the sequence and length of the chain fixed.  \label{fig:pull}}
\end{center}
\end{figure}
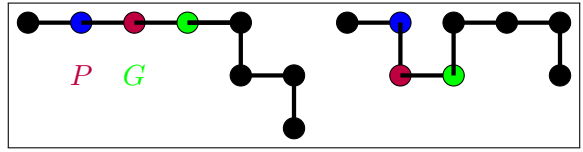

\textit{Pull Move:} A monomer is chosen at random to act as the \textit{primary monomer}, this means it is the first monomer to move to a free neighbouring position. This future position of the primary monomer is determined by the \textit{anchor monomer}, where it will move to its right or  left or above or below it depending on the availability of these positions on the lattice. If the monomer is at the end or start of the chain it can only have the penultimate monomer or second monomer as the anchor monomer respectively. If the monomer has a sequence value $s$ such that $ 1 < s < N $ , where $N$ is the total number of monomers, then the anchor monomer $s_a$ is chosen, with equal probability, between $s_a = s -1$ and $s_a = s+1$. Once the primary monomer has moved to a suitable position next to the anchor monomer, the secondary monomer (next to primary monomer on the sequence) slots into a suitable position which keeps it connected to the primary monomer. The rest of the chain 'slithers' along occupying the positions of relevant old monomers which ensures the LCSAW condition is fulfilled (figure \ref{fig:pull}).\footnote{N.B. The original pull move consisted of pulling the rest of the chain along every time, which while still effective was unnecessary and potentially unrealistic. This move algorithm was modified so that it stops once it respected the SAW condition, which makes it affect less monomers on average.}

\begin{figure}[h]
\begin{center}
\fbox{\begin{tikzpicture}[xscale=0.7,yscale=0.7]
\draw [black,fill=black](-1,0) circle [radius=0.20];
\draw [ultra thick](-1,0) -- (0,0);
\draw [black,fill=black](0,0) circle [radius=0.20];
\draw [ultra thick](0,0) -- (1,0);
\draw [black,fill=black](1,0) circle [radius=0.20];
\draw [ultra thick](1,0) -- (2,0);
\draw [black,fill=black](2,0) circle [radius=0.20];
\draw [ultra thick](2,0) -- (3,0);
\draw [ultra thick](2,0) -- (3,0);
\draw [black,fill=black](3,0) circle [radius=0.20];
\draw [ultra thick](3,0) -- (3,-1);
\draw [black,fill=blue](3,-1) circle [radius=0.20];
\draw [ultra thick](3,-1) -- (4,-1);
\draw [black,fill=red](4,-1) circle [radius=0.20];
\draw [ultra thick](4,-1) -- (4,-2);
\draw [black,fill=gray](4,-2) circle [radius=0.20];

\draw [black,fill=black](6,0) circle [radius=0.20];
\draw [ultra thick](6,0) -- (7,0);
\draw [black,fill=black](7,0) circle [radius=0.20];
\draw [ultra thick](7,0) -- (8,0);
\draw [black,fill=black](8,0) circle [radius=0.20];
\draw [ultra thick](8,0) -- (9,0);
\draw [black,fill=black](9,0) circle [radius=0.20];
\draw [ultra thick](9,0) -- (10,0);
\draw [ultra thick](10,0) -- (10,0);
\draw [black,fill=black](10,0) circle [radius=0.20];
\draw [ultra thick](10,0) -- (10,-1);
\draw [black,fill=blue](10,-1) circle [radius=0.20];
\draw [ultra thick](10,-1) -- (10,-2);
\draw [black,fill=red](10,-2) circle [radius=0.20];
\draw [ultra thick](10,-2) -- (9,-2);
\draw [black,fill=gray](9,-2) circle [radius=0.20];

\end{tikzpicture}}
\caption{An example of a $\circlearrowright$ pivot move where the anchor monomer (blue) remains fixed and the red and grey monomers move according to the change in directionality relative to the preceding monomer. Initially red was to the right of blue and now it is below it also grey was below red and now it is to the left of it. \label{fig:pivot}}
\end{center}
\end{figure}
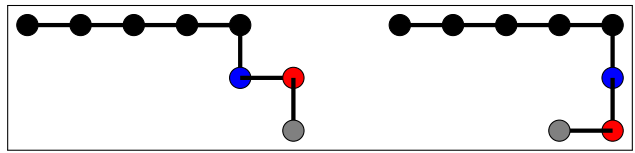

\textit{Pivot move:} This move starts by choosing a monomer at random with sequence value $1 < s < N$ to act as another anchor monomer so that it acts as an axis in which another part of the chain rotates about it. It makes no difference to the configuration of the protein chain if a rotation is executed around the 1st monomer or $N$th monomer since this doesn't change the internal structure and hence will remain stationary in energy space. So these rotations are omitted for convenience in this simulation. So when a random monomer has been chosen an acceptable move is to either rotate the part of the chain with monomers having sequence values $< s$ or monomers with sequence values $> s$. The algorithm chooses either case with probability $\frac{1}{2}$ to ensure that no biases occur. Also rotations either anticlockwise ,$\circlearrowleft$, or clockwise, $\circlearrowright$, are acceptable and hence the algorithm decides to undertake such rotations with probability $\frac{1}{2}$. The rotations then occur leaving the rotated structure internally invariant but it's relationship with the rest of the chain will change. The pivot algorithm always checks whether the future space of the monomers are available, otherwise the move is rejected. An example of a pivot move is shown in figure \ref{fig:pivot}.

The pivot move was included to accelerate convergence in the DOS computation via WL sampling \cite{oplandau}, also it ensures that the entire phase space of the system is attainable. 

The pivot algorithm proposes the future positions of monomers on the 'to be' rotated structure via operations on the \textit{directionality}. The \textit{directionality} can be defined as the relative direction that a monomer $B$ is relative to a monomer $A$. A table showing how anticlockwise and clockwise rotations affect the directionality is shown in table \ref{table:directions}.

\begin{table}[h]
\centering
\begin{tabular}{c || c | c}
 & $\circlearrowleft$ & $\circlearrowright$  \\
 \hline\hline
 $\rightarrow$ & $\uparrow$ & $\downarrow$  \\
 \hline
 $\downarrow$ & $\rightarrow$ & $\leftarrow$  \\
 \hline
 $\leftarrow$ & $\downarrow$  & $\uparrow$  \\
 \hline
 $\uparrow$ & $\leftarrow$ & $\rightarrow$  \\
 
\end{tabular}
\caption{How relative direction is changed under the two rotations. \label{table:directions}}
\end{table}

Directionality can be stored as an integer quantity $d \in \lbrace 1,2,3,4 \rbrace$ in 2D where $\rightarrow$ = 1, $\leftarrow$ = 2, $\downarrow$ = 3 and $\uparrow$ = 4. The routine \textit{buddycheck(int N,int 'position of monomer $A$', int 'position of monomer $B$')}(ref appendix of buddycheck) returns $d$ as the directionality of monomer $B$ to monomer $A$. Once the operation on the directionality has occurred successfully and the future positions are indeed available, the pivot move executes a move.

\begin{figure}[h]
\begin{center}
\fbox{\begin{tikzpicture}[xscale=0.7,yscale=0.7]
\draw [black,fill=black](0,0) circle [radius=0.20];
\draw [ultra thick](0,0) --(1,0);
\draw [ultra thick](1,-1) --(1,0);
\draw [black,fill=orange](1,0) circle [radius=0.20];
\draw [black,fill=black](1,-1) circle [radius=0.20];
\node () at (0,-1) [orange]{$O$};
\draw [dotted, ultra thick] (0,-1) --(1,0);

\draw [black,fill=black](4,0) circle [radius=0.20];
\draw [ultra thick](4,0) --(4,-1);
\draw [ultra thick](4,-1) --(5,-1);
\draw [black,fill=orange](4,-1) circle [radius=0.20];
\draw [black,fill=black](5,-1) circle [radius=0.20];
\node () at (5,0) [orange]{$O$};
\draw [dotted, ultra thick] (5,0) --(4,-1);

\draw [black,fill=black](8,-1) circle [radius=0.20];
\draw [ultra thick](9,0) --(9,-1);
\draw [ultra thick](8,-1) --(9,-1);
\draw [black,fill=orange](9,-1) circle [radius=0.20];
\draw [black,fill=black](9,0) circle [radius=0.20];
\node () at (8,0) [orange]{$O$};
\draw [dotted, ultra thick] (8,0) --(9,-1);

\draw [black,fill=black](12,-1) circle [radius=0.20];
\draw [ultra thick](13,0) --(12,0);
\draw [ultra thick](12,-1) --(12,0);
\draw [black,fill=orange](12,0) circle [radius=0.20];
\draw [black,fill=black](13,0) circle [radius=0.20];
\node () at (13,-1) [orange]{$O$};
\draw [dotted, ultra thick] (12,0) --(13,-1);

\end{tikzpicture}}
\caption{ The four possible scenarios for a kink flip move, the orange $O$ represents the future position of the primary monomer (orange). From left to right the names of the moves are as follows: \textit{bottom left quadrant move}, \textit{top right quadrant move}, \textit{top left quadrant move} and \textit{bottom right quadrant move}.  \label{fig:kink2}}
\end{center}
\end{figure}
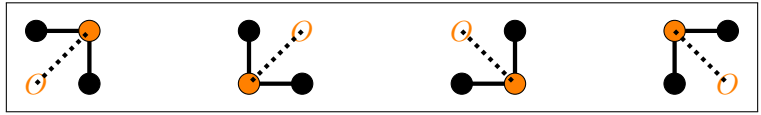

\textit{Kink flip move:} As seen in figure \ref{fig:kink} the kink flip move only affects a monomer at a corner. There are 4 possible scenarios which allow a kink flip move to be performed shown in figure \ref{fig:kink2}.

The kinkflip(\textit{int $N$,int $L$ (lattice side length),int $L^{2}-1$}) routine (insert ref to appendix for kinkflip code) searches for kinks in the chain beginning at $s_i = 2$ and ending at $s_f = N-1$. If there is a kink it will execute one of the four possible moves depending on whether the relevant future position is available or not. If a move gets rejected it continues along the chain looking for more kinks, this helps keep the rejection rate at a minimum. If any move is executed properly the routine closes.

\subsubsection*{\textcolor{black}{Bond Re-bridging}}

These moves do not change the position of the chain on the lattice, i.e. the array positions remain constant, but changes a number of the bonds of the chain and then re uploads the sequence onto it as to dramatically change the configuration. This move becomes useful in sampling low temperature phase space where compact configurations lead to high rejection rates for local moves like pull, pivot and kink flip.

There are two types of bond re-bridging moves used in this work namely chain-terminal and type II (see \cite{longrange} for an in depth discussion).

\textit{Chain Terminal:} This move consists of destroying a bond between monomers and then recreating a bond with a topological neighbour of N and 1 respectively. The destruction of a bond must only occur between the topological neighbour of N and 1 and a connected neighbour on the sequence with dependence on whether N or 1 has been chosen. For example if 1 was chosen then its topological neighbour say, m, can only destroy its bond with m-1 on the sequence. If N was chosen then its topological neighbour, m, can only destroy its bond with m+1 on the sequence.

The chain terminal algorithm implemented here first chooses (with 50$\%$ chance) monomer 1 or N and then searches for a topological neighbour for which it can form a new sequence bond. After this search for topological neighbours the step in the procedure is pictorially shown in \ref{fig:beforect}.

\begin{figure}
\centering
\fbox{
\subcaptionbox{Configuration just before the execution of the chain terminal move. The red cross on the bond 3-4 will be destroyed and the bond 1-4 will be created. \label{fig:beforect}}[6cm]{
\begin{tikzpicture}[xscale=1.5,yscale=1.5]
\draw [black,fill=white](0,0) circle [radius=0.20];
\draw [black,fill=white](-1,0) circle [radius=0.20];
\draw [black,fill=white](-1,-1) circle [radius=0.20];
\draw [black,fill=white](0,-1) circle [radius=0.20];
\draw [black,fill=white](1,-1) circle [radius=0.20];
\draw [black,fill=white](1,-2) circle [radius=0.20];
\draw [black,fill=white](0,-2) circle [radius=0.20];
\draw [black,fill=white](-1,-2) circle [radius=0.20];
\draw [black,fill=white](-1,-3) circle [radius=0.20];
\draw [black,fill=white](0,-3) circle [radius=0.20];
\draw [ultra thick](-0.25,0) --(-0.75,0);
\draw [ultra thick](-1,-0.25) --(-1,-0.75);
\draw [ultra thick](-0.75,-1) --(-0.25,-1);
\draw [ultra thick](0.25,-1) --(0.75,-1);
\draw [ultra thick](1,-1.25) --(1,-1.75);
\draw [ultra thick](0.25,-2) --(0.75,-2);
\draw [ultra thick](-0.25,-2) --(-0.75,-2);
\draw [ultra thick](-1,-2.25) --(-1,-2.75);
\draw [ultra thick](-0.25,-3) --(-0.75,-3);
\draw [red][dotted,ultra thick](0,-0.25) --(0,-0.75);
\node () at (-0.5,-1)[xscale=1,yscale=1] [red]{X};
\node () at (0,0)[xscale=0.8,yscale=0.8] {1};
\node () at (-1,0)[xscale=0.8,yscale=0.8] {2};
\node () at (-1,-1)[xscale=0.8,yscale=0.8] {3};
\node () at (0,-1)[xscale=0.8,yscale=0.8] {4};
\node () at (1,-1)[xscale=0.8,yscale=0.8] {5};
\node () at (1,-2)[xscale=0.8,yscale=0.8] {6};
\node () at (0,-2)[xscale=0.8,yscale=0.8] {7};
\node () at (-1,-2)[xscale=0.8,yscale=0.8] {8};
\node () at (-1,-3)[xscale=0.8,yscale=0.8] {9};
\node () at (0,-3)[xscale=0.8,yscale=0.8] {10};

\end{tikzpicture}}
\subcaptionbox{Resulting configuration. \label{fig:afterct}}[6cm]{
\begin{tikzpicture}[xscale=1.5,yscale=1.5]
\draw [black,fill=white](0,0) circle [radius=0.20];
\draw [black,fill=white](-1,0) circle [radius=0.20];
\draw [black,fill=white](-1,-1) circle [radius=0.20];
\draw [black,fill=white](0,-1) circle [radius=0.20];
\draw [black,fill=white](1,-1) circle [radius=0.20];
\draw [black,fill=white](1,-2) circle [radius=0.20];
\draw [black,fill=white](0,-2) circle [radius=0.20];
\draw [black,fill=white](-1,-2) circle [radius=0.20];
\draw [black,fill=white](-1,-3) circle [radius=0.20];
\draw [black,fill=white](0,-3) circle [radius=0.20];
\draw [ultra thick](-0.25,0) --(-0.75,0);
\draw [ultra thick](-1,-0.25) --(-1,-0.75);
\draw [ultra thick](0.25,-1) --(0.75,-1);
\draw [ultra thick](1,-1.25) --(1,-1.75);
\draw [ultra thick](0.25,-2) --(0.75,-2);
\draw [ultra thick](-0.25,-2) --(-0.75,-2);
\draw [ultra thick](-1,-2.25) --(-1,-2.75);
\draw [ultra thick](-0.25,-3) --(-0.75,-3);
\draw [black][ultra thick](0,-0.25) --(0,-0.75);
\node () at (0,0)[xscale=0.8,yscale=0.8] {3};
\node () at (-1,0)[xscale=0.8,yscale=0.8] {2};
\node () at (-1,-1)[xscale=0.8,yscale=0.8] {1};
\node () at (0,-1)[xscale=0.8,yscale=0.8] {4};
\node () at (1,-1)[xscale=0.8,yscale=0.8] {5};
\node () at (1,-2)[xscale=0.8,yscale=0.8] {6};
\node () at (0,-2)[xscale=0.8,yscale=0.8] {7};
\node () at (-1,-2)[xscale=0.8,yscale=0.8] {8};
\node () at (-1,-3)[xscale=0.8,yscale=0.8] {9};
\node () at (0,-3)[xscale=0.8,yscale=0.8] {10};

\end{tikzpicture}}}
\end{figure}
Since positions of the monomers are stored in a 1-dimensional array space (see section \ref{sec:latticesystem}) the chain terminal algorithm simply swaps the position of the connected neighbour, who is connected topologically to 1 or N, with 1 or N.

For example as with the before and after in figures \ref{fig:beforect} and \ref{fig:afterct} respectively the position of 3 is swapped with the position of 1. The rest of the chain is unaffected. In general, for the '1' case, the algorithm is outlined as:

\begin{enumerate}
\item Call topological neighbour m and connected neighbour m-1.
\item POS[1]=OLDPOS[m-1] 
\item FOR(i=2;j=m-2;i$<$m-2;j$>$1;i++;j- -)\newline[POS[i]=OLDPOS[j];] 
\end{enumerate}

The steps are similar for the 'N' case.

\begin{figure}[h]
\begin{center}
\fbox{\begin{tikzpicture}[xscale=1.5,yscale=1.5]
\draw [black,fill=white](0,0) circle [radius=0.20];
\draw [black,fill=white](1,0) circle [radius=0.20];
\draw [black,fill=white](0,-1) circle [radius=0.20];
\draw [black,fill=white](1,-1) circle [radius=0.20];
\draw [ultra thick](0.25,0) --(0.75,0);
\node () at (0.5,0)[xscale=0.8,yscale=0.8] [red]{X};
\draw [ultra thick](0.25,-1) --(0.75,-1);
\node () at (0.5,-1)[xscale=0.8,yscale=0.8] [red]{X};
\node () at (0,0)[xscale=0.8,yscale=0.8] {k};
\node () at (1,0)[xscale=0.7,yscale=0.7] {k+1};
\node () at (0,-1)[xscale=0.8,yscale=0.8] {p};
\node () at (1,-1)[xscale=0.7,yscale=0.7] {p+1};
\draw [red][dotted,ultra thick](0,-0.25) --(0,-0.75);
\draw [red][dotted,ultra thick](1,-0.25) --(1,-0.75);
\end{tikzpicture}}
\caption{ The red crosses represent bonds that will be destroyed and dotted lines represent future connected bonds. $k$ and $p$ are monomer values within the set $\lbrace 1,...,N \rbrace$ (Ignoring other monomers for clarity). \label{fig:quadex}}
\end{center}
\end{figure}
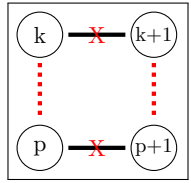

Type II Move: This move is not restricted to the ends of the chain and destroys and recreates 2 bonds in contrast to only 1 in the chain terminal move. We define the 'quad' as the sub square which contains the monomers that dictate the implementation of the move and hence swapping of position values. An example of a 'quad' can be seen in figure \ref{fig:quadex}.

Using figure \ref{fig:quadex} as a reference, we note that the monomer numbers from $k$ and $p$ increase in the same direction. This creates a linear topology where, if we cement the new proposed bonds, no part of the sequence will ever be cut off. This means it will obey the LCSAW condition. 

The pseudo algorithm, used here, for the type II move is as follows:

\begin{enumerate}
\item Select a monomer,p, at random.
\item Choose, at random, a connected neighbour,j, of p i.e. p-1 or p+1 (unless p = 1 or N).
\item Look for topological neighbours of p and j with the same relative direction (see figure \ref{fig:quadex} for reference).
\item IF (they form a quad)\newline DO step 5 \newline else go back to step 1.
\item Find the smallest and largest monomer number in the quad set.
\item Impose constant positions: \newline POS[smallest]=OLDPOS[smallest] \newline POS[largest]=OLDPOS[largest] \newline POS[1]=OLDPOS[1] \newline POS[N]=OLDPOS[N].
\item Re-upload the other monomers correctly: \newline FOR(i=smallest+1;h=largest-1;i$<$=largest-1;h$>$=smallest+1;i++;h--)

(POS[i]=POS[h];)

\end{enumerate}

\begin{figure}[h]
\centering
\captionbox{An example of a type II bond-rebridging move on a small lattice protein chain. Note the re-uploading of the HP sequence.  \label{fig:type2}}[0.5\textwidth]
{\includegraphics[scale=0.4]{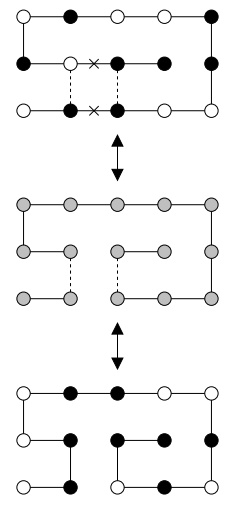}}
\end{figure}
\newpage
\subsubsection{\textcolor{black}{Fragment Random Walk}}
\label{sec:frw}
It is vital that the trial move set allows rapid coverage of configuration space but still gather a detailed picture of the rough energy and conformational landscape. This is to ensure that the DOS can be estimated quickly and accurately. Local moves such as pull, kink-flip and in some cases pivot moves only displace a relatively small amount of monomers which allows the Wang-Landau sampling scheme to gather detailed information. To aid with pivot moves (in cases where a large number of monomers are displaced) in accelerating global conformational changes \cite{oplandau}, I have supplemented the trial set with the \textit{fragment random walk} (FRW) move.

This move has not been implemented in \cite{oplandau}, \cite{wust} or any other work since it has been invented here. Hence the inclusion of this move makes the trial move set used here unique. 

\begin{figure}[h]
\begin{center}
\fbox{\includegraphics[scale=0.4]{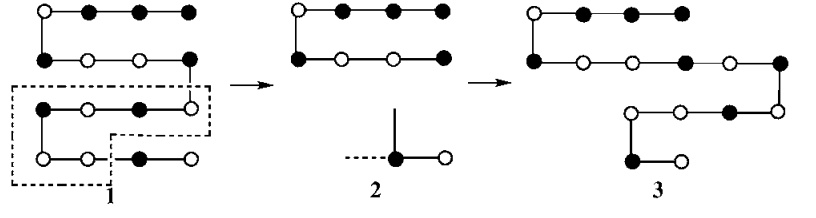}}
\caption{ 1) The initial configuration with a randomly selected fragment. 2) Fragments are being produced at random, any fragments which do not connect to the fixed points are rejected. 3) A successfully regrown fragment and its resulting configuration. (\textit{Picture originally published in \cite{frmc} and gratitude goes to Zhang, Kou and Liu.) } }
\label{fig:FRESS}
\end{center}
\end{figure}

The FRW move is partly inspired by FRESS (fragment regrowth Monte Carlo) \cite{frmc} where an \textbf{internal} segment of the protein chain (of chosen length) is chosen at random and a new fragment is 'regrown' to replace it, hence causing a conformational change. The move used in FRESS is illustrated in figure \ref{fig:FRESS}. In FRESS a fragment regrowth move is only accepted if it obeys the Metropolis - Hastings criterion \textit{see section \ref{sec:method}} \cite{frmc}.

FRW is different to the regrowth of fragments used in \cite{frmc} in that the fragments are not internal and are not necessarily of fixed size. Internal is defined as: the fragment having two fixed points that are monomers $\in \lbrace 2, N-1 \rbrace$, so the FRW has only one fixed point in the same set of monomers.

The pseudo algorithm for the FRW move is as follows:

\begin{enumerate}
\item Pick a random monomer $m \in \lbrace 2, N-1 \rbrace$.
\item With 50$\%$ probability monomers with sequence number $n > m$ or $n < m$ are chosen to form the fragment.
\item Start the self avoiding random walk for the fragment.
\item If the walk violates the LCSAW condition and not all local positions have been tried, then try another position.
\item If all local positions have been exhausted then return to old configuration.
\item Else if the fragment random walk has been completed successfully exit move algorithm
\end{enumerate}

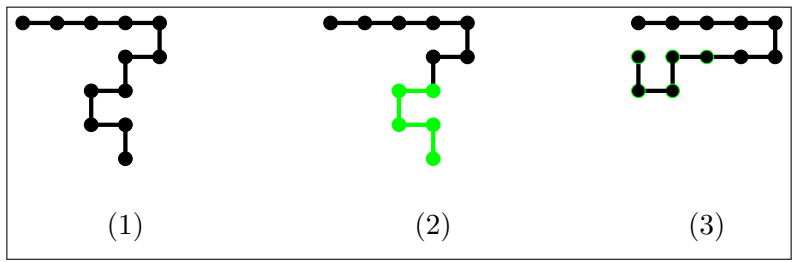
\begin{figure}[h]
\begin{center}
\fbox{\begin{tikzpicture}[xscale=0.45,yscale=0.45]
\draw [black,fill=black](0,0) circle [radius=0.20];
\draw [ultra thick](0,0) -- (1,0);
\draw [black,fill=black](1,0) circle [radius=0.20];
\draw [ultra thick](1,0) -- (2,0);
\draw [black,fill=black](2,0) circle [radius=0.20];
\draw [ultra thick](2,0) -- (3,0);
\draw [black,fill=black](3,0) circle [radius=0.20];
\draw [ultra thick](3,0) -- (4,0);
\draw [black,fill=black](4,0) circle [radius=0.20];
\draw [ultra thick](4,0) -- (4,-1);
\draw [black,fill=black](4,-1) circle [radius=0.20];
\draw [ultra thick](4,-1) -- (3,-1);
\draw [black,fill=black](3,-1) circle [radius=0.20];
\draw [ultra thick](3,-1) -- (3,-2);
\draw [black,fill=black](3,-2) circle [radius=0.20];
\draw [ultra thick](3,-2) -- (2,-2);
\draw [black,fill=black](2,-2) circle [radius=0.20];
\draw [ultra thick](2,-2) -- (2,-3);
\draw [black,fill=black](2,-3) circle [radius=0.20];
\draw [ultra thick](2,-3) -- (3,-3);
\draw [black,fill=black](3,-3) circle [radius=0.20];
\draw [ultra thick](3,-3) -- (3,-4);
\draw [black,fill=black](3,-4) circle [radius=0.20];

\node () at (3,-6) {\textcolor{black}{(1)}};

\draw [black,fill=black](9,0) circle [radius=0.20];
\draw [ultra thick](9,0) -- (10,0);
\draw [black,fill=black](10,0) circle [radius=0.20];
\draw [ultra thick](10,0) -- (11,0);
\draw [black,fill=black](11,0) circle [radius=0.20];
\draw [ultra thick](11,0) -- (12,0);
\draw [black,fill=black](12,0) circle [radius=0.20];
\draw [ultra thick](12,0) -- (13,0);
\draw [black,fill=black](13,0) circle [radius=0.20];
\draw [ultra thick](13,0) -- (13,-1);
\draw [black,fill=black](13,-1) circle [radius=0.20];
\draw [ultra thick](13,-1) -- (12,-1);
\draw [black,fill=black](12,-1) circle [radius=0.20];
\draw [ultra thick](12,-1) -- (12,-2);
\draw [green,fill=green](12,-2) circle [radius=0.20];
\draw [ultra thick][green](12,-2) -- (11,-2);
\draw [green,fill=green](11,-2) circle [radius=0.20];
\draw [ultra thick][green](11,-2) -- (11,-3);
\draw [green,fill=green](11,-3) circle [radius=0.20];
\draw [ultra thick][green](11,-3) -- (12,-3);
\draw [green,fill=green](12,-3) circle [radius=0.20];
\draw [ultra thick][green](12,-3) -- (12,-4);
\draw [green,fill=green](12,-4) circle [radius=0.20];

\node () at (12,-6) {\textcolor{black}{(2)}};

\draw [black,fill=black](18,0) circle [radius=0.20];
\draw [ultra thick](18,0) -- (19,0);
\draw [black,fill=black](19,0) circle [radius=0.20];
\draw [ultra thick](19,0) -- (20,0);
\draw [black,fill=black](20,0) circle [radius=0.20];
\draw [ultra thick](20,0) -- (21,0);
\draw [black,fill=black](21,0) circle [radius=0.20];
\draw [ultra thick](21,0) -- (22,0);
\draw [black,fill=black](22,0) circle [radius=0.20];
\draw [ultra thick](22,0) -- (22,-1);
\draw [black,fill=black](22,-1) circle [radius=0.20];
\draw [ultra thick](22,-1) -- (21,-1);
\draw [black,fill=black](21,-1) circle [radius=0.20];
\draw [ultra thick](21,-1) -- (20,-1);
\draw [green,fill=black](20,-1) circle [radius=0.20];
\draw [ultra thick][black](20,-1) -- (19,-1);
\draw [green,fill=black](19,-1) circle [radius=0.20];
\draw [ultra thick][black](19,-1) -- (19,-2);
\draw [green,fill=black](19,-2) circle [radius=0.20];
\draw [ultra thick][black](19,-2) -- (18,-2);
\draw [green,fill=black](18,-2) circle [radius=0.20];
\draw [ultra thick][black](18,-2) -- (18,-1);
\draw [green,fill=black](18,-1) circle [radius=0.20];

\node () at (20,-6) {\textcolor{black}{(3)}};

\end{tikzpicture}}
\caption{ 1) The initial configuration before the FRW move. 2) A fragment is chosen. 3) The resulting configuration of a successful FRW move. \label{fig:FRW}}
\end{center}
\end{figure}

An example of a successful FRW move is shown in figure \ref{fig:FRW}.

Why wasn't the move used in FRESS simply utilized in this scheme? The FRESS move, while having its own advantages, will suffer from high rejection ratios in low temperature conformations and does not rapidly change the energy as much as the FRW move. This is due to the fact that FRW has no limit on the fragment size meaning a large portion of the chain can be rapidly changed. Also the constraint of having two internal fixed points also will mean less acceptance and hence a slower acceleration of global conformational change, which is what the main purpose of 'non- local' move of this nature will be used for here. 

So the potential advantages of the FRESS move, for example in possibly aiding the bond re-bridging move in accessing low temperature configurations, does not compete with the advantages of the FRW move in rapidly changing the configuration of the chain.

\subsubsection{\textcolor{black}{LCSAW and Excluded Volume Barriers}}
\label{sec:LCSAWBAR}
As highlighted in section \ref{sec:hpmodel} valid configurations of the protein chain are those which respect the conditions that only one monomer can  occupy a lattice site and that the chain is simply connected. It is absolutely essential to the Wang Landau sampling scheme and native state search that only valid configurations of the system are taken into consideration, since the resulting density of states will be wrong for the assumed system. Hence barriers which block any illegal configurations from entering the Wang Landau scheme have been imposed, if such an illegal configuration is found it is rejected and the last valid configuration becomes the present one. Once the old configuration has to be used again its corresponding histogram and density of states is updated accordingly. 

\subsubsection{Trial Move Testing}

To ensure the trial move algorithms we operating as intended and were implemented in a somewhat random way, tests on each trial move algorithm were conducted. The tests involved running the move algorithms on their own (except the kink flip algorithm \footnote{This is due to the fact the chain started out as 'linear' where there were no kinks in the chain. To create new kinks the pullmove or pivotmove needed to be present and the kink move would act on any existing kinks. This allowed me to see if it actually worked. }) and manually checking the coordinates of the monomers after each move. 

The random number generator used in this testing procedure and throughout this simulation is outlined in random.C (ref appendix). 

Some brief example chain pathway diagrams, which show the configuration of the chain in increments of move time, are presented for the move algorithms. I programmed the test so that the user screen prints out the 1D array values $\epsilon_i$ for $i \in \lbrace 1,....,N \rbrace$ which I then drew out the corresponding chain diagram. 

\subsubsection*{Pivot Move Tests}

\begin{figure}[h]
\begin{center}
\fbox{\begin{tikzpicture}[xscale=0.45,yscale=0.45]
\draw [black,fill=black](0,0) circle [radius=0.20];
\draw [ultra thick](0,0) --(1,0);
\draw [black,fill=black](1,0) circle [radius=0.20];
\draw [ultra thick](1,0) --(2,0);
\draw [ultra thick](2,0) --(3,0);
\draw [black,fill=white](2,0) circle [radius=0.20];
\draw [black,fill=black](3,0) circle [radius=0.20];
\draw [ultra thick](3,0) --(4,0);
\draw [black,fill=black](4,0) circle [radius=0.20];
\draw [ultra thick](4,0) --(5,0);
\draw [black,fill=white](5,0) circle [radius=0.20];

\draw [black,fill=black](7,0) circle [radius=0.20];
\draw [ultra thick](7,0) --(8,0);
\draw [black,fill=black](8,0) circle [radius=0.20];
\draw [ultra thick](8,0) --(9,0);
\draw [ultra thick](9,0) --(10,0);
\draw [black,fill=white](9,0) circle [radius=0.20];
\draw [black,fill=black](10,0) circle [radius=0.20];
\draw [ultra thick](10,0) --(10,1);
\draw [black,fill=black](10,1) circle [radius=0.20];
\draw [ultra thick](10,1) --(10,2);
\draw [black,fill=white](10,2) circle [radius=0.20];

\draw [black,fill=black](13,-1) circle [radius=0.20];
\draw [ultra thick](13,-1) --(13,0);
\draw [black,fill=black](13,0) circle [radius=0.20];
\draw [ultra thick](13,0) --(14,0);
\draw [ultra thick](14,0) --(15,0);
\draw [black,fill=white](14,0) circle [radius=0.20];
\draw [black,fill=black](15,0) circle [radius=0.20];
\draw [ultra thick](15,0) --(15,1);
\draw [black,fill=black](15,1) circle [radius=0.20];
\draw [ultra thick](15,1) --(15,2);
\draw [black,fill=white](15,2) circle [radius=0.20];

\draw [black,fill=black](17,0) circle [radius=0.20];
\draw [ultra thick](17,0) --(18,0);
\draw [black,fill=black](18,0) circle [radius=0.20];
\draw [ultra thick](18,0) --(18,-1);
\draw [ultra thick](18,-1) --(19,-1);
\draw [black,fill=white](18,-1) circle [radius=0.20];
\draw [black,fill=black](19,-1) circle [radius=0.20];
\draw [ultra thick](19,-1) --(19,0);
\draw [black,fill=black](19,0) circle [radius=0.20];
\draw [ultra thick](19,1) --(19,0);
\draw [black,fill=white](19,1) circle [radius=0.20];
\draw [green][dashed,ultra thick](18,0) --(19,0);

\draw [black,fill=black](22,0) circle [radius=0.20];
\draw [ultra thick](22,0) --(23,0);
\draw [black,fill=black](23,0) circle [radius=0.20];
\draw [ultra thick](23,0) --(23,-1);
\draw [ultra thick](23,-1) --(24,-1);
\draw [black,fill=white](23,-1) circle [radius=0.20];
\draw [black,fill=black](24,-1) circle [radius=0.20];
\draw [ultra thick](24,-1) --(24,0);
\draw [black,fill=black](24,0) circle [radius=0.20];
\draw [ultra thick](25,0) --(24,0);
\draw [black,fill=white](25,0) circle [radius=0.20];
\draw [green][dashed,ultra thick](23,0) --(24,0);

\end{tikzpicture}}
\caption{The evolution of this chain goes from left to right. (H) monomers are represented as black circles and (P) monomers are represented as white circles. The green dashed lines represent H-H contacts and for this chain represent native states, since the maximum number of H-H contacts is 1.   \label{fig:pivottest}}
\end{center}
\end{figure}
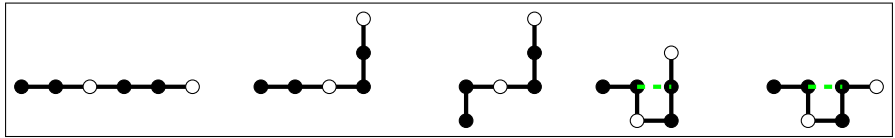

These chain pathways (figure \ref{fig:pivottest}) represent pivot move operations only on a 6-mer (HHPHHP), with 29 total attempt move operations and with random seed $\#$ 9062. Lattice side length $L$ = 300. The configurations shown are the ones that actually changed the configuration as many were rejected. The acceptance ratio, for this test, was 5/29.

One can see that for the chain pathways in figure \ref{fig:pivottest} in 5/29 successful moves the pivot move algorithm on its own has found 2 native degenerate states for the 6-mer. 

\begin{figure}[h]
\begin{center}
\fbox{\begin{tikzpicture}[xscale=0.45,yscale=0.45]
\draw [black,fill=black](0,0) circle [radius=0.20];
\draw [ultra thick](0,0) -- (1,0);
\draw [black,fill=black](1,0) circle [radius=0.20];
\draw [ultra thick](1,0) -- (2,0);
\draw [black,fill=white](2,0) circle [radius=0.20];
\draw [ultra thick](2,0) -- (3,0);
\draw [black,fill=white](2,0) circle [radius=0.20];
\draw [black,fill=white](3,0) circle [radius=0.20];
\draw [ultra thick](3,0) -- (4,0);
\draw [black,fill=white](3,0) circle [radius=0.20];
\draw [black,fill=black](4,0) circle [radius=0.20];
\draw [ultra thick](4,0) -- (5,0);
\draw [black,fill=black](5,0) circle [radius=0.20];
\draw [ultra thick](5,0) -- (6,0);
\draw [black,fill=white](6,0) circle [radius=0.20];
\draw [ultra thick](6,0) -- (7,0);
\draw [black,fill=white](6,0) circle [radius=0.20];
\draw [black,fill=white](7,0) circle [radius=0.20];
\draw [ultra thick](7,0) -- (8,0);
\draw [black,fill=white](7,0) circle [radius=0.20];
\draw [black,fill=black](8,0) circle [radius=0.20];
\draw [ultra thick](8,0) -- (9,0);
\draw [black,fill=black](9,0) circle [radius=0.20];

\draw [black,fill=black](12,0) circle [radius=0.20];
\draw [ultra thick](12,0) -- (13,0);
\draw [black,fill=black](13,0) circle [radius=0.20];
\draw [ultra thick](13,0) -- (13,-1);
\draw [ultra thick](13,-1) -- (13,-2);
\draw [black,fill=white](13,-1) circle [radius=0.20];
\draw [ultra thick](13,-2) -- (14,-2);
\draw [black,fill=white](13,-2) circle [radius=0.20];
\draw [black,fill=black](14,-2) circle [radius=0.20];
\draw [ultra thick](14,-2) -- (15,-2);
\draw [black,fill=black](15,-2) circle [radius=0.20];
\draw [ultra thick](15,-2) -- (15,-1);
\draw [ultra thick](15,-1) -- (16,-1);
\draw [black,fill=white](15,-1) circle [radius=0.20];
\draw [ultra thick](16,-1) -- (17,-1);
\draw [black,fill=white](16,-1) circle [radius=0.20];
\draw [black,fill=black](17,-1) circle [radius=0.20];
\draw [ultra thick](17,-1) -- (18,-1);
\draw [black,fill=black](18,-1) circle [radius=0.20];

\end{tikzpicture}}
\caption{ The 10-mer before any moves (left) and after 500 pivotmoves (right). The HP sequence (HHPPHHPPHH) of monomers remain invariant and the chain remains intact which means the moves respect the conditions of the HP model and LCSAW.  \label{fig:pivottest2}}
\end{center}
\end{figure}
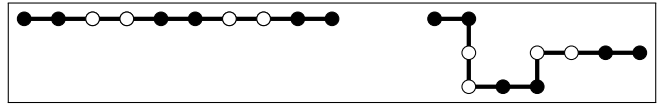

A more thorough test was conducted which comprised of 500 moves and only the starting and ending configuration was recorded to check the chain was still intact. The test was on a 10-mer (HHPPHHPPHH) using a random seed $\#$ 9062. The lattice side length $\L$ = 300. The pictorial results of the test are shown in figure \ref{fig:pivottest2}.

\subsubsection*{Pull Move Tests}

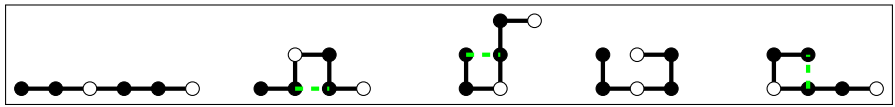
\begin{figure}[h]
\begin{center}
\fbox{\begin{tikzpicture}[xscale=0.45,yscale=0.45]
\draw [black,fill=black](0,0) circle [radius=0.20];
\draw [ultra thick](0,0) --(1,0);
\draw [black,fill=black](1,0) circle [radius=0.20];
\draw [ultra thick](1,0) --(2,0);
\draw [ultra thick](2,0) --(3,0);
\draw [black,fill=white](2,0) circle [radius=0.20];
\draw [black,fill=black](3,0) circle [radius=0.20];
\draw [ultra thick](3,0) --(4,0);
\draw [black,fill=black](4,0) circle [radius=0.20];
\draw [ultra thick](4,0) --(5,0);
\draw [black,fill=white](5,0) circle [radius=0.20];

\draw [black,fill=black](7,0) circle [radius=0.20];
\draw [ultra thick](7,0) --(8,0);
\draw [black,fill=black](8,0) circle [radius=0.20];
\draw [ultra thick](8,0) --(8,1);
\draw [ultra thick](8,1) --(9,1);
\draw [black,fill=white](8,1) circle [radius=0.20];
\draw [black,fill=black](9,1) circle [radius=0.20];
\draw [ultra thick](9,1) --(9,0);
\draw [black,fill=black](9,0) circle [radius=0.20];
\draw [ultra thick](9,0) --(10,0);
\draw [black,fill=white](10,0) circle [radius=0.20];
\draw [green][dashed,ultra thick](8,0) --(9,0);

\draw [black,fill=black](13,1) circle [radius=0.20];
\draw [ultra thick](13,1) --(13,0);
\draw [black,fill=black](13,0) circle [radius=0.20];
\draw [ultra thick](13,0) --(14,0);
\draw [ultra thick](14,0) --(14,1);
\draw [black,fill=white](14,0) circle [radius=0.20];
\draw [black,fill=black](14,1) circle [radius=0.20];
\draw [ultra thick](14,1) --(14,2);
\draw [black,fill=black](14,2) circle [radius=0.20];
\draw [ultra thick](14,2) --(15,2);
\draw [black,fill=white](15,2) circle [radius=0.20];
\draw [green][dashed,ultra thick](13,1) --(14,1);

\draw [black,fill=black](17,1) circle [radius=0.20];
\draw [ultra thick](17,1) --(17,0);
\draw [black,fill=black](17,0) circle [radius=0.20];
\draw [ultra thick](18,0) --(19,0);
\draw [ultra thick](17,0) --(18,0);
\draw [black,fill=white](18,0) circle [radius=0.20];
\draw [black,fill=black](19,0) circle [radius=0.20];
\draw [ultra thick](18,1) --(19,1);
\draw [black,fill=black](19,1) circle [radius=0.20];
\draw [ultra thick](19,1) --(19,0);
\draw [black,fill=white](18,1) circle [radius=0.20];

\draw [black,fill=black](22,1) circle [radius=0.20];
\draw [ultra thick](22,1) --(23,1);
\draw [black,fill=black](23,1) circle [radius=0.20];
\draw [ultra thick](22,1) --(22,0);
\draw [ultra thick](22,0) --(23,0);
\draw [black,fill=white](22,0) circle [radius=0.20];
\draw [black,fill=black](23,0) circle [radius=0.20];
\draw [ultra thick](23,0) --(24,0);
\draw [black,fill=black](24,0) circle [radius=0.20];
\draw [ultra thick](25,0) --(24,0);
\draw [black,fill=white](25,0) circle [radius=0.20];
\draw [green][dashed,ultra thick](23,0) --(23,1);

\end{tikzpicture}}
\caption{The first five configurations representing the evolution of the 6-mer via pull moves. One can see that three unique native states have been found.  \label{fig:pulltest}}
\end{center}
\end{figure}

Using a 6-mer (HHPHHP) a sequence of chain pathways was produced using pull moves only, with 29 total attempt move operations and with a random seed $\#$ 9062. Lattice side length $L$ = 300. The configurations shown in figure \ref{fig:pulltest} are the first 5 configurations from the sequence (for illustration purposes) as all pull moves were successfully done.

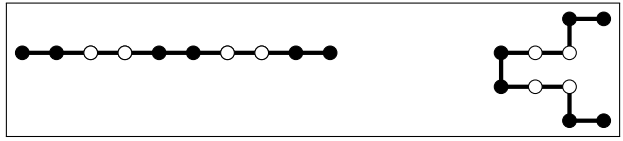
\begin{figure}[h]
\begin{center}
\fbox{\begin{tikzpicture}[xscale=0.45,yscale=0.45]
\draw [black,fill=black](0,0) circle [radius=0.20];
\draw [ultra thick](0,0) -- (1,0);
\draw [black,fill=black](1,0) circle [radius=0.20];
\draw [ultra thick](1,0) -- (2,0);
\draw [black,fill=white](2,0) circle [radius=0.20];
\draw [ultra thick](2,0) -- (3,0);
\draw [black,fill=white](2,0) circle [radius=0.20];
\draw [black,fill=white](3,0) circle [radius=0.20];
\draw [ultra thick](3,0) -- (4,0);
\draw [black,fill=white](3,0) circle [radius=0.20];
\draw [black,fill=black](4,0) circle [radius=0.20];
\draw [ultra thick](4,0) -- (5,0);
\draw [black,fill=black](5,0) circle [radius=0.20];
\draw [ultra thick](5,0) -- (6,0);
\draw [black,fill=white](6,0) circle [radius=0.20];
\draw [ultra thick](6,0) -- (7,0);
\draw [black,fill=white](6,0) circle [radius=0.20];
\draw [black,fill=white](7,0) circle [radius=0.20];
\draw [ultra thick](7,0) -- (8,0);
\draw [black,fill=white](7,0) circle [radius=0.20];
\draw [black,fill=black](8,0) circle [radius=0.20];
\draw [ultra thick](8,0) -- (9,0);
\draw [black,fill=black](9,0) circle [radius=0.20];

\draw [ultra thick](17,-2) -- (16,-2);
\draw [ultra thick](17,1) -- (16,1);
\draw [ultra thick](16,1) -- (16,0);
\draw [ultra thick](16,-1) -- (16,-2);
\draw [ultra thick](15,-1) -- (16,-1);
\draw [ultra thick](14,-1) -- (15,-1);
\draw [ultra thick](14,-1) -- (14,0);
\draw [ultra thick](14,0) -- (15,0);
\draw [ultra thick](15,0) -- (16,0);
\draw [black,fill=black](16,1) circle [radius=0.20];
\draw [black,fill=black](14,0) circle [radius=0.20];
\draw [black,fill=white](15,0) circle [radius=0.20];
\draw [black,fill=white](16,0) circle [radius=0.20];
\draw [black,fill=black](16,-2) circle [radius=0.20];
\draw [black,fill=black](14,-1) circle [radius=0.20];
\draw [black,fill=white](15,-1) circle [radius=0.20];
\draw [black,fill=white](16,-1) circle [radius=0.20];
\draw [black,fill=black](17,-2) circle [radius=0.20];
\draw [black,fill=black](17,1) circle [radius=0.20];
\end{tikzpicture}}
\caption{ The 10-mer before any moves (left) and after 500 pull moves (right). The HP sequence (HHPPHHPPHH) of monomers remain invariant and the chain remains intact which means the moves respect the conditions of the HP model and LCSAW.  \label{fig:pulltest2}}
\end{center}
\end{figure}

For the pull move algorithm, as was done with the pivot move algorithm, a test was performed consisting of 500 moves in which only the starting and ending configuration was recorded. The test was on a 10-mer (HHPPHHPPHH) using a random seed $\#$ 9062 and with the usual lattice side length $L$ = 300. The pictorial results of the test are shown in figure \ref{fig:pulltest2}.

\subsubsection*{Kink Flip Move Tests}

The kink flip move was described in section \ref{sec:trialmove}. Since, in these tests, the chain starts as a linear one where no kinks are present it was necessary to include another move to create kinks to see if the kink flip move was functioning correctly. This does not affect the quality of the testing since the configuration coordinates were printed after every pull and kink move with clear labelling as to what move caused the resulting configuration. 

A series of chain pathways, as was done for the pull and pivot move algorithms, were generated. The chain starts linear and then a pull move is executed, then a kink flip move. This is done throughout the testing: first pull then perform a kink flip move. 

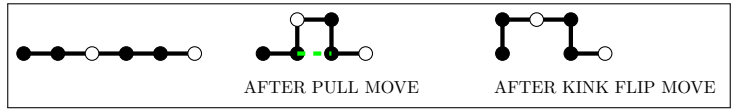
\begin{figure}[h]
\begin{center}
\fbox{\begin{tikzpicture}[xscale=0.45,yscale=0.45]
\draw [black,fill=black](0,0) circle [radius=0.20];
\draw [ultra thick](0,0) --(1,0);
\draw [black,fill=black](1,0) circle [radius=0.20];
\draw [ultra thick](1,0) --(2,0);
\draw [ultra thick](2,0) --(3,0);
\draw [black,fill=white](2,0) circle [radius=0.20];
\draw [black,fill=black](3,0) circle [radius=0.20];
\draw [ultra thick](3,0) --(4,0);
\draw [black,fill=black](4,0) circle [radius=0.20];
\draw [ultra thick](4,0) --(5,0);
\draw [black,fill=white](5,0) circle [radius=0.20];

\draw [black,fill=black](7,0) circle [radius=0.20];
\draw [ultra thick](7,0) --(8,0);
\draw [black,fill=black](8,0) circle [radius=0.20];
\draw [ultra thick](8,0) --(8,1);
\draw [ultra thick](9,0) --(9,1);
\draw [black,fill=black](9,1) circle [radius=0.20];
\draw [ultra thick](8,1) --(9,1);
\draw [black,fill=white](8,1) circle [radius=0.20];
\draw [black,fill=black](9,0) circle [radius=0.20];
\draw [ultra thick](9,0) --(10,0);
\draw [black,fill=white](10,0) circle [radius=0.20];
\draw [dashed,ultra thick][green] (8,0) --(9,0);

\node () at (9,-1)[xscale=0.6,yscale=0.6] {AFTER PULL MOVE};

\draw [black,fill=black](14,0) circle [radius=0.20];
\draw [ultra thick](14,0) --(14,1);
\draw [black,fill=black](14,1) circle [radius=0.20];
\draw [ultra thick](14,1) --(15,1);
\draw [ultra thick](16,0) --(16,1);
\draw [black,fill=black](16,1) circle [radius=0.20];
\draw [ultra thick](15,1) --(16,1);
\draw [black,fill=white](15,1) circle [radius=0.20];
\draw [black,fill=black](16,0) circle [radius=0.20];
\draw [ultra thick](16,0) --(17,0);
\draw [black,fill=white](17,0) circle [radius=0.20];

\node () at (17,-1)[xscale=0.6,yscale=0.6] {AFTER KINK FLIP MOVE};

\end{tikzpicture}}
\caption{ Simple chain pathway from linear chain $\rightarrow$ pull moved chain $\rightarrow$ kink of chain being flipped successfully. \label{fig:kinktest}}
\end{center}
\end{figure}

\begin{figure}[h]
\begin{center}
\fbox{\begin{tikzpicture}[xscale=0.45,yscale=0.45]
\draw [black,fill=white](2,2) circle [radius=0.20];
\draw [ultra thick](2,2) --(2,1);
\draw [black,fill=black](2,1) circle [radius=0.20];
\draw [ultra thick](2,0) --(2,1);
\draw [ultra thick](2,0) --(3,0);
\draw [black,fill=black](2,0) circle [radius=0.20];
\draw [black,fill=white](3,0) circle [radius=0.20];
\draw [ultra thick](3,0) --(4,0);
\draw [black,fill=black](4,0) circle [radius=0.20];
\draw [ultra thick](4,0) --(5,0);
\draw [black,fill=black](5,0) circle [radius=0.20];

\draw [black,fill=white](9,2) circle [radius=0.20];
\draw [ultra thick](9,2) --(9,1);
\draw [black,fill=black](9,1) circle [radius=0.20];
\draw [ultra thick](10,1) --(9,1);
\draw [ultra thick](10,1) --(10,0);
\draw [black,fill=black](10,1) circle [radius=0.20];
\draw [black,fill=white](10,0) circle [radius=0.20];
\draw [ultra thick](10,0) --(11,0);
\draw [black,fill=black](11,0) circle [radius=0.20];
\draw [ultra thick](11,0) --(12,0);
\draw [black,fill=black](12,0) circle [radius=0.20];
\node () at (11,-1)[xscale=0.6,yscale=0.6] {AFTER KINK MOVE};

\draw [black,fill=white](16,1) circle [radius=0.20];
\draw [ultra thick](16,1) --(17,1);
\draw [black,fill=black](17,1) circle [radius=0.20];
\draw [ultra thick](18,1) --(17,1);
\draw [ultra thick](18,1) --(18,0);
\draw [black,fill=black](18,1) circle [radius=0.20];
\draw [black,fill=white](18,0) circle [radius=0.20];
\draw [ultra thick](18,0) --(19,0);
\draw [black,fill=black](19,0) circle [radius=0.20];
\draw [ultra thick](19,0) --(20,0);
\draw [black,fill=black](20,0) circle [radius=0.20];
\node () at (19,-1)[xscale=0.6,yscale=0.6] {AFTER PULL MOVE};

\draw [black,fill=white](24,1) circle [radius=0.20];
\draw [ultra thick](24,1) --(25,1);
\draw [black,fill=black](25,1) circle [radius=0.20];
\draw [ultra thick](26,1) --(25,1);
\draw [ultra thick](26,1) --(27,1);
\draw [black,fill=black](26,1) circle [radius=0.20];
\draw [black,fill=white](27,1) circle [radius=0.20];
\draw [ultra thick](27,1) --(27,0);
\draw [black,fill=black](27,0) circle [radius=0.20];
\draw [ultra thick](27,0) --(28,0);
\draw [black,fill=black](28,0) circle [radius=0.20];
\node () at (27,-1)[xscale=0.6,yscale=0.6] {AFTER KINK MOVE};

\end{tikzpicture}}
\caption{A longer sequence of successful pull and kink flip moves.  \label{fig:kinktesttwo}}
\end{center}
\end{figure}
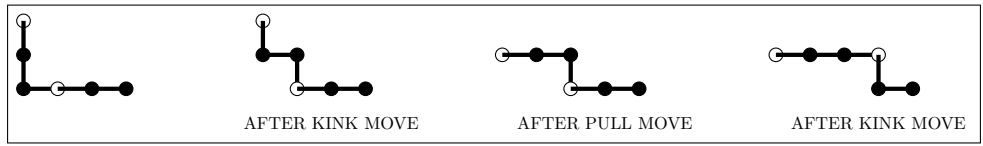

The chain pathways presented in figures \ref{fig:kinktest} and \ref{fig:kinktesttwo} are snippets of the sequence of 29 moves where the kink flip move was not rejected. The 6-mer (HHPHHP) was used aswell as with $L$ = 300 and random seed $\#$ 9062. 

\begin{figure}[h]
\begin{center}
\fbox{\begin{tikzpicture}[xscale=0.45,yscale=0.45]
\draw [black,fill=black](0,0) circle [radius=0.20];
\draw [ultra thick](0,0) -- (1,0);
\draw [black,fill=black](1,0) circle [radius=0.20];
\draw [ultra thick](1,0) -- (2,0);
\draw [black,fill=white](2,0) circle [radius=0.20];
\draw [ultra thick](2,0) -- (3,0);
\draw [black,fill=white](2,0) circle [radius=0.20];
\draw [black,fill=white](3,0) circle [radius=0.20];
\draw [ultra thick](3,0) -- (4,0);
\draw [black,fill=white](3,0) circle [radius=0.20];
\draw [black,fill=black](4,0) circle [radius=0.20];
\draw [ultra thick](4,0) -- (5,0);
\draw [black,fill=black](5,0) circle [radius=0.20];
\draw [ultra thick](5,0) -- (6,0);
\draw [black,fill=white](6,0) circle [radius=0.20];
\draw [ultra thick](6,0) -- (7,0);
\draw [black,fill=white](6,0) circle [radius=0.20];
\draw [black,fill=white](7,0) circle [radius=0.20];
\draw [ultra thick](7,0) -- (8,0);
\draw [black,fill=white](7,0) circle [radius=0.20];
\draw [black,fill=black](8,0) circle [radius=0.20];
\draw [ultra thick](8,0) -- (9,0);
\draw [black,fill=black](9,0) circle [radius=0.20];

\draw [ultra thick](11,0) -- (12,0);
\draw [ultra thick](12,0) -- (12,1);
\draw [ultra thick](12,1) -- (13,1);
\draw [ultra thick](13,1) -- (13,2);
\draw [ultra thick](13,2) -- (14,2);
\draw [ultra thick](14,2) -- (14,1);
\draw [ultra thick](14,1) -- (15,1);
\draw [ultra thick](15,1) -- (15,0);
\draw [ultra thick](15,0) -- (16,0);
\draw [black,fill=black](11,0) circle [radius=0.20];
\draw [black,fill=black](12,0) circle [radius=0.20];
\draw [black,fill=white](12,1) circle [radius=0.20];
\draw [black,fill=white](13,1) circle [radius=0.20];
\draw [black,fill=black](13,2) circle [radius=0.20];
\draw [black,fill=black](14,2) circle [radius=0.20];
\draw [black,fill=white](14,1) circle [radius=0.20];
\draw [black,fill=white](15,1) circle [radius=0.20];
\draw [black,fill=black](15,0) circle [radius=0.20];
\draw [black,fill=black](16,0) circle [radius=0.20];

\end{tikzpicture}}
\caption{ The starting linear chain (left) and the resulting chain (right) after 1000 moves.\label{fig:kinktest2}}
\end{center}
\end{figure}
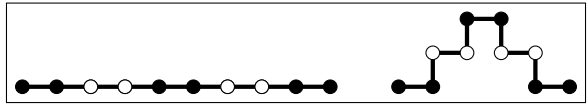

A larger test, as with the previous move tests, was conducted using a 10-mer (HHPPHHPPHH). The sequence of moves was pull $\rightarrow$ kink flip  $\rightarrow$ pull  $\rightarrow$ kink flip etc. for a total of 1000 moves. 500 kink flips and 500 pulls were conducted. In this test the same random seed $\#$ 9062 and lattice side length $L$ = 300 was used as before. The starting configuration and ending configuration shown in figure \ref{fig:kinktest2} was recorded. 

Since the FRW and bond re-bridging moves were included in latter versions of the simulation code, extensive move testing was not conducted. However small trials were run to manually check (drawing out the configurations for small chains) the robustness and execution of the move algorithms. 

\subsubsection*{Test Conclusions}

From these basic tests it is clear that the trial set moves perform their intended operations on the chain. There is a possibility that the trial move sets can produce illegal chain configurations, since no human can predict how or when this will happen it is best to place a configuration barrier as outlined in subsection \ref{sec:LCSAWBAR}.  

\subsection{Energy Computing Routine}

To register configurations which are in potential native states and to compute the total energy of the system using equation \ref{eq:Hamiltonian} for the Monte Carlo procedures, it is necessary to have an energy computing subroutine within the program. The routine needs to sum all the \textit{topological} H-H contacts and the total energy of the system, using $\epsilon_{HH}$ = 1, would simply be the negative of this sum. A monomer $B$ is said to the topological neighbour of monomer $A$ when the 1D array coordinate of B, $B_{\epsilon}$, is such that $B_{\epsilon} \in \lbrace A_{\epsilon} + 1, A_{\epsilon} - 1, A_{\epsilon} + L, A_{\epsilon} - L \rbrace$ and when the sequence value of B, $s_B$, $\neq s_A + 1$ and $\neq s_A - 1$.

The test of this routine, which was conducted early on in the development of the simulation, is shown in appendix \ref{sec:B} where native states of very short chained proteins are found, using a simple scoring system. 

\newpage
\section{Energy Interval Experiment for WLS}
\label{sec:EIEW}

Since in the Wang Landau sampling regime the reduction of the modification factor, $f$, directly dictates approximate convergence to the correct density of states, it is important to consider the energy ranges for the histogram. This consideration is unique to systems in which the difficulty of sampling configuration space grows with decreasing temperature. 

In this protein folding model the difficulty in sampling dense low temperature configurations is known \cite{wustlandau} \cite{oplandau} \cite{longrange} \cite{wust} and when exploring the thermodynamic behaviour of folding and unfolding processes one has to strike a balance between convergence and exploring very deep wells in the energy landscape. This balance is a conflict between computational time and desire for detail.

Five simulations were run for the sequence 2D64 with different seeds and energy ranges see table \ref{table:enetable}.

\begin{table}[h]
\centering
\begin{tabular}{c | c | c | c}
 Run Number & Energy Range & $ln[f_{final}]$ & Seed   \\
 \hline\hline
1 & 0:(-38)& $\approx$ 0.0002 &  591418 \\
 \hline
2 & 0:(-30)& $\approx$ 2 $\cdot 10^{-180}$ &  655512 \\
 \hline
 3 & 0:(-40)& 0.5 &  40824 \\
 \hline
4 & 0:(-25)& $\approx 9 \cdot 10^{-1324}$ &   197881\\
\hline
5 & 0:(-37)& $\approx 0.001$ & 251351\\
 
\end{tabular}
\caption{ For the energy ranges 0 is the upper bound also note that $ln[f_{initial}] = 1$ as outlined in section \ref{sec:wlmc}.\label{table:enetable}}
\end{table}

\begin{figure}[h]
\begin{center}
\fbox{\includegraphics[scale=0.3]{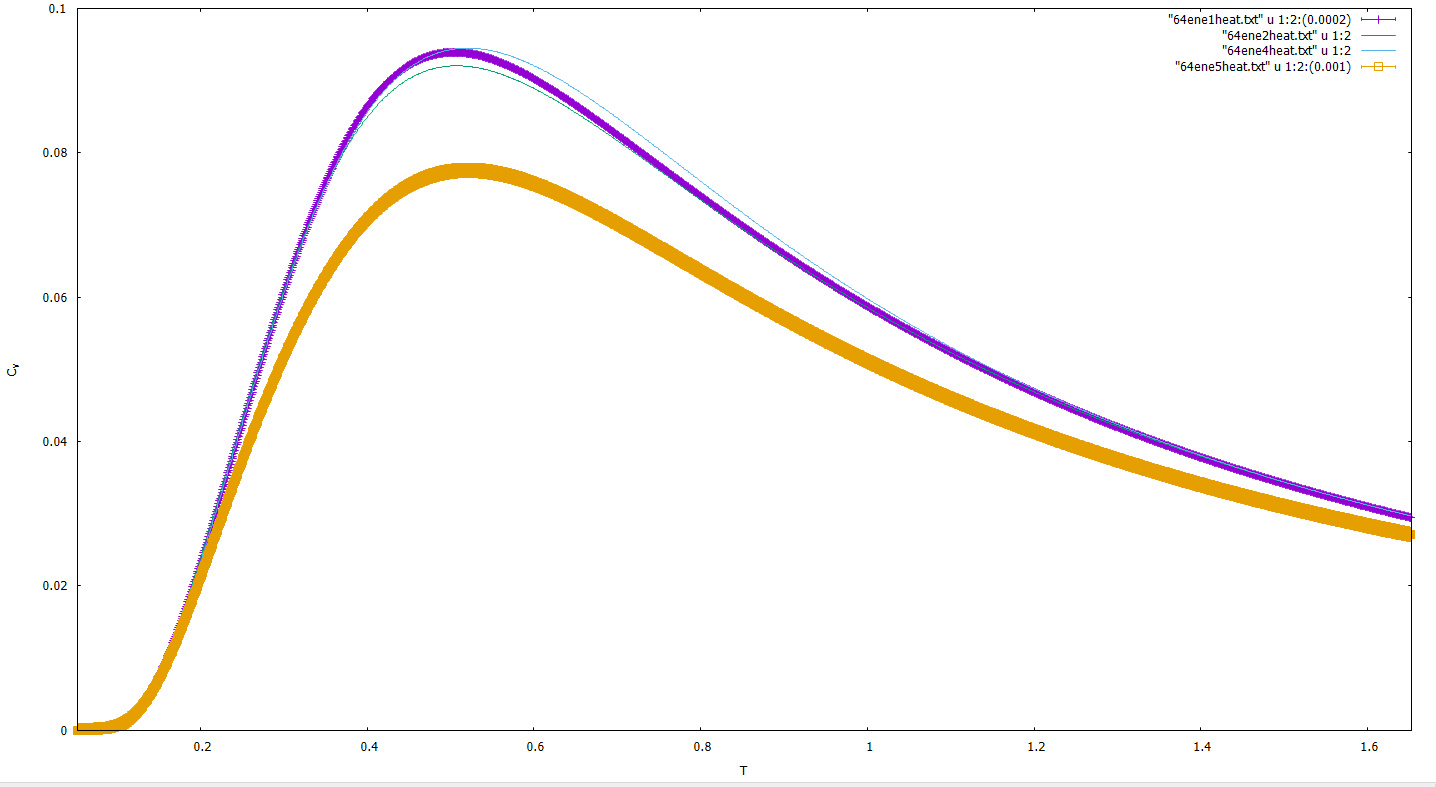}}
\caption{Purple = run 1, green = run 2, light blue = run 4 and gold = run 5. The error in $C_v$ is the final modification factor shown in table \ref{table:enetable}, the errors for run 2 and 4 were omitted since they are smaller than the data points. }
\label{fig:64eneheats}
\end{center}
\end{figure}

The specific heat results of run 1, 2, 4 and 5 are shown in figure \ref{fig:64eneheats}. Observables from run 3 were omitted due to their drastic nature as the error bars were orders of magnitude larger than the results.

\begin{figure}[h]
\begin{center}
\fbox{\includegraphics[scale=0.3]{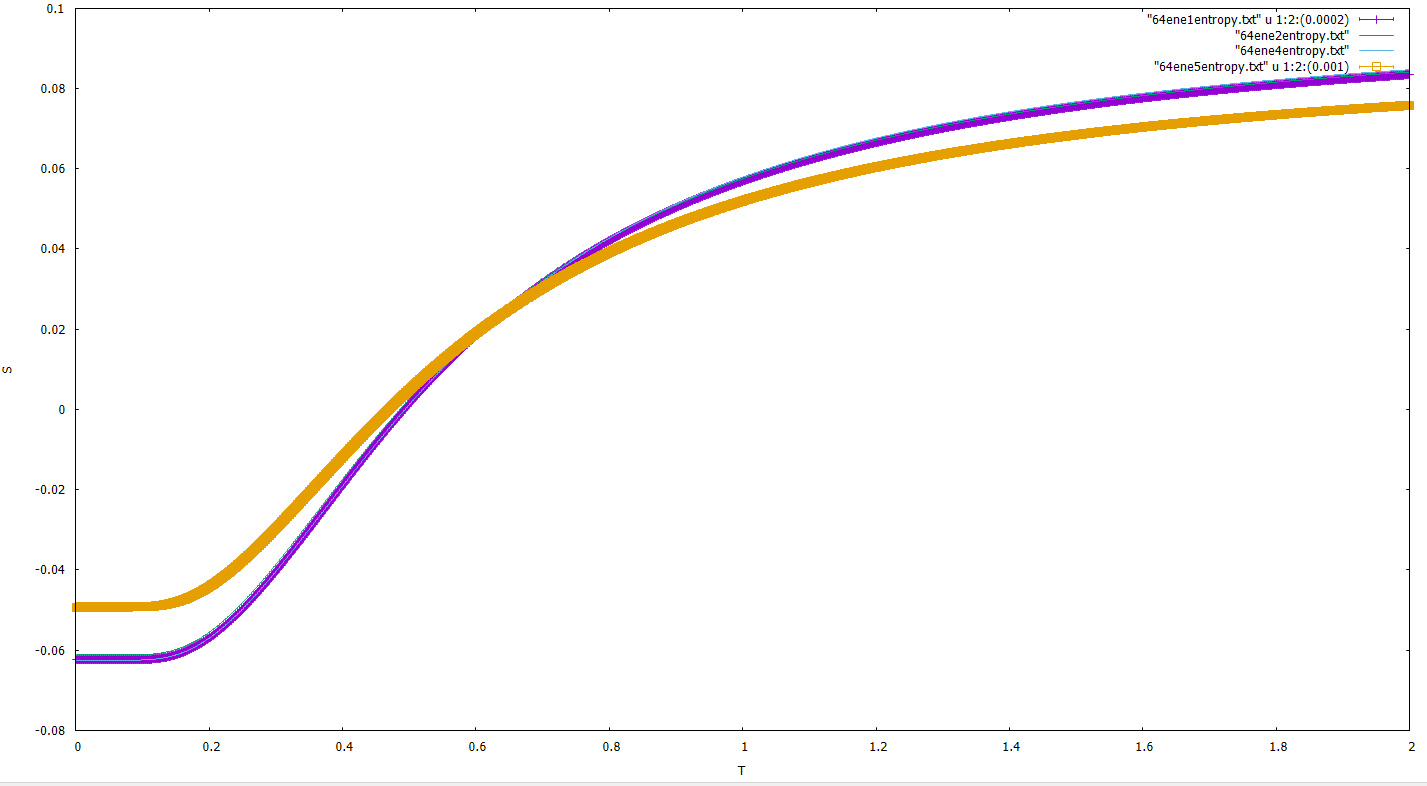}}
\caption{Purple = run 1, green = run 2, light blue = run 4 and gold = run 5. The error in $S$ is the final modification factor shown in table \ref{table:enetable}, the errors for run 2 and 4 were omitted since they are smaller than the data points. }
\label{fig:64eneentropys}
\end{center}
\end{figure}

The entropy for the same runs is shown in figure \ref{fig:64eneentropys}. 

\subsection{Discussions and Remarks}
The final modification factor is a sign of how well the WL sampling converged and as expected run 4 resulted in the lowest factor. Run 3 only had its modification factor reduced only once which reflects the difficulty WL sampling faces when encompassing the low temperature regions. Run 1 had a sub-par resulting modification factor and run 2 converged extremely well. It is not established whether the modification factor of run 4 took on 1/t functionality. Run 5 has a final modification factor which is greater than run 1 whilst having a lower energy range. This fact could hint towards the inevitability of having more than a Gaussian threshold amount of runs for results to be statistically meaningful. 

One can see that the worst converged simulation (run 5) in figure \ref{fig:64eneheats} underestimates the specific heat capacity for the sequence, even though the energy range is larger compared to run 2 and 4. The other curves are strikingly similar despite having varying sampling energy ranges. This could imply that cutting off the difficult, near native region, in the sampling is not as detrimental to the observables as was assumed.

Not surprisingly, for the entropy (see figure \ref{fig:64eneentropys}), one can see the difference between the observable computed from run 5 to the others, also note that all runs produced  $S < 0$ at very low temperatures (before $T_{c} \approx 0.51$) which of course is not physically viable. This occurrence could be itself due to the lack of sampling of low temperature configurations mixed in with poor WL convergence. 

This experiment has highlighted the need to take care in deciding the ultimate energy range for the WLS scheme for protein sequences. One needs to allow low temperature behaviour to be explored but without too much cost in accuracy. Also to attain decent modification factor reduction the routine must be run for a significant amount of time.

The lessons acquired from this small experiment were used in obtaining the final results.

\clearpage
\section{\textcolor{black}{Results}}

\subsection{\textcolor{black}{Native State Search}}

With the following move set ratios: 65$\%$ pull, 19$\%$ bond re-bridging (of which 70$\%$ is type II), 10$\%$ fragment random walk, 4$\%$ pivot and 2$\%$ kink-flip, simulation runs were implemented with the specific aim of finding the native state of some benchmark sequences.

The sequences used in these runs were 2D50, 2D64 and 2D85. The (H)(P) sequence of these proteins are as follows:

\begin{description}
\begin{footnotesize}
\item[2D50 (S1-6)] HHPHPHPHPHHHHPHPPPHPPPHPPPPHPPPHPPPHPHHHHPHPHPHPHH 
\item[2D60 (S1-7)] PPHHHPHHHHHHHHPPPHHHHHHHHHHPHPPPHHHHHH\\HHHHHHPPPPHHHHHHPHHPHP 
\item[2D64 (S1-8)] HHHHHHHHHHHHPHPHPPHHPPHHPPHPPHHPPHHPPHPPHHPPHHP\\PHPHPHHHHHHHHHHHH 
\item[2D85 (S1-9)] HHHHPPPPHHHHHHHHHHHHPPPPPPHHHHHHHHHHHHPPPHHHHHHH\\HHHHHPPPHHHHHHHHHHHHPPPHPPHHPPHHPPHPH
\item[2D100a (S1-10)] PPPPPPHPHHPPPPPHHHPHHHHHPHHPPPPHHPPHHPHHHHHPHHHHH\\HHHHHPHHPHHHHHHHPPPPPPPPPPPHHHHHHHPPHPHHHPPPPPPHPHH
\item[2D100b (S1-11)] PPPPPPHPHHPPPPPHHHPHHHHHPHHPPPPHHPPHHPHHH\\HHPHHHHHHHHHHPHHPHHHHHHHPPPPPPPPPPPHHHHHHHPPHPHHHPPPPPPHPHH
\end{footnotesize}
\end{description}

Results for the best minimum energy ($E_{min}$) found compared to the best known native states from other methods are shown in table \ref{table:nativeresults}.

\begin{table}[h]
\centering
\resizebox{\columnwidth}{!}{%
\begin{tabular}{c | c | c | c | c | c | c | c | c | c}
 \textbf{Sequence} & \textcolor{blue}{$E_{min}$}& \textbf{WLS} \cite{oplandau} & \textbf{EMC} \cite{EMC} & \textbf{SISPER} \cite{SISPER} & \textbf{GSA}\cite{GSA} & \textbf{nPERMis} \cite{nPERMIS} & \textbf{EES} \cite{EES} & \textbf{FRESS} \cite{frmc} & \textbf{ACO} \cite{ACO} \\
 \hline\hline
2D50 & \textcolor{blue}{-21} & N/A &-21&-21& N/A& N/A& -21& -21& -21\\
 \hline
2D60 & \textcolor{blue}{-36} & N/A &-35&-36&-36&-36 &-36 & -36& -36\\
 \hline
 2D64 & \textcolor{blue}{-42} &-42 &-39 &-39& -42&-42 & -42& -42& -42\\
\hline
2D85 & \textcolor{blue}{-52} &-53 &N/A &-52& -52& -53& -53& -53 & -53\\
\hline
2D100a & \textcolor{blue}{-47} & -48&N/A&-48& -48& -48& -48&  -48& -47 \\
\hline
2D100b & \textcolor{blue}{-49} & -50&N/A &-49& -50& -50&-49 & -50& -49\\
\hline
\end{tabular}}
\caption{ Comparison of native states found in this work (blue) with different methods . \label{table:nativeresults}}
\end{table}

The configuration for the native state of 2D50 and 2D64, found in this work, are shown in figures \ref{fig:native2D50} and \ref{fig:native2D64} respectively.

\begin{figure}[h]
\centering
\fbox{\subcaptionbox{Native state of 2D50 where black monomers are hydrophobic and white monomers are polar.For illustration purposes dotted green lines shown represent contributions to the energy. \label{fig:native2D50}}[10cm]{\begin{tikzpicture}[xscale=1,yscale=1]
\draw [thick](0,0) -- (0,-1);
\draw [dotted,ultra thick][green](0,0) -- (-1,0);
\draw [dotted,ultra thick][green](0,0) -- (1,0);
\draw [dotted,ultra thick][green](0,0) -- (0,1);
\draw [black,fill=black](0,0) circle [radius=0.20];
\draw [thick](0,-1) -- (-1,-1);
\draw [dotted,ultra thick][green](0,-1) -- (1,-1);
\draw [dotted,ultra thick][green](0,-1) -- (0,-2);
\draw [black,fill=black](0,-1) circle [radius=0.20];
\draw [thick](-1,-1) -- (-1,0);
\draw [black,fill=white](-1,-1) circle [radius=0.20];
\draw [dotted,ultra thick][green](-1,0) -- (-2,0);
\draw [thick](-1,0) -- (-1,1);
\draw [black,fill=black](-1,0) circle [radius=0.20];
\draw [thick](-1,1) -- (0,1);
\draw [black,fill=white](-1,1) circle [radius=0.20];
\draw [thick](0,1) -- (1,1);
\draw [black,fill=black](0,1) circle [radius=0.20];
\draw [thick](1,1) -- (1,0);
\draw [black,fill=white](1,1) circle [radius=0.20];
\draw [thick](1,0) -- (2,0);
\draw [dotted,ultra thick][green](1,0) -- (1,-1);
\draw [black,fill=black](1,0) circle [radius=0.20];
\draw [thick](2,0) -- (2,-1);
\draw [black,fill=white](2,0) circle [radius=0.20];
\draw [thick](2,-1) -- (1,-1);
\draw [dotted,ultra thick][green](2,-1) -- (2,-2);
\draw [black,fill=black](2,-1) circle [radius=0.20];
\draw [thick](1,-1) -- (1,-2);
\draw [black,fill=black](1,-1) circle [radius=0.20];
\draw [thick](1,-2) -- (0,-2);
\draw [dotted,ultra thick][green](1,-2) -- (1,-3);
\draw [dotted,ultra thick][green](1,-2) -- (2,-2);
\draw [black,fill=black](1,-2) circle [radius=0.20];
\draw [thick](0,-2) -- (-1,-2);
\draw [dotted,ultra thick][green](0,-2) -- (0,-3);
\draw [black,fill=black](0,-2) circle [radius=0.20];
\draw [thick](-1,-2) -- (-1,-3);
\draw [black,fill=white](-1,-2) circle [radius=0.20];
\draw [thick](-1,-3) -- (-2,-3);
\draw [dotted,ultra thick][green](-1,-3) -- (0,-3);
\draw [dotted,ultra thick][green](-1,-3) -- (-1,-4);
\draw [black,fill=black](-1,-3) circle [radius=0.20];
\draw [thick](-2,-3) -- (-2,-2);
\draw [black,fill=white](-2,-3) circle [radius=0.20];
\draw [thick](-2,-2) -- (-2,-1);
\draw [black,fill=white](-2,-2) circle [radius=0.20];
\draw [thick](-2,-1) -- (-2,0);
\draw [black,fill=white](-2,-1) circle [radius=0.20];
\draw [thick](-2,0) -- (-2,1);
\draw [dotted,ultra thick][green](-2,0) -- (-1,0);
\draw [black,fill=black](-2,0) circle [radius=0.20];
\draw [thick](-2,1) -- (-3,1);
\draw [black,fill=white](-2,1) circle [radius=0.20];
\draw [thick](-3,1) -- (-3,-1);
\draw [black,fill=white](-3,1) circle [radius=0.20];
\draw [black,fill=white](-3,0) circle [radius=0.20];
\draw [thick](-3,-1) -- (-4,-1);
\draw [dotted,ultra thick][green](-3,-1) -- (-3,-2);
\draw [black,fill=black](-3,-1) circle [radius=0.20];
\draw [thick](-4,-1) -- (-5,-1);
\draw [black,fill=white](-4,-1) circle [radius=0.20];
\draw [thick](-5,-1) -- (-5,-2);
\draw [black,fill=white](-5,-1) circle [radius=0.20];
\draw [thick](-5,-2) -- (-4,-2);
\draw [black,fill=white](-5,-2) circle [radius=0.20];
\draw [thick](-4,-2) -- (-3,-2);
\draw [black,fill=white](-4,-2) circle [radius=0.20];
\draw [thick](-3,-2) -- (-3,-3);
\draw [black,fill=black](-3,-2) circle [radius=0.20];
\draw [thick](-3,-3) -- (-3,-4);
\draw [black,fill=white](-3,-3) circle [radius=0.20];
\draw [thick](-3,-4) -- (-2,-4);
\draw [black,fill=white](-3,-4) circle [radius=0.20];
\draw [thick](-2,-4) -- (-1,-4);
\draw [black,fill=white](-2,-4) circle [radius=0.20];
\draw [thick](-1,-4) -- (-1,-5);
\draw [black,fill=black](-1,-4) circle [radius=0.20];
\draw [thick](-1,-5) -- (0,-5);
\draw [black,fill=white](-1,-5) circle [radius=0.20];
\draw [thick](0,-5) -- (1,-5);
\draw [black,fill=white](0,-5) circle [radius=0.20];
\draw [thick](1,-5) -- (1,-4);
\draw [black,fill=white](1,-5) circle [radius=0.20];
\draw [thick](1,-4) -- (0,-4);
\draw [dotted,ultra thick][green](1,-4) -- (2,-4);
\draw [dotted,ultra thick][green](1,-4) -- (1,-3);
\draw [black,fill=black](1,-4) circle [radius=0.20];
\draw [thick](0,-4) -- (0,-3);
\draw [thick](0,-3) -- (1,-3);
\draw [black,fill=white](0,-4) circle [radius=0.20];
\draw [black,fill=black](0,-3) circle [radius=0.20];
\draw [thick](1,-3) -- (2,-3);
\draw [black,fill=black](1,-3) circle [radius=0.20];
\draw [thick](2,-3) -- (2,-2);
\draw [dotted,ultra thick][green](2,-3) -- (2,-4);
\draw [black,fill=black](2,-3) circle [radius=0.20];
\draw [thick](2,-2) -- (3,-2);
\draw [black,fill=black](2,-2) circle [radius=0.20];
\draw [thick](3,-2) -- (3,-3);
\draw [black,fill=white](3,-2) circle [radius=0.20];
\draw [thick](3,-3) -- (4,-3);
\draw [dotted,ultra thick][green](3,-3) -- (2,-3);
\draw [dotted,ultra thick][green](3,-3) -- (3,-4);
\draw [black,fill=black](3,-3) circle [radius=0.20];
\draw [thick](4,-3) -- (4,-4);
\draw [black,fill=white](4,-3) circle [radius=0.20];
\draw [thick](4,-4) -- (4,-5);
\draw [dotted,ultra thick][green](4,-4) -- (3,-4);
\draw [black,fill=black](4,-4) circle [radius=0.20];
\draw [thick](4,-5) -- (3,-5);
\draw [thick](4,-5) -- (3,-5);
\draw [black,fill=white](4,-5) circle [radius=0.20];
\draw [thick](3,-5) -- (2,-5);
\draw [dotted,ultra thick][green](3,-5) -- (3,-4);
\draw [thick](2,-5) -- (2,-4);
\draw [thick](2,-4) -- (3,-4);
\draw [black,fill=black](3,-5) circle [radius=0.20];
\draw [black,fill=white](2,-5) circle [radius=0.20];
\draw [black,fill=black](2,-4) circle [radius=0.20];
\draw [black,fill=black](3,-4) circle [radius=0.20];

\end{tikzpicture}}}

\fbox{\subcaptionbox{Native structure of 2D64 found using this method. \label{fig:native2D64}}[10cm]{\begin{tikzpicture}[xscale=1,yscale=1]
\draw [black,fill=black](0,0) circle [radius=0.20];
\draw [thick](0,0) -- (-1,0);
\draw [black,fill=black](-1,0) circle [radius=0.20];
\draw [thick](-1,0) -- (-1,1);
\draw [black,fill=black](-1,1) circle [radius=0.20];
\draw [thick](-1,1) -- (0,1);
\draw [black,fill=black](0,1) circle [radius=0.20];
\draw [thick](0,1) -- (1,1);
\draw [black,fill=black](1,1) circle [radius=0.20];
\draw [thick](1,1) -- (1,0);
\draw [black,fill=black](1,0) circle [radius=0.20];
\draw [thick](1,0) -- (2,0);
\draw [black,fill=black](2,0) circle [radius=0.20];
\draw [thick](2,0) -- (2,1);
\draw [black,fill=black](2,1) circle [radius=0.20];
\draw [thick](2,1) -- (3,1);
\draw [black,fill=black](3,1) circle [radius=0.20];
\draw [thick](3,1) -- (3,0);
\draw [black,fill=black](3,0) circle [radius=0.20];
\draw [thick](3,0) -- (4,0);
\draw [black,fill=black](4,0) circle [radius=0.20];
\draw [thick](4,0) -- (5,0);
\draw [black,fill=black](5,0) circle [radius=0.20];
\draw [thick](5,0) -- (5,1);
\draw [thick](5,1) -- (4,1);
\draw [black,fill=white](5,1) circle [radius=0.20];
\draw [black,fill=black](4,1) circle [radius=0.20];
\draw [thick](4,1) -- (4,2);
\draw [thick](4,2) -- (3,2);
\draw [black,fill=white](4,2) circle [radius=0.20];
\draw [black,fill=black](3,2) circle [radius=0.20];
\draw [thick](3,2) -- (3,3);
\draw [thick](3,3) -- (2,3);
\draw [black,fill=white](3,3) circle [radius=0.20];
\draw [thick](2,3) -- (2,2);
\draw [black,fill=white](2,3) circle [radius=0.20];
\draw [black,fill=black](2,2) circle [radius=0.20];
\draw [thick](2,2) -- (1,2);
\draw [black,fill=black](1,2) circle [radius=0.20];
\draw [thick](1,2) -- (1,3);
\draw [thick](1,3) -- (0,3);
\draw [thick](0,3) -- (0,2);
\draw [black,fill=white](1,3) circle [radius=0.20];
\draw [black,fill=white](0,3) circle [radius=0.20];
\draw [black,fill=black](0,2) circle [radius=0.20];
\draw [thick](0,2) -- (-1,2);
\draw [black,fill=black](-1,2) circle [radius=0.20];
\draw [thick](-1,2) -- (-1,3);
\draw [thick](-1,3) -- (-2,3);
\draw [thick](-2,3) -- (-2,2);
\draw [black,fill=white](-1,3) circle [radius=0.20];
\draw [black,fill=white](-2,3) circle [radius=0.20];
\draw [black,fill=black](-2,2) circle [radius=0.20];
\draw [thick](-2,2) -- (-3,2);
\draw [thick](-3,2) -- (-3,1);
\draw [thick](-3,1) -- (-2,1);
\draw [thick](-2,1) -- (-2,0);
\draw [thick](-2,0) -- (-3,0);
\draw [thick](-3,0) -- (-3,-1);
\draw [thick](-3,-1) -- (-2,-1);
\draw [thick](-2,-1) -- (-2,-2);
\draw [thick](-2,-2) -- (-3,-2);
\draw [thick](-3,-2) -- (-3,-3);
\draw [thick](-3,-3) -- (-2,-3);
\draw [thick](-2,-3) -- (-2,-4);
\draw [thick](-2,-4) -- (-1,-4);
\draw [thick](-1,-4) -- (-1,-3);
\draw [thick](-1,-3) -- (0,-3);
\draw [thick](0,-3) -- (0,-4);
\draw [thick](0,-4) -- (1,-4);
\draw [thick](1,-4) -- (1,-3);
\draw [thick](1,-3) -- (2,-3);
\draw [thick](2,-3) -- (2,-4);
\draw [thick](2,-4) -- (3,-4);
\draw [thick](3,-4) -- (3,-3);
\draw [thick](3,-3) -- (4,-3);
\draw [thick](4,-3) -- (4,-2);
\draw [thick](4,-2) -- (5,-2);
\draw [thick](5,-2) -- (5,-1);
\draw [thick](5,-1) -- (4,-1);
\draw [thick](4,-1) -- (3,-1);
\draw [thick](3,-1) -- (3,-2);
\draw [thick](3,-2) -- (2,-2);
\draw [thick](2,-2) -- (2,-1);
\draw [thick](2,-1) -- (1,-1);
\draw [thick](1,-1) -- (0,-1);
\draw [thick](0,-1) -- (-1,-1);
\draw [thick](-1,-1) -- (-1,-2);
\draw [thick](-1,-2) -- (0,-2);
\draw [thick](0,-2) -- (1,-2);
\draw [black,fill=white](-3,2) circle [radius=0.20];
\draw [black,fill=white](-3,1) circle [radius=0.20];
\draw [black,fill=white](-3,0) circle [radius=0.20];
\draw [black,fill=white](-3,-1) circle [radius=0.20];
\draw [black,fill=white](-3,-2) circle [radius=0.20];
\draw [black,fill=white](-3,-3) circle [radius=0.20];
\draw [black,fill=white](-2,-4) circle [radius=0.20];
\draw [black,fill=white](-1,-4) circle [radius=0.20];
\draw [black,fill=white](0,-4) circle [radius=0.20];
\draw [black,fill=white](1,-4) circle [radius=0.20];
\draw [black,fill=white](2,-4) circle [radius=0.20];
\draw [black,fill=white](3,-4) circle [radius=0.20];
\draw [black,fill=white](4,-3) circle [radius=0.20];
\draw [black,fill=white](5,-2) circle [radius=0.20];
\draw [black,fill=white](5,-2) circle [radius=0.20];
\draw [black,fill=black](5,-1) circle [radius=0.20];
\draw [black,fill=black](4,-1) circle [radius=0.20];
\draw [black,fill=black](3,-1) circle [radius=0.20];
\draw [black,fill=black](2,-1) circle [radius=0.20];
\draw [black,fill=black](1,-1) circle [radius=0.20];
\draw [black,fill=black](0,-1) circle [radius=0.20];
\draw [black,fill=black](-1,-1) circle [radius=0.20];
\draw [black,fill=black](-2,-1) circle [radius=0.20];
\draw [black,fill=black](-2,-2) circle [radius=0.20];
\draw [black,fill=black](-1,-2) circle [radius=0.20];
\draw [black,fill=black](0,-2) circle [radius=0.20];
\draw [black,fill=black](1,-2) circle [radius=0.20];
\draw [black,fill=black](2,-2) circle [radius=0.20];
\draw [black,fill=black](3,-2) circle [radius=0.20];
\draw [black,fill=black](4,-2) circle [radius=0.20];
\draw [black,fill=black](-2,-3) circle [radius=0.20];
\draw [black,fill=black](-1,-3) circle [radius=0.20];
\draw [black,fill=black](0,-3) circle [radius=0.20];
\draw [black,fill=black](1,-3) circle [radius=0.20];
\draw [black,fill=black](2,-3) circle [radius=0.20];
\draw [black,fill=black](3,-3) circle [radius=0.20];
\draw [black,fill=black](-2,1) circle [radius=0.20];
\draw [black,fill=black](-2,0) circle [radius=0.20];

\end{tikzpicture}}}
\caption{Example native structures.}
\end{figure}
\clearpage
\subsection{\textcolor{black}{Wang Landau Sampling}}

\subsubsection{\textcolor{black}{2D50}}

For the sequence 2D50 thermodynamic behaviour was investigated via the computation of $C_{V}/N$, $U/N$,$S/N$ and $F/N$. The flatness criterion for this simulation was $p=0.8$ and the move ratios were $70\%$,$19\%$, $5\%$,$4\%$ and $2\%$ for pull, bond re-bridging, FRW, pivot and kink-flip moves respectively.

The 'critical' temperature was found to be $T_{C} = 0.576001$. The final modification factor for each process is shown in table \ref{table:50processtable1}. Apart from process 2, 12 and 10 all reached the native state ($E_{min}=-21$) and sampled it well. The energy range for the WLS was set to $[-20:0]$. 

\begin{table}[h]
\centering
\resizebox{0.30\columnwidth}{!}{%
\begin{tabular}{| c | c |}
 Process ID & $ln(f_{final})$ \\
 \hline
 \hline
 0 & $\approx$ 2.38 $\cdot 10^{-7}$ \\
 \hline
 1 & $\approx$ 2.98 $\cdot 10^{-8}$ \\
 \hline
 2 & $\approx$ 2.98 $\cdot 10^{-8}$ \\
 \hline
 3 & $\approx$ 4.76 $\cdot 10^{-7}$ \\
 \hline
 4 & $\approx$ 2.38 $\cdot 10^{-7}$ \\
 \hline
 5 & $\approx$ 2.98 $\cdot 10^{-8}$ \\
 \hline
 6 & $\approx$ 1.49 $\cdot 10^{-8}$ \\
 \hline
 7 & $\approx$ 2.98 $\cdot 10^{-8}$ \\
 \hline
 8 & $\approx$ 1.19 $\cdot 10^{-7}$ \\
 \hline
 9 & $\approx$ 2.98 $\cdot 10^{-8}$ \\
 \hline
 10 & $\approx$ 1.49 $\cdot 10^{-8}$ \\
 \hline
 11 & $\approx$ 1.86 $\cdot 10^{-9}$ \\
 \hline
 12 & $\approx$ 7.45 $\cdot 10^{-9}$ \\
 \hline
 13 & $\approx$ 1.19 $\cdot 10^{-7}$ \\
 \hline\
 14 & $\approx$ 1.19 $\cdot 10^{-7}$ \\
 \hline
\end{tabular}}
\caption{ The right column reflects the convergence of the intrinsic DOS for each process, the majority are $\leqslant 10^{-7}$, this convergence is adequate for the results shown in figure \ref{fig:50obser} \label{table:50processtable1}}
\end{table}

The Monte Carlo simulation for the following results completed $\approx 2.7 \times 10^{9}$ iterations.

\begin{figure}[h]
\centering
\fbox{\subcaptionbox{The free energy $F/N$ for 2D50. Error bars computed as described in \ref{sec:errors}.}[7.5cm]{\includegraphics[scale=0.45]{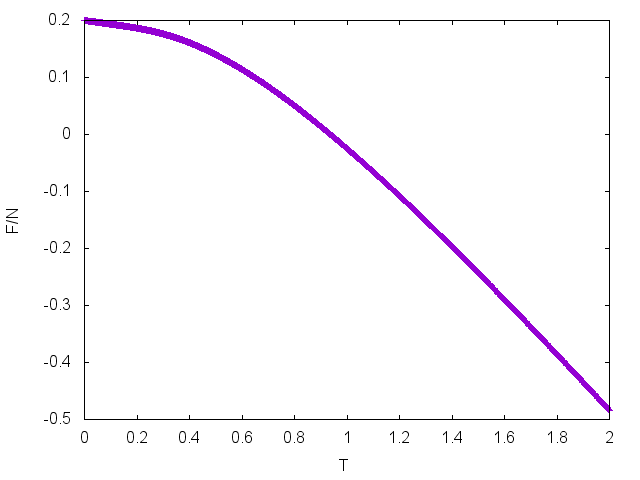}}}
\fbox{\subcaptionbox{The specific heat capacity $C_{V}/N$ for 2D50.  Error bars computed as described in \ref{sec:errors}.}[7.5cm]{\includegraphics[scale=0.45]{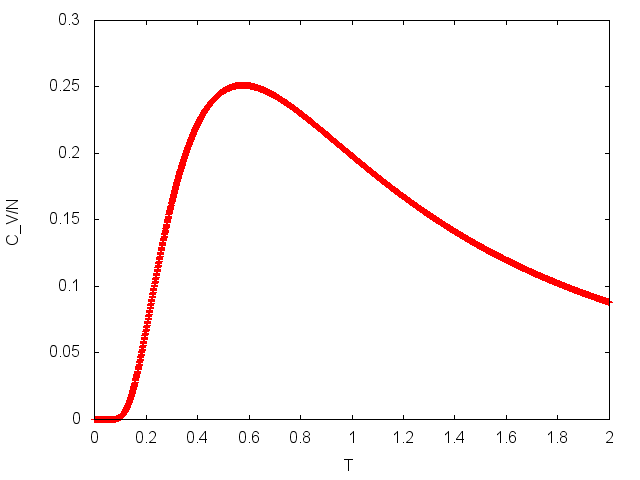}}}
\fbox{\subcaptionbox{The entropy $S/N$ for 2D50.  Error bars computed as described in \ref{sec:errors}.}[7.5cm]{\includegraphics[scale=0.45]{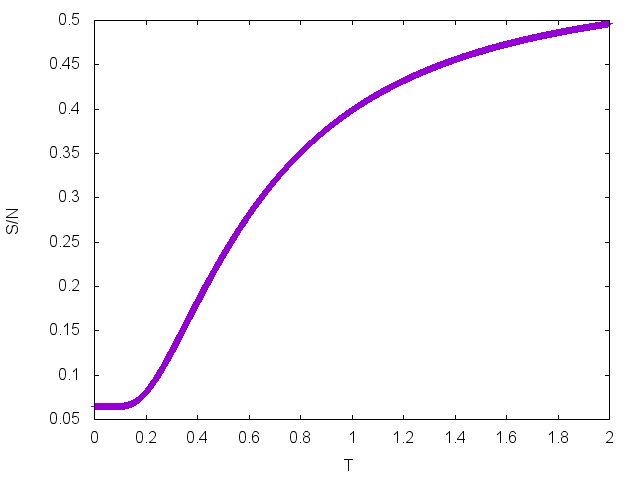}}}
\fbox{\subcaptionbox{The internal energy $U/N$ for 2D50.  Error bars computed as described in \ref{sec:errors}.}[7.5cm]{\includegraphics[scale=0.45]{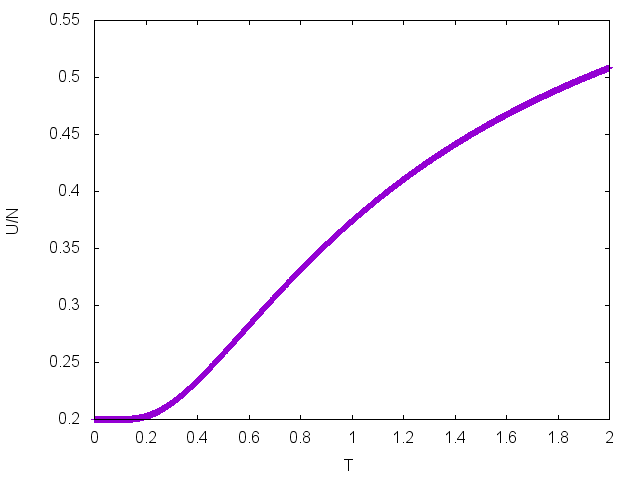}}}
\caption{Computed thermodynamic observables for 2D50. \label{fig:50obser}}
\end{figure}
\clearpage

\subsubsection{\textcolor{black}{2D60}}
For the sequence 2D60 thermodynamic behaviour was investigated via the computation of $C_{V}/N$, $U/N$,$S/N$ and $F/N$. The flatness criterion for this simulation was $p=0.8$ and the move ratios were $70\%$,$19\%$, $5\%$,$4\%$ and $2\%$ for pull, bond re-bridging, FRW, pivot and kink-flip moves respectively.

The 'critical' temperature was found to be $T_{C} = 0.42$. The final modification factor for each process is shown in table \ref{table:60processtable1}. The achievement of accessing the native state (-36) of 2D60 was accomplished during this run. Only process 6 achieved this state and the others reached a minimum of (-35). The energy range for the sampling was set to [-34:00] for these results.

\begin{table}[h]
\centering
\resizebox{0.30\columnwidth}{!}{%
\begin{tabular}{| c | c |}
 Process ID & $ln(f_{final})$ \\
 \hline
 \hline
 0 & $\approx$ 3.50 $\cdot 10^{-46}$ \\
 \hline
 1 & $\approx$ 8.55 $\cdot 10^{-50}$ \\
 \hline
 2 & $\approx$ 4.7 $\cdot 10^{-38}$ \\
 \hline
 3 & $\approx$ 4.27 $\cdot 10^{-50}$ \\
 \hline
 4 & $\approx$ 1.34 $\cdot 10^{-51}$ \\
 \hline
 5 & $\approx$ 7.45 $\cdot 10^{-9}$ \\
 \hline
 6 & $\approx$ 8.55 $\cdot 10^{-50}$ \\
 \hline
 7 & $\approx$ 3.85 $\cdot 10^{-34}$ \\
 \hline
 8 & $\approx$ 1.15 $\cdot 10^{-41}$ \\
 \hline
 9 & $\approx$ 3.58 $\cdot 10^{-43}$ \\
 \hline
 10 & $\approx$ 9.183 $\cdot 10^{-41}$ \\
 \hline
 11 & $\approx$ 5.72 $\cdot 10^{-42}$ \\
 \hline
 12 & $\approx$ 8.758 $\cdot 10^{-47}$ \\
 \hline
 13 & $\approx$ 1.54 $\cdot 10^{-33}$ \\
 \hline\
 14 & $\approx$ 2.08 $\cdot 10^{-53}$ \\
 \hline
\end{tabular}}
\caption{ The right column reflects the convergence of the intrinsic DOS for each process, the majority are $\leqslant 10^{-30}$, this convergence is adequate for the results shown in figure \ref{fig:60obser}.\label{table:60processtable1}}
\end{table}

The Monte Carlo simulation for the following results completed $\approx 5.8 \times 10^{8}$ iterations.

\begin{figure}[h]
\centering
\fbox{\subcaptionbox{The free energy $F/N$ for 2D60. Error bars computed as described in \ref{sec:errors}.}[7.5cm]{\includegraphics[scale=0.45]{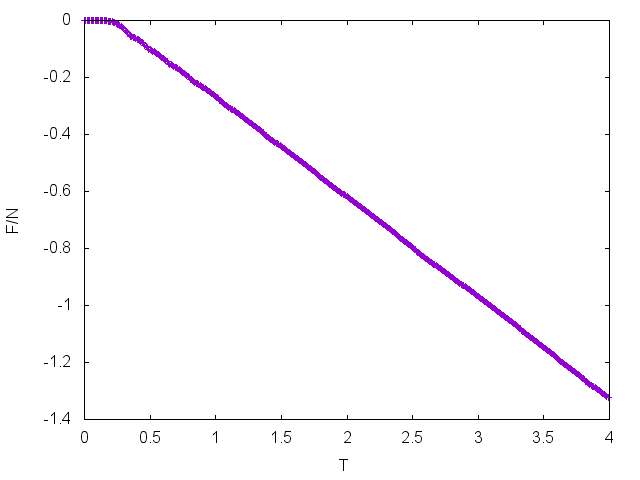}}}
\fbox{\subcaptionbox{The specific heat capacity $C_{V}/N$ for 2D60.  Error bars computed as described in \ref{sec:errors}.}[7.5cm]{\includegraphics[scale=0.45]{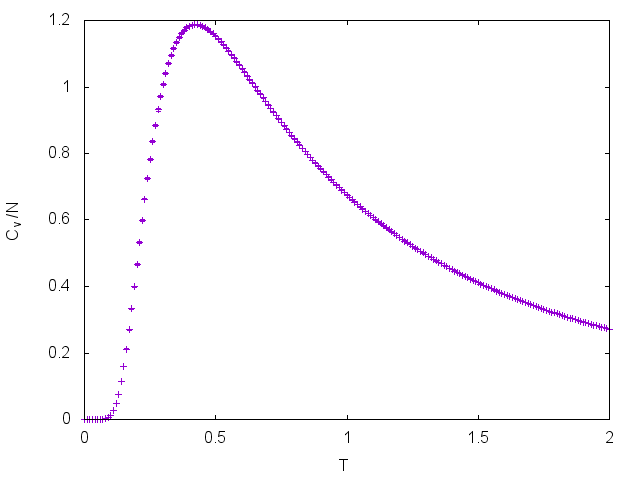}}}
\fbox{\subcaptionbox{The entropy $S/N$ for 2D60.  Error bars computed as described in \ref{sec:errors}.}[7.5cm]{\includegraphics[scale=0.45]{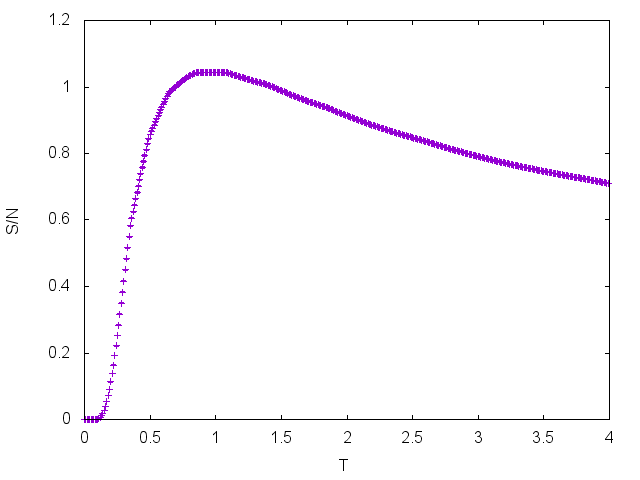}}}
\fbox{\subcaptionbox{The internal energy $U/N$ for 2D60.  Error bars computed as described in \ref{sec:errors}.}[7.5cm]{\includegraphics[scale=0.45]{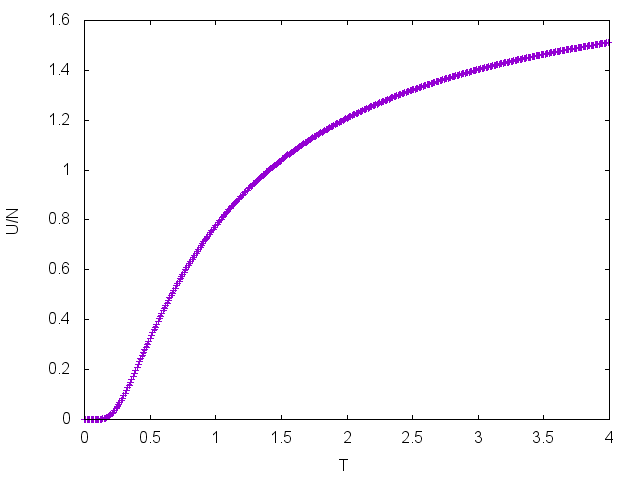}}}
\caption{Computed thermodynamic observables for 2D60. \label{fig:60obser}}
\end{figure}

\clearpage
\subsubsection{\textcolor{black}{2D64}}
For the sequence 2D64 thermodynamic behaviour was investigated via the computation of $C_{V}/N$, $U/N$,$S/N$ and $F/N$. The flatness criterion for this simulation was $p=0.8$ and the move ratios were $70\%$,$19\%$, $5\%$,$4\%$ and $2\%$ for pull, bond re-bridging, FRW, pivot and kink-flip moves respectively.

The 'critical' temperature was found to be $T_{C} = 0.39$. The final modification factor for each process is shown in table \ref{table:60processtable1}. During this short simulation every process attained the minimum energy of -40 which was set to the lower bound of the WL energy range. 

\begin{table}[h]
\centering
\resizebox{0.30\columnwidth}{!}{%
\begin{tabular}{| c | c |}
 Process ID & $ln(f_{final})$ \\
 \hline
 \hline
 0 & $\approx$ 1.563 $\cdot 10^{-2}$ \\
 \hline
 1 & $\approx$ 0.313 \\
 \hline
 2 & $\approx$ 0.01563 \\
 \hline
 3 & $\approx$ 0.01563\\
 \hline
 4 & $\approx$ 0.0078\\
 \hline
 5 & $\approx$ 0.01563 \\
 \hline
 6 & $\approx$ 0.01563 \\
 \hline
 7 & $\approx$ 0.007813 \\
 \hline
 8 & $\approx$ 0.03125 \\
 \hline
 9 & $\approx$ 0.015625 \\
 \hline
 10 & $\approx$ 0.007813 \\
 \hline
 11 & $\approx$ 0.0313\\
 \hline
 12 & $\approx$ 0.0313 \\
 \hline
 13 & $\approx$  0.007813\\
 \hline\
 14 & $\approx$ 0.0313 \\
 \hline
\end{tabular}}
\caption{ The right column reflects the convergence of the intrinsic DOS for each process. This convergence is questionably adequate for the results shown in figure \ref{fig:64obser} (see section \ref{sec:disresults} for an explanation.\label{table:64processtable1}}
\end{table}

The Monte Carlo simulation for the following results completed $\approx 808 \times 10^{6}$ iterations.

\begin{figure}[h]
\centering
\fbox{\subcaptionbox{The free energy $F/N$ for 2D64. Error bars computed as described in \ref{sec:errors}.}[7.5cm]{\includegraphics[scale=0.45]{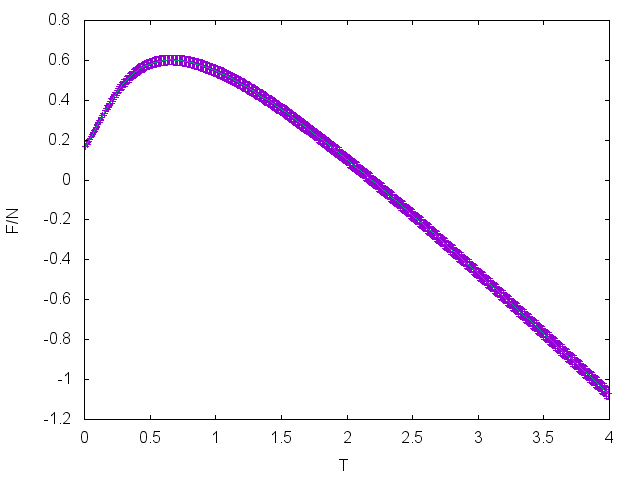}}}
\fbox{\subcaptionbox{The specific heat capacity $C_{V}/N$ for 2D64.  Error bars computed as described in \ref{sec:errors}.}[7.5cm]{\includegraphics[scale=0.45]{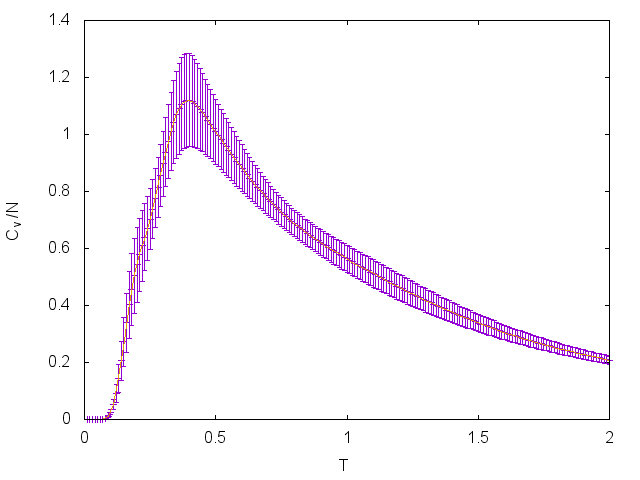}}}
\fbox{\subcaptionbox{The entropy $S/N$ for 2D64.  Error bars computed as described in \ref{sec:errors}.}[7.5cm]{\includegraphics[scale=0.45]{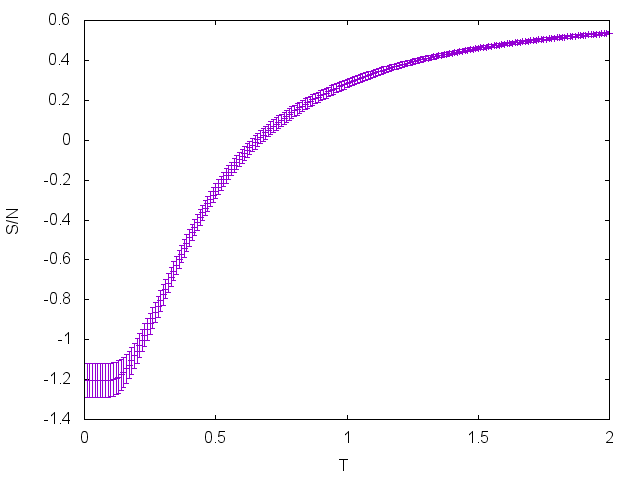}}}
\fbox{\subcaptionbox{The internal energy $U/N$ for 2D64.  Error bars computed as described in \ref{sec:errors}.}[7.5cm]{\includegraphics[scale=0.45]{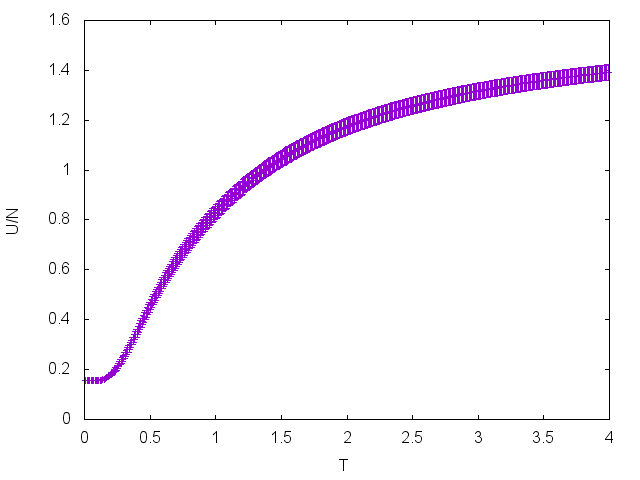}}}
\caption{Computed thermodynamic observables for 2D64. \label{fig:64obser}}
\end{figure}

\clearpage
\subsubsection{\textcolor{black}{2D85}}

For the sequence 2D85 thermodynamic behaviour was investigated via the computation of $C_{V}/N$, $U/N$,$S/N$ and $F/N$. The flatness criterion for this simulation was $p=0.8$ and the move ratios were $70\%$,$19\%$, $5\%$,$4\%$ and $2\%$ for pull, bond re-bridging, FRW, pivot and kink-flip moves respectively.

The 'critical' temperature, at which $C_V/N$ is a maximum, was found to be $T_C$ = 0.545001. The final modification factor for each process is shown in table \ref{table:85processtable1}. All processes reached a minimum of -51 which was used as the lower limit of the energy range.

\begin{table}[h]
\centering
\resizebox{0.30\columnwidth}{!}{%
\begin{tabular}{| c | c |}
 Process ID & $ln(f_{final})$ \\
 \hline
 \hline
 0 & $\approx$ 1.22 $\cdot 10^{-4}$ \\
 \hline
 1 & $\approx$ 1.91 $\cdot 10^{-6}$ \\
 \hline
 2 & $\approx$ 7.63 $\cdot 10^{-6}$ \\
 \hline
 3 & $\approx$ 7.63 $\cdot 10^{-6}$ \\
 \hline
 4 & $\approx$ 1.91 $\cdot 10^{-6}$ \\
 \hline
 5 & $\approx$ 3.81 $\cdot 10^{-6}$ \\
 \hline
 6 & $\approx$ 1.91 $\cdot 10^{-6}$ \\
 \hline
 7 & $\approx$ 9.54 $\cdot 10^{-7}$ \\
 \hline
 8 & $\approx$ 1.91$\cdot 10^{-6}$ \\
 \hline
 9 & $\approx$ 3.81 $\cdot 10^{-6}$ \\
 \hline
 10 & $\approx$ 1.91 $\cdot 10^{-6}$ \\
 \hline
 11 & $\approx$ 3.81 $\cdot 10^{-6}$ \\
 \hline
 12 & $\approx$ 4.7 $\cdot 10^{-7}$ \\
 \hline
 13 & $\approx$ 1.91 $\cdot 10^{-6}$ \\
 \hline\
 14 & $\approx$ 1.22 $\cdot 10^{-4}$ \\
 \hline
\end{tabular}}
\caption{ The right column reflects the convergence of the intrinsic DOS for each process, the majority are $< 10^{-5}$, this convergence is adequate for the results shown in figures \ref{fig:85obs}. \label{table:85processtable1}}
\end{table}

The Monte Carlo iterations for this simulation run was = 1347840000.
 
\begin{figure}[h]
\centering
\fbox{\subcaptionbox{Specific heat capacity, $C_V$, divided by the number of monomers, $N$, against $T$. \label{fig:2D85heat} Error bars computed as described in \ref{sec:errors}.}[7.5cm]{\includegraphics[scale=0.45]{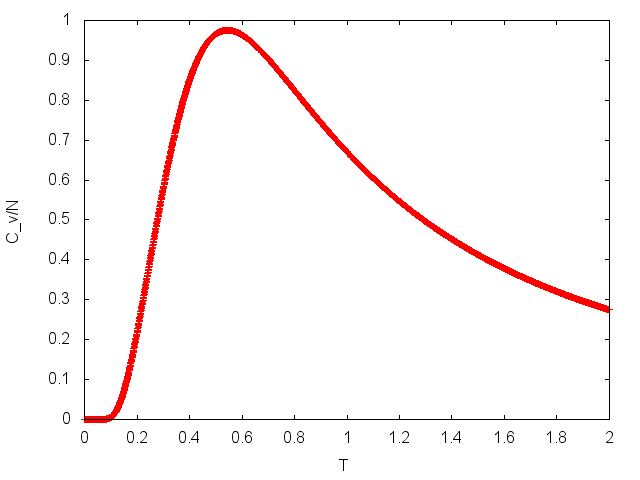}}}
\fbox{\subcaptionbox{Internal energy, $U$, divided by the number of monomers, $N$, against $T$. \label{fig:2D85internal} Error bars computed as described in \ref{sec:errors}.}[7.5cm]{\includegraphics[scale=0.45]{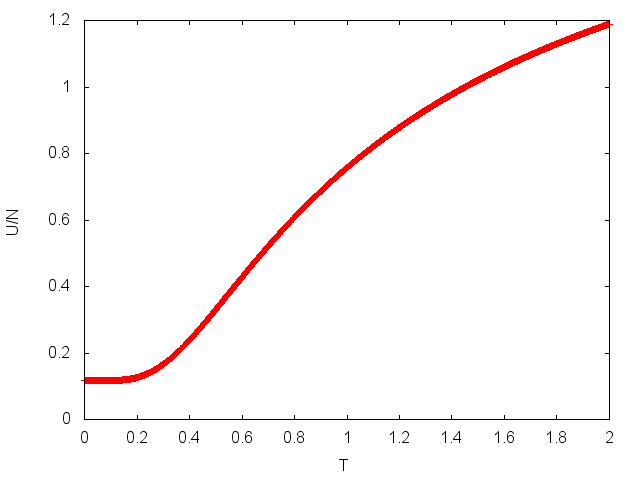}}}
\fbox{\subcaptionbox{Free energy, $F$, divided by the number of monomers, $N$, against $T$. \label{fig:2D85free} Error bars computed as described in \ref{sec:errors}.}[7.5cm]{\includegraphics[scale=0.45]{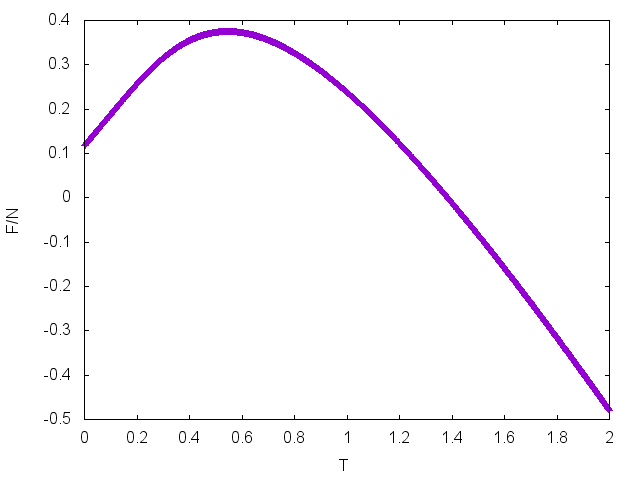}}}
\fbox{\subcaptionbox{Entropy, $S$, divided by the number of monomers, $N$, against $T$. \label{fig:2D85entropy} Error bars computed as described in \ref{sec:errors}.}[7.5cm]{\includegraphics[scale=0.45]{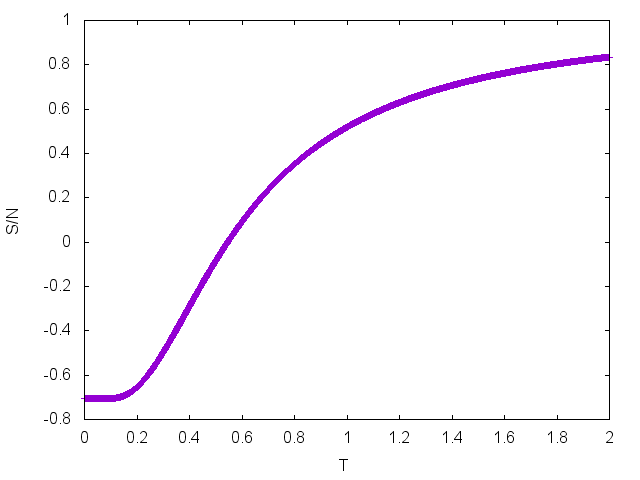}}}
\caption{Thermodynamic observables for 2D85. \label{fig:85obs}}
\end{figure}
\clearpage
\subsubsection{\textcolor{black}{2D100a}}

For the sequence 2D100a thermodynamic behaviour was investigated via the computation of $C_{V}/N$, $U/N$,$S/N$ and $F/N$. The flatness criterion for this simulation was $p=0.8$ and the move ratios were $70\%$,$19\%$, $5\%$,$4\%$ and $2\%$ for pull, bond re-bridging, FRW, pivot and kink-flip moves respectively.

The 'critical' temperature was found to be $T_{C} = 0.535$. The final modification factor for each process is shown in table \ref{table:100aprocesstable1}. Every process attained a minimum energy = -47 which is a unit of energy greater than the lowest known (see \ref{table:nativeresults}). The energy range was then, automatically, set to [0:-47] (the global range being [0:-48] but the processes only attained -47).

\begin{table}[h]
\centering
\resizebox{0.30\columnwidth}{!}{%
\begin{tabular}{| c | c |}
 Process ID & $ln(f_{final})$ \\
 \hline
 \hline
 0 & $\approx$ 3.82 $\cdot 10^{-6}$ \\
 \hline
 1 & $\approx$ 3.82 $\cdot 10^{-6}$ \\
 \hline
 2 & $\approx$ 7.63 $\cdot 10^{-6}$ \\
 \hline
 3 & $\approx$ 3.05 $\cdot 10^{-5}$ \\
 \hline
 4 & $\approx$ 1.53 $\cdot 10^{-5}$ \\
 \hline
 5 & $\approx$ 7.63 $\cdot 10^{-6}$ \\
 \hline
 6 & $\approx$ 3.82 $\cdot 10^{-6}$ \\
 \hline
 7 & $\approx$ 3.82 $\cdot 10^{-6}$ \\
 \hline
 8 & $\approx$ 3.82 $\cdot 10^{-6}$ \\
 \hline
 9 & $\approx$ 3.052 $\cdot 10^{-5}$ \\
 \hline
 10 & $\approx$ 7.63 $\cdot 10^{-6}$ \\
 \hline
 11 & $\approx$ 1.53 $\cdot 10^{-5}$ \\
 \hline
 12 & $\approx$ 1.91 $\cdot 10^{-6}$ \\
 \hline
 13 & $\approx$ 3.052 $\cdot 10^{-5}$ \\
 \hline\
 14 & $\approx$ 3.815 $\cdot 10^{-6}$ \\
 \hline
\end{tabular}}
\caption{ The right column reflects the convergence of the intrinsic DOS for each process, the majority are $< 10^{-5}$, this convergence is adequate for the results shown in figure \ref{fig:100aobs} \label{table:100aprocesstable1}}
\end{table}
The Monte Carlo iterations for this simulation run was = 1347840000.
\begin{figure}[h]
\centering
\fbox{\subcaptionbox{Specific heat capacity, $C_V$, divided by the number of monomers, $N$, against $T$. \label{fig:2D100aheat} Error bars computed as described in \ref{sec:errors}.}[7.5cm]{\includegraphics[scale=0.45]{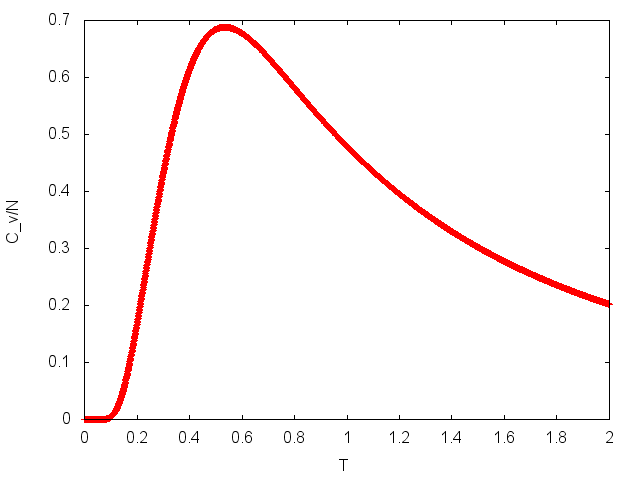}}}
\fbox{\subcaptionbox{Internal energy, $U$, divided by the number of monomers, $N$, against $T$. \label{fig:2D100ainternal} Error bars computed as described in \ref{sec:errors}.}[7.5cm]{\includegraphics[scale=0.45]{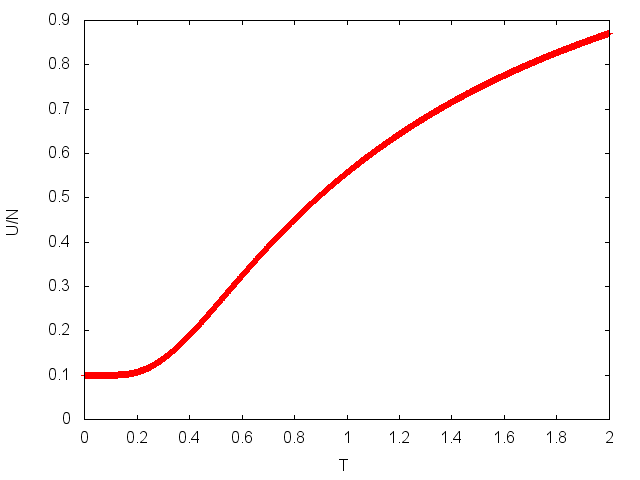}}}
\fbox{\subcaptionbox{Free energy, $F$, divided by the number of monomers, $N$, against $T$. \label{fig:2D100afree} Error bars computed as described in \ref{sec:errors}.}[7.5cm]{\includegraphics[scale=0.45]{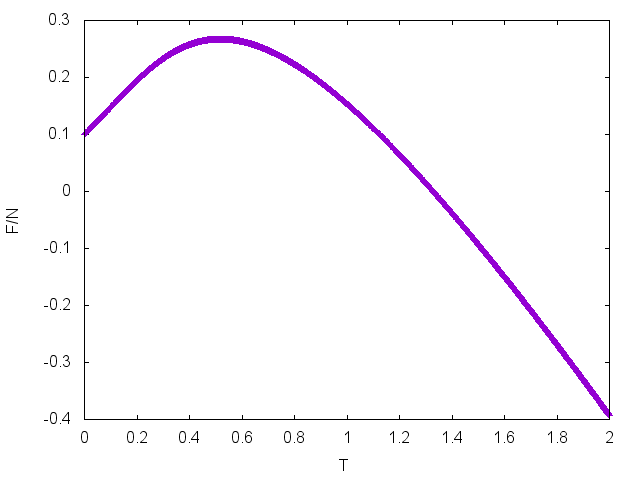}}}
\fbox{\subcaptionbox{Entropy, $S$, divided by the number of monomers, $N$, against $T$. \label{fig:2D100aentropy} Error bars computed as described in \ref{sec:errors}.}[7.5cm]{\includegraphics[scale=0.45]{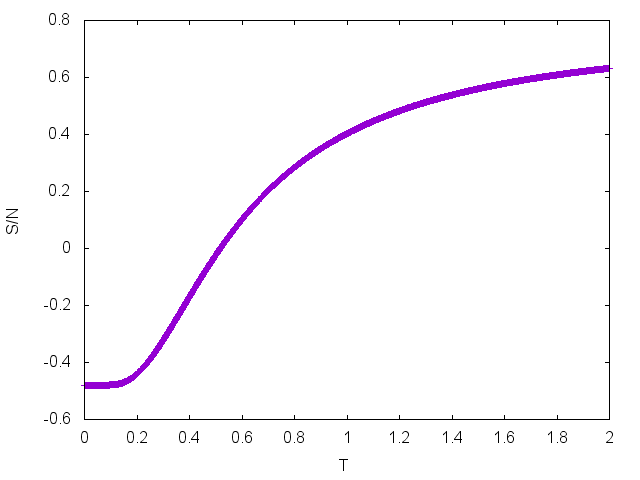}}}
\caption{Thermodynamic observables for 2D100a. \label{fig:100aobs}}
\end{figure}

\clearpage
\subsubsection{\textcolor{black}{2D100b}}

For the sequence 2D100b thermodynamic behaviour was investigated via the computation of $C_{V}/N$, $U/N$,$S/N$ and $F/N$. The flatness criterion for this simulation was $p=0.8$ and the move ratios were $70\%$,$19\%$, $5\%$,$4\%$ and $2\%$ for pull, bond re-bridging, FRW, pivot and kink-flip moves respectively.

The 'critical' temperature was found to be $T_{C} = 0.5765 \pm 0.02$. The final modification factor for each process is shown in table \ref{table:100bprocesstable1}. Each process attained the energy of -46 which is 4 more than the known native state of 100b. 

\begin{table}[h]
\centering
\resizebox{0.30\columnwidth}{!}{%
\begin{tabular}{| c | c |}
 Process ID & $ln(f_{final})$ \\
 \hline
 \hline
 0 & $\approx$ 3.91 $\cdot 10^{-3}$ \\
 \hline
 1 & $\approx$ 0.0039 \\
 \hline
 2 & $\approx$ 0.0078 \\
 \hline
 3 & $\approx$ 0.00097 \\
 \hline
 4 & $\approx$  0.0039\\
 \hline
 5 & $\approx$ 0.0039 \\
 \hline
 6 & $\approx$ 0.00195 \\
 \hline
 7 & $\approx$  0.0039\\
 \hline
 8 & $\approx$ 0.0039 \\
 \hline
 9 & $\approx$  0.0078\\
 \hline
 10 & $\approx$ 0.00098\\
 \hline
 11 & $\approx$ 0.00195\\
 \hline
 12 & $\approx$ 0.00195 \\
 \hline
 13 & $\approx$ 0.0039 \\
 \hline\
 14 & $\approx$ 0.0039\\
 \hline
\end{tabular}}
\caption{ The right column reflects the convergence of the intrinsic DOS for each process, the majority are $< 0.01$. Observables for this run are shown in figure \ref{fig:100bobs}. \label{table:100bprocesstable1}}
\end{table}
The Monte Carlo iterations for this simulation run was = 1347840000.
\begin{figure}[h]
\centering
\fbox{\subcaptionbox{Specific heat capacity, $C_V$, divided by the number of monomers, $N$, against $T$. \label{fig:2D100bheat} Error bars computed as described in \ref{sec:errors}.}[7.5cm]{\includegraphics[scale=0.45]{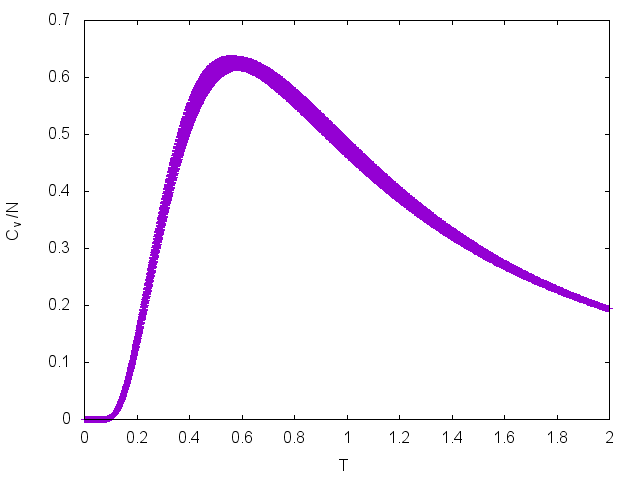}}}
\fbox{\subcaptionbox{Internal energy, $U$, divided by the number of monomers, $N$, against $T$. \label{fig:2D100binternal} Error bars computed as described in \ref{sec:errors}.}[7.5cm]{\includegraphics[scale=0.45]{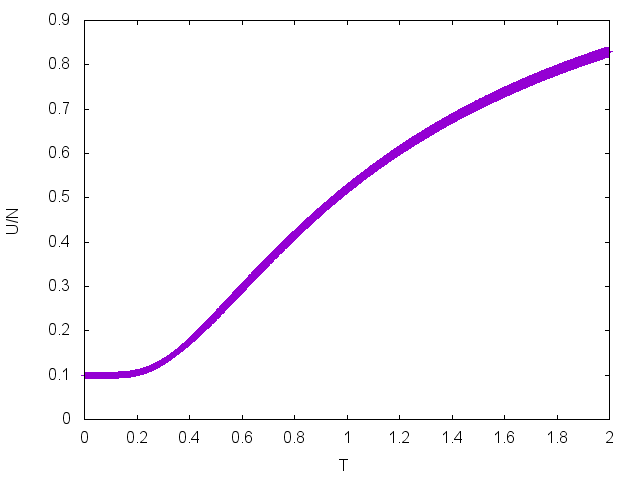}}}
\fbox{\subcaptionbox{Free energy, $F$, divided by the number of monomers, $N$, against $T$. \label{fig:2D100bfree} Error bars computed as described in \ref{sec:errors}.}[7.5cm]{\includegraphics[scale=0.45]{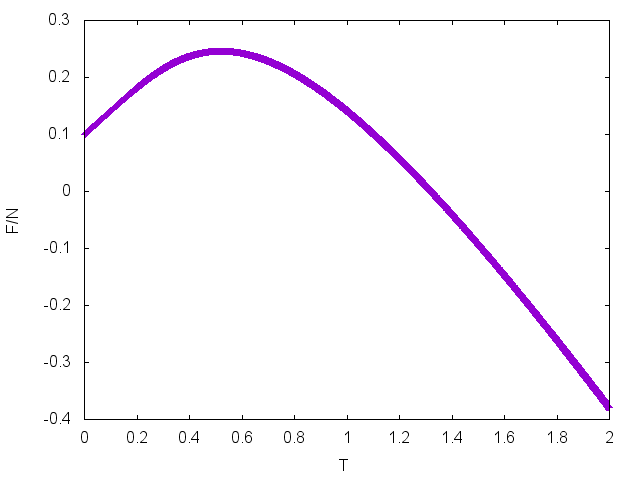}}}
\fbox{\subcaptionbox{Entropy, $S$, divided by the number of monomers, $N$, against $T$. \label{fig:2D100bentropy} Error bars computed as described in \ref{sec:errors}.}[7.5cm]{\includegraphics[scale=0.45]{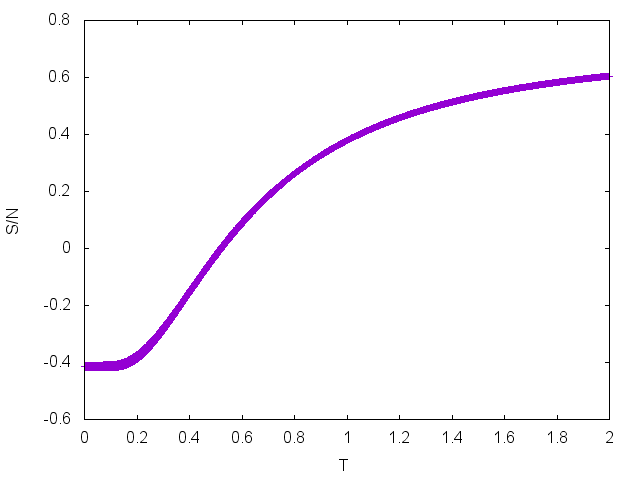}}}
\caption{Thermodynamic observables for 2D100b. \label{fig:100bobs}}
\end{figure}

\clearpage
\subsection{ISAWs}

Homopolymers, which consist of identical sub units, are found in industrial plastics and biology. Using the methodology (see section \ref{sec:method}) which was used to investigate the behaviour of HP proteins, ISAWs (Interacting Self Avoiding Walks) of Homopolymers were also studied. Each sub unit of the homopolymer is assumed to be hydrophobic and the energy function given by equation \ref{eq:Hamiltonian}.

\begin{figure}[h]
\centering
\fbox{{\includegraphics[scale=0.8]{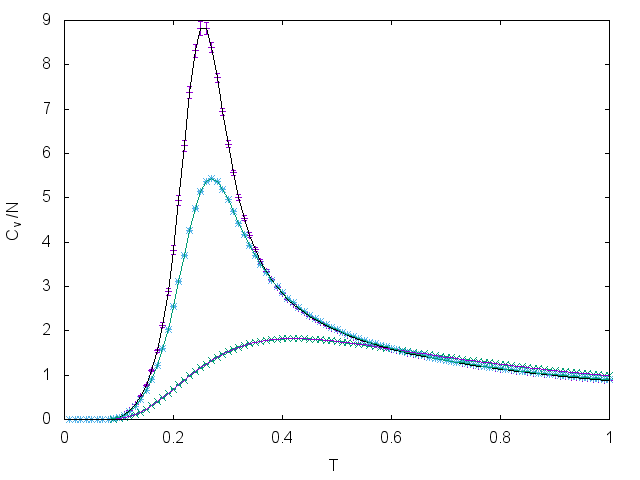}}}
\caption{Specific heat capacity per monomer, $C_{v}/N$, against temperature $T$. Length 25 (purple, green error bars) = lower curve, length 64 (green, blue errorbars) = middle curve and length 100 (black, purple error bars) = highest curve. \label{fig:isawheat}}
\end{figure}

\begin{figure}[h]
\centering
\fbox{{\includegraphics[scale=0.8]{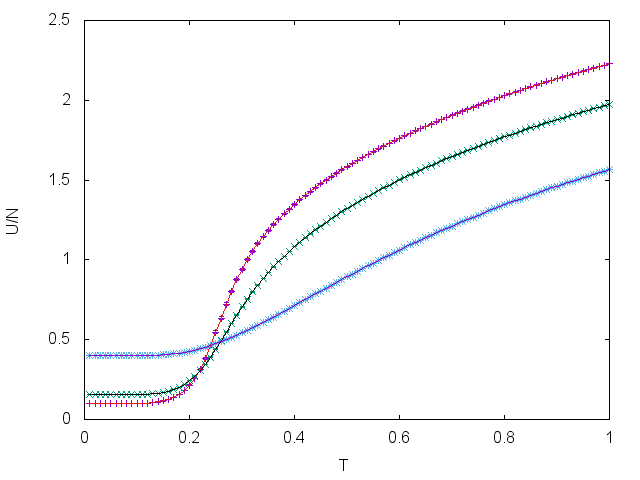}}}
\caption{Internal energy per monomer, $U/N$, against temperature $T$. Length 25 (blue error bars), length 64 (green errorbars) and length 100 ( purple error bars) = highest curve. \label{fig:isawinternal}}
\end{figure}

Homopolymers of lengths: 25, 64, 100, 144, 225 and 400 were studied and the convergence of the modification factors for each sequence and the comparison of thermodynamic behaviour is given in table \ref{table:isawmodi} and figures \ref{fig:isawheat} and \ref{fig:isawinternal} respectively. The ratios of the trial moves were exactly the same as in the previous section. The flatness criterion was $p=0.8$ for lengths 25,64, 100, 144 and $p=0.7$ for lengths 225 and 400 due to difficulty in traversing the conformational space in the given time length. 

\begin{table}[h]
\centering
\begin{subtable}[b]{7.5cm}
\centering
\begin{tabular}{| c | c |}
 Process ID & $ln(f_{final})$ \\
 \hline
 \hline
 0 & $\approx$  3.953 $\cdot 10^{-4327}$ \\
 \hline
 1 & $\approx$ 3.287 $\cdot 10^{-4438}$ \\
 \hline
 2 & $\approx$ 1.412 $\cdot 10^{-3785}$ \\
 \hline
 3 & $\approx$ 4.707 $\cdot 10^{-4188}$ \\
 \hline
 4 & $\approx$ 3.298 $\cdot 10^{-4233}$ \\
 \hline
 5 & $\approx$ 9.246 $\cdot 10^{-2992}$ \\
 \hline
 6 & $\approx$ 6.324 $\cdot 10^{-4326}$ \\
 \hline
 7 & $\approx$ 2.871 $\cdot 10^{-3403}$ \\
 \hline
 8 & $\approx$ 1.867 $\cdot 10^{-4305}$ \\
 \hline
 9 & $\approx$ 2.815 $\cdot 10^{-3990}$\\
 \hline
\end{tabular}
\caption{Convergence for each process for length \textbf{25}.}
\end{subtable}
\begin{subtable}[b]{7.5cm}
\centering
\begin{tabular}{| c | c |}
 Process ID & $ln(f_{final})$ \\
 \hline
 \hline
 0 & $\approx$  2.328 $\cdot 10^{-10}$ \\
 \hline
 1 & $\approx$ 2.910 $\cdot 10^{-11}$ \\
 \hline
 2 & $\approx$ 2.91 $\cdot 10^{-11}$ \\
 \hline
 3 & $\approx$ 3.725 $\cdot 10^{-9}$ \\
 \hline
 4 & $\approx$ 1.863 $\cdot 10^{-9}$ \\
 \hline
 5 & $\approx$ 7.451 $\cdot 10^{-9}$ \\
 \hline
 6 & $\approx$ 4.768 $\cdot 10^{-7}$ \\
 \hline
 7 & $\approx$ 1.863 $\cdot 10^{-9}$ \\
 \hline
 8 & $\approx$ 4.6567 $\cdot 10^{-10}$ \\
 \hline
 9 & $\approx$ 4.6567 $\cdot 10^{-10}$\\
 \hline
\end{tabular}
\caption{Convergence for each process for length \textbf{64}.}
\end{subtable}
\begin{subtable}[b]{7.5cm}
\centering
\begin{tabular}{| c | c |}
 Process ID & $ln(f_{final})$ \\
 \hline
 \hline
 0 & $\approx$  6.104 $\cdot 10^{-5}$ \\
 \hline
 1 & $\approx$ 9.313 $\cdot 10^{-10}$ \\
 \hline
 2 & $\approx$ 1.192 $\cdot 10^{-7}$ \\
 \hline
 3 & $\approx$ 2.980 $\cdot 10^{-8}$ \\
 \hline
 4 & $\approx$ 1.526 $\cdot 10^{-5}$ \\
 \hline
 5 & $\approx$ 9.313 $\cdot 10^{-10}$ \\
 \hline
 6 & $\approx$ 9.313 $\cdot 10^{-10}$ \\
 \hline
 7 & $\approx$ 2.980 $\cdot 10^{-8}$ \\
 \hline
 8 & $\approx$ 9.131 $\cdot 10^{-10}$ \\
 \hline
 9 & $\approx$ 3.725 $\cdot 10^{-9}$\\
 \hline
\end{tabular}
\caption{Convergence for each process for length \textbf{100}.}
\end{subtable}
\begin{subtable}[b]{7.5cm}
\centering
\begin{tabular}{| c | c |}
 Process ID & $ln(f_{final})$ \\
 \hline
 \hline
 0 & $\approx$  3.125 $\cdot 10^{-2}$ \\
 \hline
 1 & $\approx$ 0.03125 \\
 \hline
 2 & $\approx$ 0.03125  \\
 \hline
 3 & $\approx$ 0.007813 \\
 \hline
 4 & $\approx$ 0.015625  \\
 \hline
 5 & $\approx$ 0.015625  \\
 \hline
 6 & $\approx$0.015625 \\
 \hline
 7 & $\approx$ 0.0625 \\
 \hline
 8 & $\approx$ 0.03125  \\
 \hline
 9 & $\approx$ 0.03215 \\
 \hline
\end{tabular}
\caption{Convergence for each process for length \textbf{144}.}
\end{subtable}
\end{table}
\begin{table}[h]
\centering
\begin{subtable}[b]{7.5cm}
\centering
\begin{tabular}{| c | c |}
 Process ID & $ln(f_{final})$ \\
 \hline
 \hline
 0 & $\approx$  7.813 $\cdot 10^{-3}$ \\
 \hline
 1 & $\approx$ 0.007813 \\
 \hline
 2 & $\approx$ 0.007813 \\
 \hline
 3 & $\approx$ 0.007813 \\
 \hline
 4 & $\approx$ 0.01563  \\
 \hline
 5 & $\approx$ 0.01563 \\
 \hline
 6 & $\approx$ 0.01563 \\
 \hline
 7 & $\approx$ 0.01563 \\
 \hline
 8 & $\approx$ 0.01563  \\
 \hline
 9 & $\approx$ 0.03125 \\
 \hline
\end{tabular}
\caption{Convergence for each process for length \textbf{225}.}
\end{subtable}
\begin{subtable}[b]{7.5cm}
\centering
\begin{tabular}{| c | c |}
 Process ID & $ln(f_{final})$ \\
 \hline
 \hline
 0 & $\approx$  1\\
 \hline
 1 & $\approx$ 1 \\
 \hline
 2 & $\approx$ 1  \\
 \hline
 3 & $\approx$ 1 \\
 \hline
 4 & $\approx$ 1  \\
 \hline
 5 & $\approx$ 1  \\
 \hline
 6 & $\approx$ 1 \\
 \hline
 7 & $\approx$ 1 \\
 \hline
 8 & $\approx$ 1 \\
 \hline
 9 & $\approx$ 1 \\
 \hline
\end{tabular}
\caption{ 'Convergence' for each process for length \textbf{400}.}
\end{subtable}
\caption{Final modification factors for ISAW length simulations. \label{table:isawmodi}}
\end{table}

\begin{table}[h]
\centering 
\begin{tabular}{| c | c | c |}
ISAW length & Total MC iterations & Duration (s)\\
\hline
\hline

25 & 2695 $\times$ $10^{6}$ & 80386  \\
\hline
64 &  439 $\times$ $10^{6}$ & 89450 \\
\hline
100 & 236 $\times$ $10^{6}$ & 92119 \\
\hline
144 &  124 $\times$ $10^{6}$ & 100427 \\
\hline
225 & 48 $\times$ $10^{6}$ & 90609 \\
\hline
400 &  31 $\times$ $10^{6}$ & 171526\\
\hline

\end{tabular}
\caption{Monte Carlo iterations and duration of simulation runs. \label{table:isawtime} }
\end{table}
\newpage
A comparison of $C_{v}/N$ and $U/N$ for all lengths except 144, 225 and 400, since the simulations did not converge adequately (error bars greater than the results), as a function of temperature is shown in figures \ref{fig:isawheat} and \ref{fig:isawinternal} respectively. The errors are computed following the descriptions in section \ref{sec:errors}.

\begin{figure}[h]
\centering
\fbox{{\includegraphics[scale=0.8]{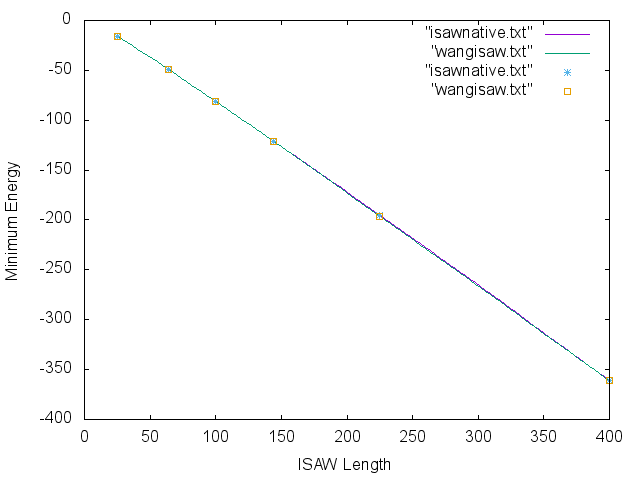}}}
\caption{Graphical comparison of minimum energies found for benchmark ISAWs. N.B. simulation timings for the Wust-Landau results unknown. 'isawnative' are results from this work and 'wangisaw' are results taken from \cite{ISAW}. \label{figure:natcompare}}
\end{figure}

The minimum energies attained, which represent the lower boundary of the WL energy range for the ISAWs, are compared to that found by \cite{ISAW} in figure \ref{figure:natcompare}.

\clearpage
\subsubsection{Error Analysis}
\label{sec:errors}
It has been stated that the general uncertainty in the computed DOS is $\propto$ $f$ (the modification factor) \ref{sec:wlmc}. However since thermodynamic observables were computed using the parallel-trajectory- swapping scheme where many random walkers are used computing their own observables and averages were taken, it is necessary to consider statistical variations centred about the mean\footnote{Results obtained from single walkers only have error bars centred around the modification factor.}.

The variance, $s^2$, of $n$ observations $\lbrace x_1, x_2, \ldots ,x_n \rbrace$ is:

\begin{equation}
s^{2} = \frac{1}{n - 1} [(x_{1} - \overline{x})^{2} + (x_{2} - \overline{x})^{2} + \ldots + (x_{n} - \overline{x})^{2}]
\label{eq:variance}
\end{equation}

where $\overline{x}$ is the sample mean of the data set \cite{statistics}.
The standard deviation, $\sigma$, is simply given by the square root of $s^2$. 

When the standard deviation of a statistic is estimated from the data this is the standard error, $SE$, \cite{statistics}, which is the error used here:

\begin{equation}
SE = \frac{\sigma}{\sqrt{n}}
\label{eq:SE}
\end{equation}

For each temperature the thermodynamic observables e.g. $C_V$ from each WL walker were averaged and the resulting error was computed using equation \ref{eq:SE} \cite{newwlapp}.

\clearpage
\section{\textcolor{black}{Discussion of Results}}
\label{sec:disresults}
\subsubsection{\textcolor{black}{Native State Search}}

\begin{figure}[h]
\centering
\subcaptionbox{2D64 native structure found in this work.}[7.5cm]{\includegraphics[scale=0.3]{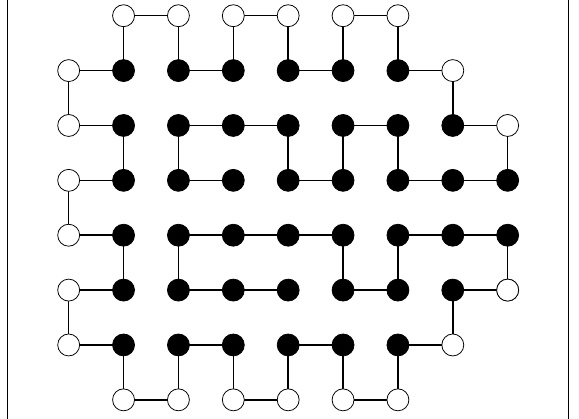}}
\subcaptionbox{2D64 native structure found in WLS (Wust-Landau)\cite{oplandau}.}[7.5cm]{\includegraphics[scale=0.5]{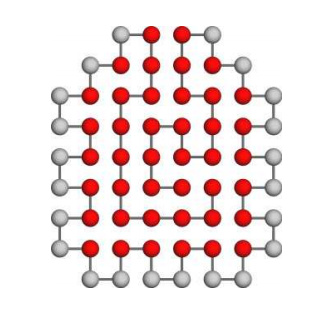}}
\subcaptionbox{2D64 native structure found in the ACO method \cite{ACO}.}[7.5cm]{\includegraphics[scale=0.5]{s18}}
\caption{ Comparison of native structures for 2D64. The similarity of the external polar amino acid placement is striking, also each sequence has the same line of symmetry (externally). The difference between the native structures is found within the hydrophobic core, however this will not necessarily alter the function of the protein since it interacts with others via its external structure. \label{fig:64compare}}
\end{figure}

Attaining the native state of 2D64 is known to be difficult, for example the computational methods of EMC and SISPER only could attain $E_{min}=-39$ \cite{frmc} (see table \ref{table:nativeresults}). However this simulation method has not only reached the lowest known energy of this sequence but also found a unique hydrophobic core for the native structure of 2D64 (see figure \ref{fig:native2D64}). A visual comparison of the native structures found here and with WLS \cite{oplandau} and ACO \cite{ACO} is shown in figure \ref{fig:64compare}. One can then see that the particular external structure of 2D64 is exact, which explains why it is difficult to access the native region since there are few native structures.

This striking similarity between found native structures for 2D64 reinforces the notion that proteins fold into specific structures in order to execute the same physiological function.

The native structures for 2D50 and 2D64 found in this particular run (see figure \ref{fig:native2D50}) has confirmed the expectation that most of the hydrophobic amino acids push towards the center of the compact configuration, leaving the polar amino acids to bond with the external aqueous solution. Also compare the native state found in the previous section to appendix \ref{sec:B} (2D50) where the best $E_{min}$ was -13. This vast improvement arose due to the inclusion of the bond re-bridging and FRW move and an improved move ratios. 

Note that the native structure for 2D50 found here (see figure \ref{fig:native2D50}) contains a few polar amino acids 'locked' in the hydrophobic center. Recall that ionic bonds are formed as amino acids bearing opposite electrical charge are close together within the hydrophobic core of proteins. Ionic bonding in the hydrophobic core is rare since most charged amino acids are polar, which are normally pushed towards the edge of the protein surface due to the hydrophobic effect. Although rare, ionic bonds can play an important role in stabilizing the native structure that can approach the strength of covalent bonding.

The inability to access the lowest known energies of 2D85 (-53), 2D100a (-48) and 2D100b (-50) does not reflect an intrinsic limitation of this method or trial set. Since WLS \cite{oplandau} attained these native structures with a similar but ultimately distinct trial move set, it is very likely that this method can also in principle access these low temperature structures. I believe it is a matter of insufficient computational effort and time that limited accessing these energies. The simulations here attained energies  1 unit more than the native state for the sequences just mentioned, which in terms of exploring global thermodynamic behaviour is not a complete downfall. The reader is encouraged to see section \ref{sec:EIEW} where it was demonstrated that thermodynamic observables are not greatly changed if low energies become unavailable to the WL sampler.

Further optimisation and computational 'tinkering' could improve the efficiency of the code used here which would be able to directly compete with WLS \cite{oplandau}.

\newpage
\begin{center}
\textbf{ISAWs}
\end{center}

As shown in figure \ref{figure:natcompare} the lowest energy states for the same length of ISAWs in this work compared well with \cite{ISAW}. For $N> 144$ the lowest energy states found in this work were slightly higher than \cite{ISAW}, this is notably to the huge computational burden long chains present. Since all ISAW sequences were run for a similar amount of time in seconds (See table \ref{table:isawtime}) the amount of MC iterations for longer chains obviously decreased. This would mean potentially less coverage of conformational space and hence not accessing the native configuration. 

ISAWs present a distinct problem in accessing low temperature structures since they exist in sharp wells in the rough energy landscape, using Wang- Landau sampling, trajectory swapping and the trial move set proposed here allows quick and thorough coverage of conformational space and with enough computational effort could access extremely long ISAWs and their native structures.

\subsubsection{Thermodynamic Investigations}

The general thermodynamic observables, the specific heat capacity in particular, show a 'psuedo phase transition'\footnote{'Pseudo' since the system is finite in size.} at a particular critical temperature $T_{C}$. The following discussion of thermodynamic and protein behaviour will consist of first the high-T regime and then the low-T regime.

\begin{center}
\large\underline{$T > T_{C}$}
\end{center}

The specific heat capacity, $C_{V}/N$, computed for all 2D benchmark protein (H)(P) sequences shows a gradual increase as the temperature goes from very high to just above the critical temperature $T_{C}$ (e.g. see figure \ref{fig:50obser}(b)). This happens as the chain goes from almost a 'string-like' configuration at high T and increasing the amount of H-H bonds as the temperature decreases. Going from a denatured state $\rightarrow$ molten globule occurs during the temperatures just above the critical temperature.

All the 2D sequences, despite varying degrees of convergence, also show similar behaviour for $U/N$, which decreases at a gradually faster rate (closer to $T_{C}$). This can easily be explained due to the increase in thermal agitation of the monomers on the chain as the temperature increases. Even for the accurate results for 2D50 and 2D60 there is no obvious sudden collapse of this thermodynamic observable.

The free energy, $F/N$, in this temperature regime grows almost linearly with decreasing temperature. This linear relationship is shown almost perfectly for 2D60 (See figure \ref{fig:60obser}). Towards the critical region this growth in free energy gradually slows down for all sequences. The growth in free energy is again due to the increased number of H-H contacts and the protein becoming more globule-like, increasing the amount of thermodynamic work it can perform. 

The entropy, $S/N$, behaves very much like the internal energy with temperature for all but 2D60. The entropy for 2D50, 2D64, 2D85, 2D100a and 2D100b decreases gradually (at a faster rate towards $T_{C}$). The entropy for 2D50 (see figure \ref{fig:50obser}) can be taken as an accurate representation for this entropy behaviour since it converged better than the other 4 sequences. The gradual decrease in entropy aligns with our thermodynamic expectations since the degeneracy for configurations at lower temperatures decreases. These results also meet the expectation that the system will be in a swollen SAW state where entropy should dominate \cite{topoisaw}. 

2D60 (the best converged simulation) showed an entropy that contained a peak near $T_{C} = 0.42$ (see figure \ref{fig:60obser}). The entropy very slightly increases with decreasing temperature towards its peak, this behaviour contrasts the results for the other sequences where the maximum entropy occurs at the highest temperature for the simulation. A physical explanation for this behaviour could be that at extremely high temperatures the chain becomes close to or attains a straight line conformation on the lattice, where the degeneracy for this drops slightly. The entropy also seems to be converging to a value much $> 0$, so it is not expected that the entropy will continue to decrease with temperature which would not be physically viable. Whether this behaviour is sequence, length or convergent (final modification value) dependent unfortunately cannot be ascertained from the results here. However it does raise the question: Does the entropy always increase as T increases for every sequence?  Also the entropy for 2D60 does signal a clear psuedo phase transition which could mean that the entropy can play a role as a 'phase transition signaller' for lattice polymers.

In general the thermodynamic observables in the $T > T_{C}$ region do  confirm the expectation that at high temperatures the protein chain becomes denatured and tends toward a rigid straight line configuration. The smoothness of the increase or decrease of observables (increasing rate towards $T_{C}$) with decreasing temperature reflects the gradual increase of H-H topological contacts which are leading the chain into a globule-like configuration.

\begin{center}
\large\textbf{\underline{$T \leqslant T_{C}$}}
\end{center}

\begin{figure}[h]
\centering
\includegraphics[scale=0.095]{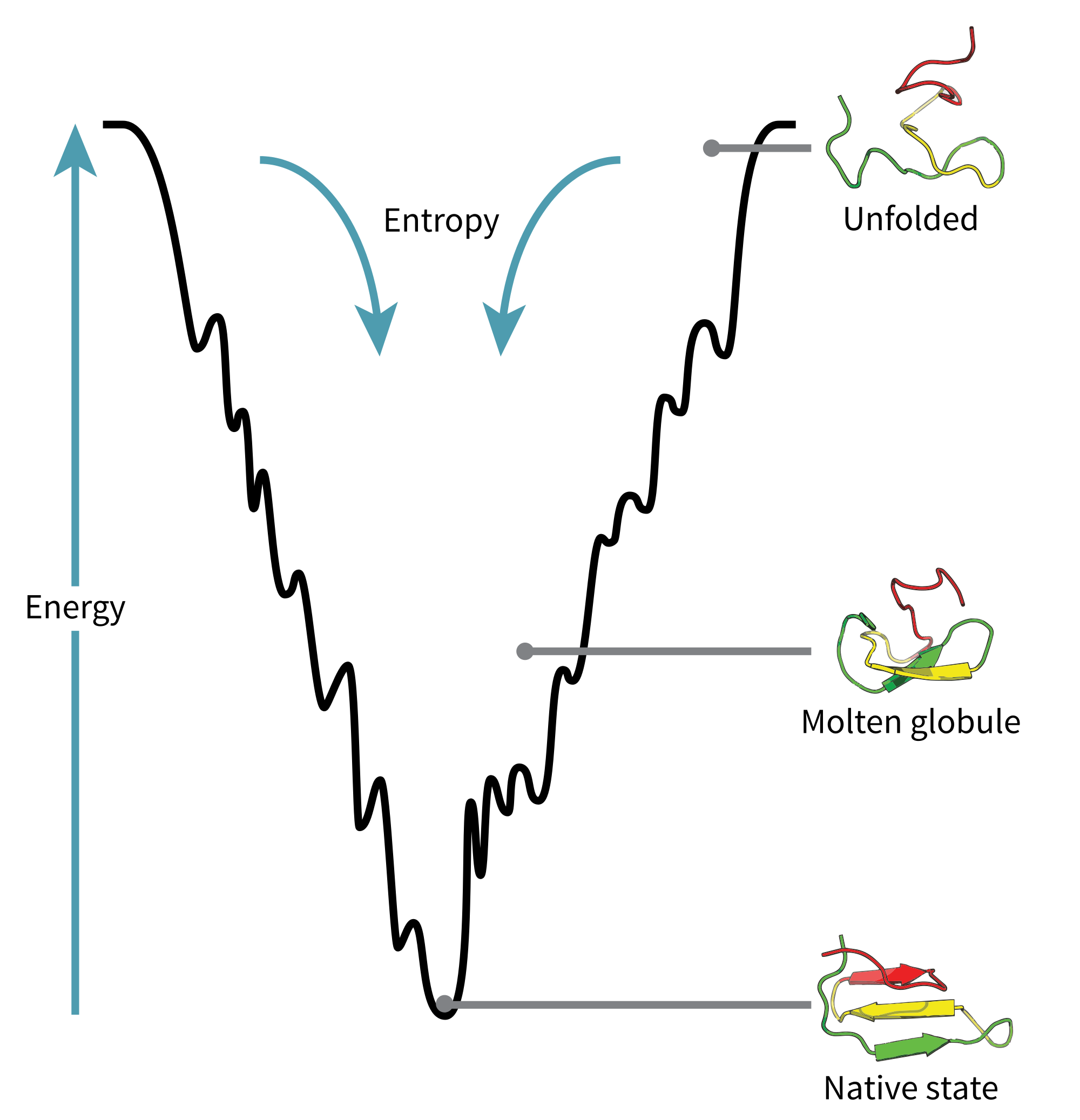}
\caption{ Diagram reflecting a rough (simplistic) 2D energy folding funnel of an arbitrary protein. At higher temperatures and energies the protein becomes denatured (unfolded) and entropy dominates. As $T \rightarrow 0$ the protein forms a molten globule and then enters the near native region. The native state is located exactly at the minimum of the energy folding funnel. The transition temperature $T_{C}$ can be located anywhere between the molten globule and native state. \label{fig:funnel}}
\end{figure}

The specific heat capacity for all sequences change drastically as the temperature passes through $T_{C}$ to lower values. The gradient of the observable at $T < T_{C}$ is of opposite sign to that at $T > T_{C}$, also the absolute magnitude of the gradient is larger due to the fast rate of decline as $T \rightarrow 0$. This behaviour reflects the chain attaining highly compact configurations in the near native region which have energy values existing in deep wells on the energy landscape (see figure \ref{fig:funnel}).

\begin{figure}[h]
\centering
\fbox{\subcaptionbox{$C_V/N$ for 2D100a found here.}[7.5cm]{
\centering
\includegraphics[scale=0.4]{100aheat}}}
\fbox{\subcaptionbox{$C_V/N$ for 2D100b found here.}[7.5cm]{
\centering
\includegraphics[scale=0.4]{100bheat}}}
\fbox{\subcaptionbox{$C_V/N$ for 2D100a found by Wust and Landau \cite{oplandau}}[7.5cm]{
\centering
\includegraphics[scale=0.5]{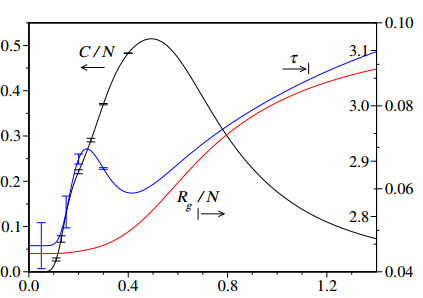}}}
\fbox{\subcaptionbox{$C_V/N$ for 2D100b foud by Wust and Landau \cite{oplandau}}[7.5cm]{
\centering
\includegraphics[scale=0.5]{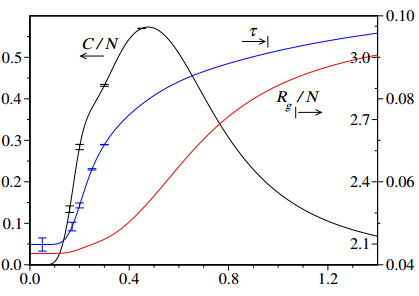}}}
\caption{Notice the similarity in values and qualitative behaviour. One noticeable difference is the 'rough' quality the $C_{V}/N$ from Wust and Landau has in the native region whereas my results are smoother. This could reflect that Wust and Landau ran their simulations for longer and hence sampled the conformational space in the native region more thoroughly. \label{fig:100compare}}
\end{figure}

The heat capacities for 2D100a and 2D100b are very similar (as with all their observables) which is not surprising as they share the same length and have similar H rations (0.55 and 0.56 respectively). For comparison these two curves are compared with those found by Wust and Landau \cite{oplandau} in figure \ref{fig:100compare}.

\newpage

The transition temperatures differed between the sequences but no accurate relationship could be deciphered between these temperatures and the chain lengths or hydrophobic ratios (see appendix \ref{sec:D} for table).

For the internal energy at low temperatures the observable continues to decrease and converge to a very small value. This is due to minimal thermal agitation and the compactness of structures in the native region.

Interestingly the free energy for sequences 2D64, 2D85, 2D100a and 2D100b all decrease with decreasing temperature past the critical temperatures, for example see figure \ref{fig:64obser}. This contrasts with the free energies of 2D50 and 2D60 which show the free energy still gradually increasing (as with 2D50) or practically constant (as with 2D60). This could be due to the quality of convergence of the simulations. 2D50 and 2D60 converged very well and hence as explained in section \ref{sec:wlmc} this relates to more accurate results. The free energies of 2D50 and 2D60 do align with expectations that the capacity to perform work increases with decreasing temperature, even in the $T < T_{C}$ region.

The entropy of 2D60 reaches a peak just beyond $T_{C} = 0.42$ and significantly drops in the low temperature region, this reflects the fast collapse of the protein chain into a compact native structure which has very low degeneracy. For 2D50 the entropy drops but not as rapidly as with 2D60, this could be due to differences in the folding funnel for the respective sequences. All the other 2D benchmark sequences, while qualitatively showing physically acceptable behaviour, had their entropies go below 0 as soon as the temperature passed from high-T through $T_{C}$ to low-T. This cannot reflect the intrinsic physics since $S/N < 0$ violates Boltzmann's entropy formula:

\begin{equation}
S = k_{B} \cdot ln [W]
\end{equation}

where $k_{B}$ is Boltzmann's constant and $W$ is the number of microstates. It is unreasonable to conclude that the entire model used here is now considered redundant because of this violation, it is simply a matter of inadequate sampling for these sequences. More computational time and effort will lead to better convergence and more realistic observables. 

In general the results for 2D benchmark (H)(P) sequences found here have unearthed the denatured $\rightarrow$ globule $\rightarrow$ native state transition as shown through the computation of thermodynamic observables. Whilst some observables 'broke' down in the low-T region beyond the critical temperature it can be corrected through more computational effort. The results of 2D60 are exemplary and characterise the thermodynamics of this sequence and lattice protein folding behaviour extremely well.

\newpage

\begin{center}
\textbf{ISAWs}
\end{center}

The specific heat capacity $C_{V}/N$ for ISAWs of length 25, 64 and 100 shown in figure \ref{fig:isawheat} shows only length specific behaviour in the temperature region of $0.1 < T < 0.5$. Beyond this range the heat capacity for the three lengths show universal behaviour. Stronger peaks can be seen with longer lengths of the homo-polymer, this is explained due to the fact that as $N \rightarrow \infty$ the psuedo phase transition resembles a real phase transition. The growth in magnitude of the observable is due to the significant increase in H-H contacts. 

The most accurate results obtained in this work was for the $L=25$ ISAW (see table \ref{table:isawmodi}), the modification factor reduction scheme seems to have taken on $1/t$ functionality, due to the rapid coverage of conformational space of a short chain.

For lengths 64 and 100 the peak widths at the transition temperature are less than their (H)(P) sequence counterparts. This can be explained via the folding funnel, the globule $\rightarrow$ native area on the folding funnel will be deeper and smoother for ISAWs than for protein (H)(P) chains.

The internal energy for ISAWs behaves similarly to that of protein sequences with a gradual decline from high-T to $T_{C}$ (with a faster rate closer to $T_{C}$) then a convergence to a small value in the native region. Interestingly $U/N$ is greater for the shortest ISAW in the native region but smallest in the $T > T_{C}$ region.

A comparison that has not been made in the literature is between the internal energy found in WLS and with the genus/energy ratio in a simulation conducted to study the topology of pseudoknotted homopolymers \cite{topoisaw}. The genus can be defined as the minimum number of handles the disk should have in order that all the cords are not intersecting\footnote{Definition: The genus of a connected, orientable surface is an integer representing the maximum number of cuttings along non-intersecting closed simple curves without rendering the resultant manifold disconnected. \cite{genus}}  (see \cite{topoisaw} for clarity). 

\begin{figure}[h]
\centering
\includegraphics[scale=0.5]{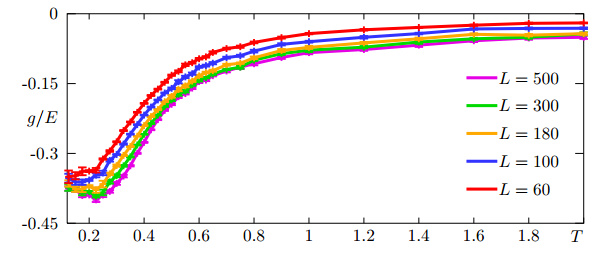}
\caption{The ratio genus/energy of a homopolymer on a cubic lattice, as a function of T, at different lengths of the chain. Thanks go to the authors of \cite{topoisaw}. \label{fig:genus}}
\end{figure}

One should compare the behaviour of $g/E$ and $U/N$ in figures \ref{fig:genus} and \ref{fig:isawinternal} respectively, they share similar behaviour especially in the low-T region. This comparison both reflects the tendency of the polymer chain to extend out into coil and 'string-like' structures with increasing temperature. I expect the genus will also share similar qualitative behaviour as the entropy also.

\section{Conclusions}

I have presented here a fully blind and straight forward parallel Monte Carlo scheme using Wang Landau sampling and the trajectory swapping method. Applying the trajectory swapping method to the problem of dynamical trapping and enhancing the efficiency of traversing conformational space to the HP model of lattice proteins has been successful. This has not been conducted within the existing literature. The development of the unique FRW trial move and its inclusion with pull, pivot, kink, pivot and bond re-bridging has enabled a different trial move set to complement WL sampling and to compare with the original work or Wust and Landau \cite{oplandau}. The trial move set is an important element of this Monte Carlo method and a huge amount of work was put in at the start of this project to develop original algorithms to implement trial moves which are described briefly in the literature. A thorough explanation and description of these moves has been presented to bridge the gap for those students and scientists willing to join the field of lattice polymer simulation.

Whilst WL is a powerful and generic Monte Carlo scheme to estimate thermodynamic behaviour its limitations were realized and an attempt to improve on the WL scheme consisted in implementing the 1/t algorithm. In this scheme it is difficult to ensure that the modification reduction operation takes on 1/t functionality, since the rate of coverage of conformational space differs for chain length and type. To ensure its regular success it will take committed tinkering of the Monte Carlo time for each sequence run. 

Thermodynamic quantities were successfully computed for typical 2D benchmark sequences, some more accurate than others, which revealed the intrinsic folding and un-folding behaviour in the $T>T_{C}$ and $T \leqslant T_{C}$. These computed observables also showed the existence of a denatured $\rightarrow$ globule $\rightarrow$ native structure pseudo phase transition. These results confirmed physical expectations and conclusions from related works. The amount of computational time and effort needed for longer sequences was under appreciated and in future it is fairly easy to obtain accurate results for these sequences (just run the simulations for a longer duration of time). 

A successful native state search for 2D50, 2D60 and 2D64 was conducted. The native structure for 2D64 found here resonated with those found in other works and confirmed the expectation that protein sequences prefer to fold into particular shapes with a stable hydrophobic core. While the native states of 2D85, 2D100a and 2D100b was not attained this method came very close in a seemingly short amount of time. The difficulty in accessing these low temperature energy states for a simple lattice model emphasizes the challenge in protein structure prediction and sampling.

This Monte Carlo scheme was also applied to lattice homopolymers and their thermodynamic behaviour was also successfully investigated, a connection to a previously unrelated observable (g/E) and 'classical' thermodynamic observables was made. 

Overall I believe this project fulfils the aims that were set out in section \ref{sec:intro}.

\section{Areas for Future Work}

To begin simple with any new area and problem of science is essential. One first asks simple questions and with certain answers one can then ask more subtle questions to gather detailed knowledge and understanding of the problem at hand. This project is the simple beginnings in exploring the physics of protein folding and lattice polymer dynamics. There are many ways one can further expand on the work conducted here. I will mention but a few.

Firstly one could try and implement the 1/t algorithm for the modification factor reduction successfully and try and find a way to code it such that it will adjust its definition of MC time so that the WL sampler will always converge asymptotically to the correct DOS. This is non-trivial and potentially time consuming, however very rewarding and ground breaking if it is done successfully.

A trivial expansion of this work is to increase the dimensions to 3 and investigate benchmark protein sequences. This would require the modification of the trial move algorithms and the lattice system. Though natural as this path is, there are more interesting pathways one could take since 3D benchmark sequences are already very well investigated. This however needs to be done at some point.

Within the same lattice dimensions and sequences used here one could study behaviour using variants of the Hamiltonian function (see equation \ref{eq:Hamiltonian}). There are various HP matrices which could be investigated using this methodology for the first time (see \cite{propre} (Oct 2015) for further details). Also variants of the HOP model \cite{HOP} could be generated and investigated and compare which HP energy function/matrix produces thermodynamic observables that best mimic the globular phase transitions seen in real proteins. 

The most interesting expansion (or deviation?) of this work would be to develop a continuous model of the HP model using rotary degrees of freedom and simulate it using the recently outlined LLR (logarithmic linear routine) method for computing the DOS (see \cite{LLR} for a detailed explanation and application). The trial move set used here could be assimilated into this continuous model. This would be novel work and it would be a great chance to compete with Wang-Landau sampling for the supreme algorithm for polymer and protein simulations. 
\clearpage
\section{Acknowledgements}
For fruitful and useful conversations on general physics theory and computation I would like to thank Professor Simon Hands and Dr. Edward Bennett. I also thank the Department of physics for granting me access to Vivian room 606 which has been my workstation throughout this project and the use of the supercomputer which has been essential.

I would also like to mention Professor Adi Armoni and Dr. Maurizio Piai for discussions on my future in physics and for playing a part in my acceptance at UCL. This has helped with keeping me motivated and to enjoy the journey of research.

A huge thank you to my project supervisor Professor Biagio Lucini who has offered this unique opportunity to work on a problem in a growing field and for his expert guidance.

Many thanks to the regular MPHYS visitors of 606 who helped generate an stimulating and lively environment to work in.
\clearpage

\begin{appendices}
\section{Amino acid HP table}
\label{sec:A}
\begin{center}
\begin{tabular}{|c|c|c|}
\hline
{\emph{AMINO ACID}} & {\emph{CODE}} & {\emph{H/P}} \\
[0.5ex]
\hline\hline
Alanine & A & H \\
Arginine & R & P \\
Asparagine & N & P \\
Aspartic Acid & D & P \\
Asparagine or Aspartic Acid & B & P \\
Cysteine & C & P \\
Glutamine & Q & P \\
Glutamic Acid & E & P \\
Glutamine or Glutamic Acid & Z & P \\
Glycine & G & P \\
Histidine & H & P \\
Isoleucine & I & H \\
Leucine & L & H \\
Lysine & K & P \\
Methionine & M & H \\
Phenylalanine & F & H \\
Proline & P & H \\
Serine & S & P \\
Threonine & T & P \\
Tryptophan & W & H \\
Tyrosine & Y & P \\
Valine & V & H \\
\hline\hline
\end{tabular}
\end{center}

\section{Preliminary Testing Results}
\label{sec:B}

\textbf{(These results were obtained during the initial procedures of the simulation development)}

After testing the trial move sets and devising an energy computing routine I decided to see whether my program could produce the native configurations of chains with $N_{monomers} < 20$.  
\subsection{Trial Move Prioritising}

Since, as hinted at in section \ref{sec:trialmove}, pivot moves will have a smaller acceptance probability than performing a pull move (especially for longer chains in more compact configurations)  so it was reasonable to impose that more pull moves were conducted on average than pivot moves. Pivot moves have the potential to drastically change the global configuration compared to pull and kink flip moves, hence conducting a sufficient amount of this move will enable rapid coverage of configuration space (see \cite{oplandau} for their implementation too) which might outweigh the low acceptance rate. Kink flip moves are handy for performing tiny movements in configuration space since they change only one monomer. In these results pull, kink flip and pivot moves were conducted 60$\%$, 15$\%$ and 25$\%$ of the time respectively.

  The probabilities assigned to this exact procedure was arbitrary and at best simple guess work, however in the future a more systematized approach will be adopted to ensure an efficient simulation.

\subsection{Energy Scoring}

The scoring system is simple: If the total energy of the configuration is less than the previous \textit{known} minimum energy (which is = 0 to begin with) then set that as the new minimum energy. The configuration related to the minimum energy is then printed to file 'native.txt' which stores the coordinates of the monomers so that the configuration can be drawn either manually or via another program. 

\begin{figure}[h]
\centering
\fbox{\includegraphics[scale=0.5]{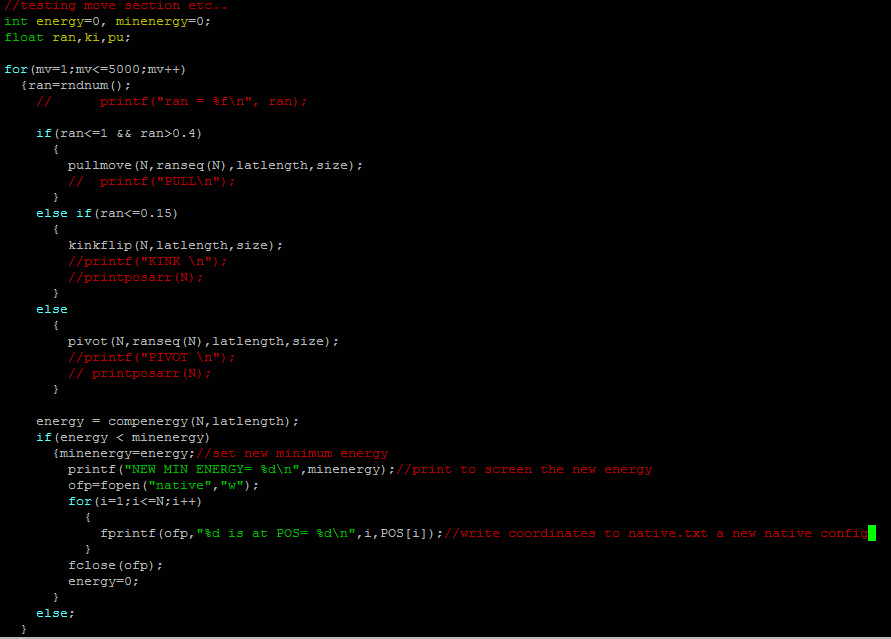}}
\caption{The routine in main which attempts 5000 moves on the chain recording minimum energy configurations. }
\label{fig:preroutine}

\end{figure}

The computer routine for the following results is shown in figure \ref{fig:preroutine}. In all cases the chain starts out as a horizontal linear one.

\newpage
\newpage

\subsection{Results}
In this section results are presented for short sequences of proteins which I have created for illustration purposes. The sequence name e.g. 2D7A represents the lattice dimension (2D), $N_{monomers}=7$ and a letter 'A' signifying its unique HP sequence. The results consist of the lowest minimum energy value and the chain diagram derived from the coordinates.

For all runs of the simulation the seed $\#$ will be specified.

\subsubsection*{2D7A}

2D7A has HP sequence: (HHPHHPH) and the results for 5 different seeds are presented in table \ref{table:2d7a}. 

\begin{table}[h]
\centering
\begin{tabular}{|| c | c | c ||}
  Seed $\#$ & $E_{min}$ & Configuration Type \\
 \hline\hline
 7412 & -2 & (i)   \\
 \hline
 8293 & -2 & (ii)   \\
 \hline
 2823 & -2 & (i)  \\
 \hline
 6902 & -2 & (ii)  \\
 \hline
 9382 & -2 & (iii) \\
 
\end{tabular}
\caption{Configuration type (i), (ii) and (iii) are shown in figure \ref{fig:2d7a}. \label{table:2d7a}}
\end{table}

The fact 3 distinct types of configuration were found only for 5 different seeds after 5000 attempted moves reflects the degeneracy of this short sequence with more (H) than (P) monomers.

\begin{figure}[h]
\begin{center}
\fbox{\begin{tikzpicture}[xscale=0.45,yscale=0.45]

\draw [ultra thick](0,0) -- (1,0);
\draw [ultra thick](1,1) -- (2,1);
\draw [ultra thick](1,1) -- (1,0);
\draw [dashed,ultra thick][green](2,1) -- (3,1);
\draw [ultra thick](3,1) -- (3,0);
\draw [ultra thick](3,0) -- (2,0);
\draw [ultra thick](2,1)--(2,0);
\draw [dashed,ultra thick][green](2,0) -- (1,0);
\node () at (2,-1) {(i)};
\draw [black,fill=black](0,0) circle [radius=0.20];
\draw [black,fill=black](1,0) circle [radius=0.20];
\draw [black,fill=black](2,0) circle [radius=0.20];
\draw [black,fill=white](3,0) circle [radius=0.20];
\draw [black,fill=black](3,1) circle [radius=0.20];
\draw [black,fill=black](2,1) circle [radius=0.20];
\draw [black,fill=white](1,1) circle [radius=0.20];

\draw [ultra thick](7,-1) -- (7,0);
\draw [ultra thick](7,1) -- (8,1);
\draw [ultra thick](7,1) -- (7,0);
\draw [dashed,ultra thick][green](2,1) -- (3,1);
\draw [ultra thick](9,1) -- (9,0);
\draw [ultra thick](9,0) -- (8,0);
\draw [ultra thick](8,1) -- (9,1);
\draw [dashed,ultra thick][green](8,1)--(8,0);
\draw [dashed,ultra thick][green](8,0) -- (7,0);
\node () at (9,-1) {(ii)};
\draw [black,fill=black](7,-1) circle [radius=0.20];
\draw [black,fill=black](7,0) circle [radius=0.20];
\draw [black,fill=black](8,0) circle [radius=0.20];
\draw [black,fill=white](9,0) circle [radius=0.20];
\draw [black,fill=black](9,1) circle [radius=0.20];
\draw [black,fill=black](8,1) circle [radius=0.20];
\draw [black,fill=white](7,1) circle [radius=0.20];

\draw [dashed,ultra thick][green](14,-1) -- (14,0);
\draw [dashed,ultra thick][green](14,0) -- (13,0);
\draw [ultra thick](13,0) -- (13,1);
\draw [ultra thick](14,1) -- (13,1);
\draw [ultra thick](14,1) -- (14,0);
\draw [ultra thick](14,0) -- (15,0);
\draw [ultra thick](15,0) -- (15,-1);
\draw [ultra thick](15,-1) -- (14,-1);
\node () at (14,-2) {(iii)};

\draw [black,fill=black](13,0) circle [radius=0.20];
\draw [black,fill=white](13,1) circle [radius=0.20];
\draw [black,fill=black](14,1) circle [radius=0.20];
\draw [black,fill=black](14,0) circle [radius=0.20];
\draw [black,fill=white](15,0) circle [radius=0.20];
\draw [black,fill=black](15,-1) circle [radius=0.20];
\draw [black,fill=black](14,-1) circle [radius=0.20];

\end{tikzpicture}}
\caption{The configurations (i),(ii) and (iii) found in the 2D7A runs. \label{fig:2d7a}}
\end{center}
\end{figure}
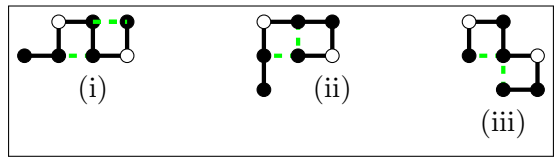

\subsubsection*{2D10A}

2D10A has HP sequence: (PHPHHPHPHH) and the results for 5 different seeds are presented in table \ref{table:2d10a}. The number of attempted moves was 5000.

\begin{table}[h]
\centering
\begin{tabular}{|| c | c ||}
  Seed $\#$ & $E_{min}$  \\
 \hline\hline
 9382 & -3\\
 \hline
 4510 & -3\\
 \hline
 6902 & -3\\
 \hline
 7523 & -3\\
 \hline
 7123 & -3\\
 
\end{tabular}
\caption{Results for 2d10a \label{table:2d10a}}
\end{table}

\subsubsection*{2D20A}

2D20A has HP sequence: (PHPHHPHPHHPHPHHPHPHH) (2 x 2D10A) and the results for 5 different seeds are presented in table \ref{table:2d20a}.  The number of attempted moves was 20000.

\begin{table}[h]
\centering
\begin{tabular}{|| c | c ||}
  Seed $\#$ & $E_{min}$  \\
 \hline\hline
 7123 & -7\\
 \hline
 6521 & -7\\
 \hline
 8715 & -8\\
 \hline
 3829 & -8\\
 \hline
 5782 & -8\\
 
\end{tabular}
\caption{Results for 2d20a \label{table:2d20a}}
\end{table}

\subsubsection*{2D50A} 

A simulation run on a real benchmark (2D50) was attempted using 100000 attempted moves. The HP sequence for 2D50 is: (HHPHPHPHPHHHHPHPPPHPPPHPPPPHPPPHPPPHPHHHHPHPHPHPHH). The minimal energy found, using seed $\#$ 6138, was = -13. This however is not the minimum found in simulations using EMC, SISPER, EES and FRESS which found $E_{min}$ = -21 (See \cite{presentation}).
\clearpage
\section{Replica Exchange Routine}
\label{sec:C}
\begin{footnotesize}
\begin{lstlisting}

  //======================= REPLICA EXCHANGING ===========
  if(mv%1000==0)
    {

      for(i=0;i<=(numprocs-1);i++)
        {if(myid==0)//if I am master thread
            {source=(int)(rndnum()*(numprocs-1));//printf("source= %d\n",source);
              dest=i;//printf("dest= %d\n",dest);
            }
          else;
          MPI_Barrier(MPI_COMM_WORLD);
          MPI_Bcast(&source,1,MPI_INT,0,MPI_COMM_WORLD);
          MPI_Bcast(&dest,1,MPI_INT,0,MPI_COMM_WORLD);
          if(source != dest)
            {
              if(myid==source)
                {
                  MPI_Send(POS,possize,MPI_INT,dest,1,MPI_COMM_WORLD);
                }
              else if(myid==dest)
                {
                  MPI_Recv(POS,possize,MPI_INT,source,1,MPI_COMM_WORLD,MPI_STATUS_IGNORE);
                }
              else;
            }
          else;
        }
    }
  else;
  //======================================================

\end{lstlisting}
\end{footnotesize}
\clearpage
\section{Critical temperatures for 2D benchmark sequences}
\label{sec:D}
\begin{table}[h]
\centering
\begin{tabular}{c | c | c }
Sequence & $T_{C} $ & (H) ratio \\
\hline \hline
2D50 & 0.576 & 0.5 \\
\hline
2D60 & 0.42 & 0.716 \\
\hline
2D64 & 0.39 & 0.656 \\
\hline
2D85 & 0.545 & 0.694 \\
\hline
2D100a & 0.535 & 0.55\\
\hline
2D100b & 0.577 & 0.56 \\
\hline
\end{tabular}

\end{table}
\end{appendices}

\clearpage
\section{\textcolor{gray}{References}}
\label{sec:ref}

\end{document}